\documentclass[article,11pt]{article}
\usepackage{jheppub}
\usepackage[T1]{fontenc}
\usepackage{amsmath,amssymb,amsfonts,graphicx,bm,bbm}
\usepackage{array}
\usepackage{float}
\usepackage{multirow,multicol}
\usepackage{subfigure}
\usepackage{titlesec}
\usepackage{lineno}

\providecommand{\tabularnewline}{\\}

\newcolumntype{x}[1]{
{\centering\hspace{0pt}}p{#1}}

\newcommand{\I}{\rm 1\kern-.24em l}

\def\lamDIS{ \Lambda_{ \gamma }^{ \rm DIS  } }
\def\lamIn{\Lambda_{ \gamma}^{ \rm Inel   } }
\def\lamEl{\Lambda_{ \gamma}^{ \rm El } }

\def\beq{\begin{equation}}
\def\eeq{\end{equation}}
\def\bea{\begin{eqnarray}}
\def\eea{\end{eqnarray}}
\def\bmat{\begin{pmatrix}}
\def\emat{\end{pmatrix}}
\def\to{\rightarrow}

\def\GeV{{\rm ~GeV}}
\def\TeV{{\rm ~TeV}}

\def\invfb{{\rm ~fb^{-1}}}
\def\invab{{\rm ~ab^{-1}}}

\def\lsim{\mathrel{\raise.3ex\hbox{$<$\kern-.75em\lower1ex\hbox{$\sim$}}}}
\def\gsim{\mathrel{\raise.3ex\hbox{$>$\kern-.75em\lower1ex\hbox{$\sim$}}}}

\newcommand{ \slashchar }[1]{\setbox0=\hbox{$#1$}   
   \dimen0=\wd0                                     
   \setbox1=\hbox{/} \dimen1=\wd1                   
   \ifdim\dimen0>\dimen1                            
      \rlap{\hbox to \dimen0{\hfil/\hfil}}          
      #1                                            
   \else                                            
      \rlap{\hbox to \dimen1{\hfil$#1$\hfil}}       
      /                                             
   \fi}                                             %

\title{\boldmath Heavy Majorana Neutrinos from $W\gamma$ Fusion at Hadron Colliders}

\author[a,b]{Daniel Alva,}
\author[b,c]{Tao Han,}
\author[b]{and Richard Ruiz}

\affiliation[a]{Centro de Ci\^{e}ncias Naturais e Humanas,\\ Universidade Federal do ABC, Santo Andr\'{e}, SP 09210-170, Brazil}
\affiliation[b]{Pittsburgh Particle Physics, Astronomy, and Cosmology Center (Pitt-PACC)\\
Department of Physics $\&$ Astronomy, University of Pittsburgh, Pittsburgh, PA 15260, USA}
\affiliation[c]{Korea Institute for Advanced Study (KIAS), Seoul 130-012, Korea}

\emailAdd{danalva@pitt.edu}
\emailAdd{than@pitt.edu}
\emailAdd{rer50@pitt.edu}

\abstract{Vector boson fusion processes become increasingly more important at higher collider energies and for probing larger mass scales due to collinear logarithmic enhancements of the cross section. 
In this context, we revisit the production of a hypothetic heavy Majorana neutrino ($N$) at hadron colliders. Particular attention is paid to the fusion process $W\gamma \to N\ell^{\pm}$. We systematically categorize the contributions from a photon initial state in the elastic, inelastic, and deeply inelastic channels.  Comparing with the leading channel via the Drell-Yan production $q \overline{q}'\rightarrow W^{*}\rightarrow N\ell^{\pm}$ at NNLO in QCD, 
we find that the $W\gamma$ fusion process becomes relatively more important at higher scales, surpassing the DY mechanism at
$m_{N} \sim 1~\text{TeV} \ (770~\text{GeV})$, at the 14 TeV LHC (100 TeV VLHC). 
We investigate the inclusive heavy Majorana neutrino signal, including QCD corrections, and quantify the Standard Model backgrounds at future hadron colliders. 
We conclude that, with the currently allowed mixing $\vert V_{\mu N}\vert ^2<6\times 10^{-3}$, 
a $5\sigma$ discovery can be made via the same-sign dimuon channel for $m_N = 530~(1070)$ GeV at the 14 TeV LHC (100 TeV VLHC) after 1 ab$^{-1}$.
Reversely, for $m_N = 500$ GeV and the same integrated luminosity, 
a mixing $\vert V_{\mu N}\vert^2$ of the order $1.1\times10^{-3}~(2.5\times10^{-4})$ may be probed.}

\arxivnumber{1411.7305}

\begin{document}

\begin{flushright}
PITT-PACC-1407
\end{flushright}
\hfill

\maketitle
\flushbottom

\section{INTRODUCTION}
\label{sec:intro}

The discovery of the Higgs boson completes the Standard Model (SM). Yet, the existence of nonzero neutrino masses 
remains one of the clearest indications of physics beyond the Standard Model 
(BSM)~\cite{Mohapatra:1998rq,Gluza:2002vs,Fukugita:2003en,Barger:2003qi,Eidelman:2004wy,GonzalezGarcia:2007ib,Mohapatra:2006gs,Strumia:2006db}
The simplest SM extension that can simultaneously explain both the existence of neutrino masses and their smallness,
the so-called Type I seesaw 
mechanism~\cite{Minkowski:1977sc,Yanagida:1979as,VanNieuwenhuizen:1979hm,Ramond:1979py,Glashow:1979nm,Mohapatra:1979ia,GellMann:1980vs,Schechter:1980gr,Shrock:1980ct,Schechter:1981cv}, 
introduces a right handed (RH) neutrino $N_{R}$.
Via a Yukawa coupling $y_{\nu}$,
the resulting Dirac mass is $m_{D}= y_{\nu}\langle \Phi\rangle$, where $\Phi$ is the SM Higgs SU$(2)_{L}$ doublet.
As $N_{R}$ is a SM-gauge singlet, 
one could assign $N_{R}$ a Majorana mass $m_{M}$ without violating any fundamental symmetry of the model. 
Requiring that $m_{M}\gg m_{D}$, the neutrino mass eigenvalues are
\begin{equation}
 m_{1}\sim m_D \frac{m_D}{m_M} \quad\text{and}\quad m_2 \sim m_M.
\end{equation}
Thus, the apparent smallness of neutrino masses compared to other fermion masses is due to the suppression by a new scale above the EW scale. Taking the Yukawa coupling to be $y_{\nu}\sim \mathcal{O}(1)$, the Majorana mass scale must be of the order $10^{13}$ GeV to recover sub-eV light neutrinos masses.
However, if the Yukawa couplings are as small as the electron Yukawa coupling, i.e., $y_{\nu}\lesssim\mathcal{O}(10^{-5})$, 
then the mass scale could be at $\mathcal{O}(1)$ TeV or lower \cite{ArkaniHamed:2000bq,Borzumati:2000mc,deGouvea:2005er,deGouvea:2006gz}. 

\begin{figure}[!t]
\centering
\subfigure[]{\includegraphics[width=.5\textwidth]{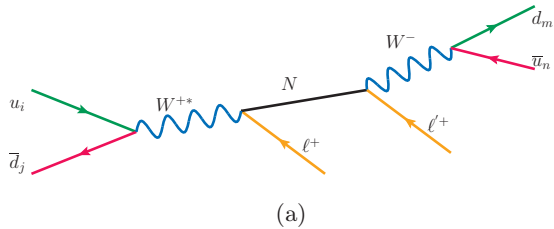}}
\caption{Diagram representing resonant heavy Majorana neutrino production through the DY process and its decay into same-sign leptons and dijet.
All diagrams drawn using JaxoDraw~\cite{Binosi:2003yf}.} 
\label{feynDYDecay.fig}
\end{figure}

Given the lack of guidance from theory of lepton flavor physics, 
searches for Majorana neutrinos must be carried out as general and model-independent as possible. 
Low-energy phenomenology of Majorana neutrinos has been studied in detail
~\cite{Keung:1983uu,Dicus:1991fk,Pilaftsis:1991ug,Datta:1993nm,deGouvea:2005er,deGouvea:2006gz,Han:2006ip,del Aguila:2007em,Atre:2009rg,Chao:2009ef,
AguilarSaavedra:2012fu,Das:2012ii,AguilarSaavedra:2012gf,Han:2012vk,Chen:2013foz,Davoudiasl:2014pya,Dev:2013wba}.
Studied first in Ref.~\cite{Keung:1983uu} and later in 
Refs.~\cite{Dicus:1991fk,Pilaftsis:1991ug,Datta:1993nm,Han:2006ip,del Aguila:2007em,Atre:2009rg},
the production channel most sensitive to heavy Majorana neutrinos $(N)$ at hadron colliders is the resonant Drell-Yan (DY) process,
\begin{equation}
 p p \rightarrow W^{\pm*} \to N ~\ell^{\pm},\ \ 
\ \ {\rm with}\ \ N \rightarrow W^{\mp} ~\ell^{'\pm},  \quad W^{\mp} \rightarrow j ~j,
 \label{dy.EQ}
\end{equation}
in which the same-sign dilepton channel violates lepton number $L$ by two units $(\Delta L=2)$; see figure~\ref{feynDYDecay.fig}.
Searches for Eq.~(\ref{dy.EQ}) are underway at LHC experiments~\cite{Chatrchyan:2012fla, ATLAS:2012yoa,Aaij:2012zr}. 
Non-observation in the dimuon channel has set a lower bound on the heavy neutrino mass of 100 (300) GeV for mixing 
$\vert V_{\mu N}\vert^{2} = 10^{-2~(-1)}$~\cite{ATLAS:2012yoa}.
Bounds on mixing from $0\nu\beta\beta$ \cite{Belanger:1995nh,Benes:2005hn} and EW precision data \cite{Nardi:1994iv,Nardi:1994nw,delAguila:2008pw,Antusch:2014woa} indicate that 
the 14 TeV LHC is sensitive to Majorana neutrinos with mass between 10 and 375 GeV after 100 $\invfb$ of data~\cite{Han:2006ip}.
Recently renewed interest in a very large hadron collider (VLHC) with a center of mass (c.m.)~energy about 100 TeV,
which will undoubtedly extend the coverage, suggests a reexamination of the search strategy at the new energy frontier.

Production channels for heavy Majorana neutrinos at higher orders of $\alpha$ were systematically cataloged in Ref.~\cite{Datta:1993nm}. 
Recently, the vector boson fusion (VBF) channel $W\gamma \to N\ell^{\pm}$ was studied at the LHC, and its $t$-channel enhancement 
to the total cross section was emphasized~\cite{Dev:2013wba}. 
Along with that, they also considered corrections to the DY process by including the tree-level QCD contributions to $N\ell^{\pm} +$jets.
Significant enhancement was claimed over both the leading order (LO) DY signal \cite{Han:2006ip,Atre:2009rg} and the expected next-to-next-to-leading order (NNLO) in QCD-corrected DY rate~\cite{Hamberg:1990np}, prompting us to revisit the issue.

\begin{figure}[!t]
\centering
\subfigure[]{\includegraphics[width=.44\textwidth]{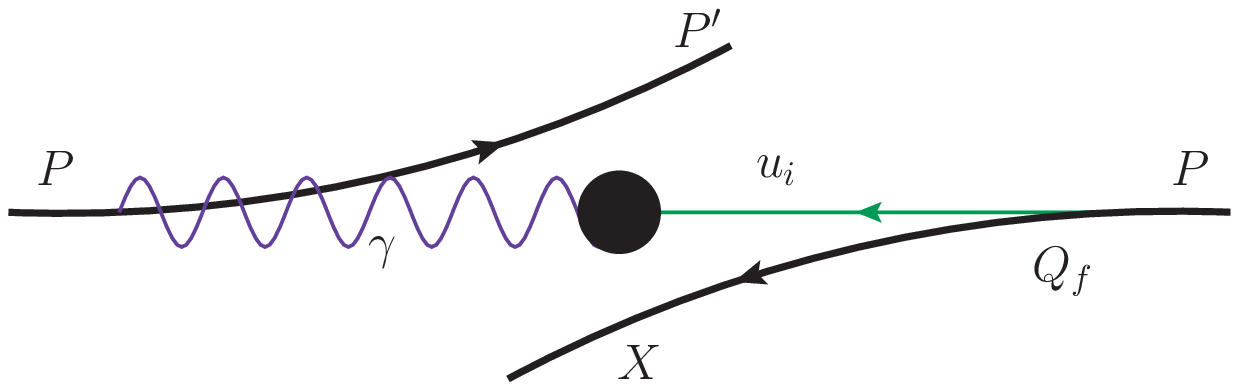}	\label{gamFromP.fig}}
\subfigure[]{\includegraphics[width=.52\textwidth]{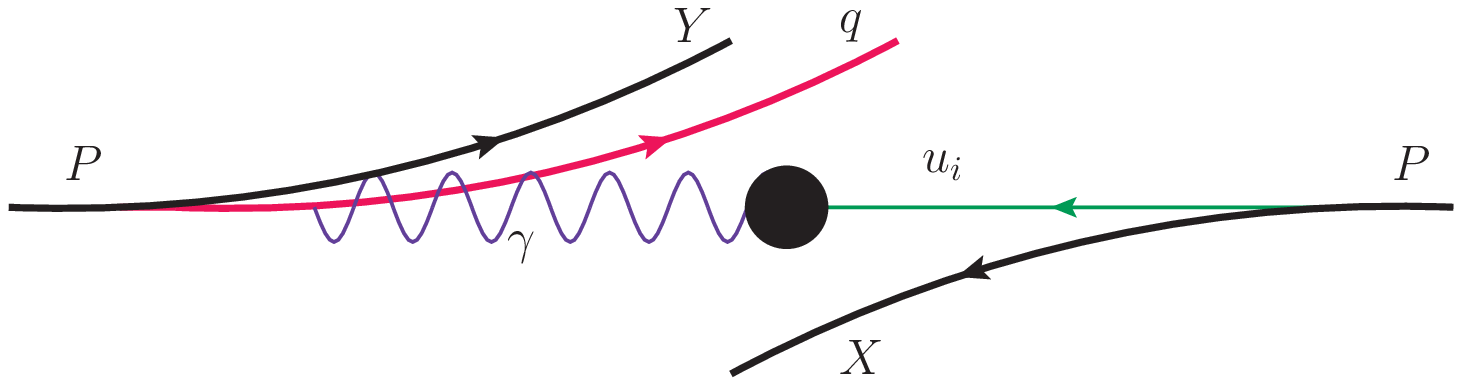}	\label{gamFromQ.fig}}
\caption{Diagrammatic representation of (a) elastic and (b) inelastic/deeply inelastic $\gamma p$ scattering.} 
\label{pascatt.FIG}
\end{figure}

We carry out a systematic treatment of the photon-initiated processes. 
The elastic emission (or photon emission off a nucleon) at colliders, as shown in figure~\ref{pascatt.FIG}(a), 
is of considerable interest for both 
SM~\cite{Budnev:1974de,Kniehl:1990iv,Block:1998hu,Gluck:2002fi,Cox:2005if,deFavereaudeJeneret:2009db,d'Enterria:2013yra} 
and 
BSM processes~\cite{Drees:1984cx,Drees:1994zx,Khoze:2001xm,Han:2007bk,Chapon:2009hh,Sahin:2010zr,Gupta:2011be,Sahin:2012zm,Sahin:2012mz},
and has been observed at 
electron~\cite{Abreu:1994vp}, hadron~\cite{Chatrchyan:2011ci,Chatrchyan:2013foa}, and lepton-hadron~\cite{Aid:1996dn,Adloff:2000vm} colliders.
The inelastic (collinear photon off a quark) and deeply inelastic (large momentum transfer off a quark) channels, as depicted in figure~\ref{pascatt.FIG}(b), may take over at higher momentum transfers~\cite{Drees:1988pp,Gluck:2002fi,Martin:2004dh}.
Comparing with the DY production $q q'\to W^{*}\to N\ell^{\pm}$, we find that the $W\gamma$ fusion process becomes relatively more important at higher scales,
taking over the QCD-corrected DY mechanism at {$\gtrsim 1\TeV~ (770\GeV)$} at the 14-TeV LHC (100 TeV VLHC).
At {$m_{N} \sim 375$} GeV, a benchmark value presented in \cite{Atre:2009rg}, 
we find the $W\gamma$ contribution to be about {20\%\ (30\%)} of the LO DY cross section.

NNLO in QCD corrections to the DY processes are well-known \cite{Hamberg:1990np} and 
the K-factor for the inclusive cross sections are about {$1.2-1.4~(1.2-1.5)$} at LHC (VLHC) energies.
Taking into account all the contributions, we present the state-of-the-art results for the inclusive production of heavy neutrinos in 14 and 100 TeV $pp$ collisions. 
We further perform a signal-versus-background analysis for a 100 TeV collider of the fully reconstructible and $L$-violating final state in Eq.~(\ref{dy.EQ}).
With the currently allowed mixing {$\vert V_{\mu N}\vert ^2<6\times 10^{-3}$,
we find that the $5\sigma$ discovery potential of Ref.~\cite{Atre:2009rg} can be extended to
{$m_N = 530~(1070)$} GeV at the 14 TeV LHC (100 TeV VLHC) after 1 ab$^{-1}$.
Reversely, for $m_N = 500$ GeV and the same integrated luminosity, 
a mixing $\vert V_{\mu N}\vert^2$ of the order {$1.1\times10^{-3}~(2.5\times10^{-4})$} may be probed.}
Our results are less optimistic than reported in \cite{Dev:2013wba}. 
We attribute the discrepancy to their significant overestimate of the signal in the tree-level QCD calculations, as quantified in section~\ref{sec:totIsPhoton}.

The rest of paper is organized as follows: 
In section~\ref{sec:inclusive}, 
we describe our treatment of the several production channels considered in this study,
address the relevant scale dependence, 
and present the inclusive $N\ell^\pm$ rate at the 14 TeV LHC and 100 TeV VLHC.
In section~\ref{sec:100TeV}, we perform the signal-versus-background analysis at a future 100 TeV $pp$ collider and report the discovery potential.
Finally summarize and conclude in section~\ref{sec:summary}. 
Appendices~\ref{sec:AppEl} and ~\ref{sec:AppIn} present the details of the photon PDF's for the elastic and inelastic channels, respectively. 
Appendix~\ref{sec:stats} gives our treatment for the Poisson statistics.


\section{HEAVY $N$ PRODUCTION AT HADRON COLLIDERS}
\label{sec:inclusive}


For the production of a heavy Majorana neutrino at hadron colliders, the leading channel is the DY process at order $\alpha^{2}$ (LO)~\cite{Keung:1983uu}
\begin{equation}
 q ~ \overline{q}' \rightarrow W^{\pm*} \to N ~\ell^\pm  .
 \label{dyDef.EQ}
\end{equation}
The QCD corrections to DY-type processes up to $\alpha_{s}^{2}$ (NNLO) are known \cite{Hamberg:1990np}, and will be included in our later analyses. 
Among other potential contributions, the next promising channel perhaps is the VBF channel \cite{Datta:1993nm}
\begin{equation}
W\ \gamma \to N\ \ell^{\pm}, 
\label{eq:fuse}
\end{equation}
due to the collinear logarithmic enhancement from $t$-channel vector boson radiation. 
Formally of order $\alpha^{2}$, there is an additional $\alpha$ suppression from the photon coupling to the radiation source. 
Collinear radiation off charged fermions (protons or quarks) leads to significant enhancement but requires proper treatment.
In our full analysis, $W$s are not considered initial-state partons~\cite{Datta:1993nm} 
and all gauge invariant diagrams, including non-VBF contributions, are included.

We write the production cross section of a heavy state $X$ in hadronic collisions as 
\begin{eqnarray}
  \sigma(pp\rightarrow X+ \text{anything}) &=& 
  \sum_{i,j}
  \int^{1}_{\tau_{0}}d\xi_{a}\int^{1}_{\tau_{0}\over \xi_a}d\xi_{b} 
  \left[f_{i/p}(\xi_a,Q_{f}^{2})f_{j/p}(\xi_b,Q_{f}^{2}) \hat{\sigma}(i j \rightarrow X) + (i\leftrightarrow j)\right]~~
    \label{factTheorem.EQ}
  \\
   &=& \int_{\tau_{0}}^{1} d\tau  \sum_{ij}\frac{d\mathcal{L}_{ij}}{d\tau}\  \hat{\sigma}(ij\rightarrow X).
\end{eqnarray}
where 
$\xi_{a,b}$ are the fractions of momenta carried by initial partons $(i,j)$, $Q_{f}$ is the parton factorization scale,
and $\tau = \hat s/s$ with 
$\sqrt{s}\ (\sqrt{\hat{s}})$ the proton beam (parton) c.m.~energy. For heavy neutrino production, the threshold is 
$\tau_{0} = m_{N}^{2}/s$. 
Parton luminosities are given in terms of the parton distribution functions (PDFs) $f_{i,j/p}$ by the expression
\begin{equation}
\Phi_{ij}(\tau) \equiv  \frac{d\mathcal{L}_{ij}}{d\tau}  = 
  \frac{1}{1+\delta_{ij}}  \int^{1}_{\tau} \frac{d\xi}{\xi} 
 \left[  f_{i/p}(\xi, Q_{f}^{2})f_{j/p}\left(\frac{\tau}{\xi},Q_{f}^{2} \right) + (i \leftrightarrow j) \right].
 \label{partonLumi.EQ}
\end{equation}
We include the light quarks $(u,d,c,s)$ and adopt the 2010 update of the CTEQ6L PDFs\cite{Pumplin:2002vw}.  
Unless stated otherwise, all quark (and gluon) factorization scales are set to half the c.m.~energy:
\begin{equation}
 Q_{f} = \sqrt{\hat s} / 2 .
 \label{QfScale.EQ}
\end{equation}
For the processes with initial state photons $(\gamma)$, their treatment and associated scale choices are given in section~\ref{sec:isPhoton}.

Our formalism and notation follow Ref.~\cite{Atre:2009rg}.
For the heavy neutrino production via the SM charged current coupling, 
the cross section is proportional to the mixing parameter (squared) between the mass eigenstate $N$ and the charged lepton $\ell\ (e,\mu,\tau)$. 
Thus it is convenient to factorize out the model-dependent parameter $\vert V_{\ell N}\vert^2$
\begin{equation}
 \sigma(pp\rightarrow N\ell^{\pm}) \equiv  \sigma_{0}(pp\rightarrow N\ell^{\pm}) ~\times~  \vert V_{\ell N}\vert^2,
 \label{bareProd.EQ}
\end{equation}
where $\sigma_{0}$ will be called the ``bare cross section''.
The branching fraction of a heavy neutrino to a particular lepton flavor $\ell$ 
is proportional to $\vert V_{N\ell}\vert^{2} / \sum_{\ell'}\vert V_{N\ell'}\vert^{2}$.
Thus for neutrino production and decay into same-sign leptons with dijet, it is similarly convenient to factorize out this ratio~\cite{Han:2006ip}:
\begin{eqnarray}
 \sigma(pp\rightarrow \ell^{\pm} \ell^{'\pm}+2j) 
 &\equiv& \sigma_{0}(pp\rightarrow \ell^{\pm} \ell^{'\pm} +2j) ~\times ~ S_{\ell\ell'},
  \label{bareXSecDef.EQ}\\
  S_{\ell\ell'} & = & \frac{\vert V_{\ell N}\vert^2\vert V_{\ell' N}\vert^{2}}{\sum_{\ell''} \vert V_{\ell'' N}\vert^{2}}.
 \label{sellell.EQ}  
\end{eqnarray}
The utility of this approach is that all the flavor-model dependence is encapsulated into a single, measurable number. Factorization into a bare rate and mixing coefficient holds generally for QCD and EW corrections as well.


\subsection{Constraints on Heavy Neutrino mixing}
\label{sec:constraints}

As seen above in Eq.~(\ref{bareProd.EQ}), one of the most important model-dependent parameters to control the signal production rate is the neutrino mixing $V_{\ell N}$. 
Addressing the origin of lepton flavor is beyond the scope of this study, so 
masses and mixing factors are taken as independent, phenomenological parameters.
We consider only the lightest, heavy neutrino mass eigenstate and require it to be kinematically accessible.
Updates on heavy neutrino constraints can be found elsewhere~\cite{Atre:2009rg,Han:2012vk,Dev:2013oxa}.
Here we list only the most stringent bounds relevant to our analysis.
\begin{itemize}
 \item \textbf{Bounds from $0\nu\beta\beta$}: 
 For heavy Majorana neutrinos with $M_i \gg 1\GeV$, the absence of $0\nu\beta\beta$ 
 decay restricts the mixing between 
 heavy mass and electron-flavor eigenstates~\cite{Belanger:1995nh,Benes:2005hn}:
 \begin{equation}
  \sum_{m'} \frac{\vert V_{em'}\vert^{2} }{M_{m'}} < 5 \times 10^{-5} \TeV^{-1}.
 \end{equation}
 \item \textbf{Bounds from EW Precision Data}: Mixing between a SM singlet above a few hundred GeV in mass and lepton flavor eigenstates 
is constrained by EW data~\cite{delAguila:2008pw}:
\begin{equation}
 \vert V_{\mu N}\vert^2  < 3.2 \times 10^{-3}, \quad 
 \vert V_{\tau N}\vert^2 < 6.2 \times 10^{-3} \quad \text{at}\quad 90\%~\text{C.L.}
\end{equation}
\end{itemize}
We consider the existence of only the lightest heavy Majorana neutrino,
which is equivalent to the decoupling limit where heavier eigenstates are taken to have infinite mass.
Thus, for representative neutrino masses
\begin{equation}
 m_N = 300~(500)~[1000]\GeV,
 \label{massParam.EQ}
\end{equation}
we use the following mixing coefficients
\begin{equation}
 \vert V_{e N}\vert^{2}    = 1.5~(2.5)~[5]\times 10^{-5}, \qquad
 \vert V_{\mu N}\vert^{2}  = 3.2 \times 10^{-3}, \qquad
 \vert V_{\tau N}\vert^{2} = 6.2 \times 10^{-3},
 \label{mixingParam.EQ}
\end{equation}
corresponding to a total neutrino width of
\begin{equation}
 \Gamma_{N} = 0.303~(1.50)~[12.3]\GeV.
  \label{widthParam.EQ}
\end{equation}
As $\Gamma_t / m_{N} \approx 0.1\% - 1\%$, the heavy neutrino resonance is very narrow and application of the narrow width approximation (NWA) is justified.
For $S_{\ell\ell}$, these mixing parameters imply
\begin{eqnarray}
 S_{ee} 	=  2.4~(6.6)~[26] \times 10^{-8} &\quad\text{for}\quad& m_N= 300~(500)~[1000]\GeV
 \label{see.EQ}
 \\
 S_{e\mu} 	= S_{\mu e} = 5.1~(8.5)~[17] \times 10^{-6} &\quad\text{for}\quad& m_N= 300~(500)~[1000]\GeV 
 \label{semu.EQ}
 \\
 S_{\mu\mu} 	= 1.1 \times 10^{-3} &\quad\text{for}\quad& m_N\in[100,1000]\GeV
 \label{smumu.EQ}
\end{eqnarray}
Though the bound on $\vert V_{eN}\vert$ varies with $m_N$, 
$S_{\mu\mu}$ changes at the per mil level over the masses we investigate and is taken as constant.
The allowed sizes of $S_{e\mu}$, $S_{\mu\mu}$, and $S_{\tau\ell}$
demonstrate the complementarity to searches for $L$-violation at $0\nu\beta\beta$ experiments afforded by hadron colliders.
To make an exact comparison with Ref.~\cite{Atre:2009rg}, we also consider the bound~\cite{Nardi:1994iv,Nardi:1994nw}
\begin{equation}
 S_{\mu\mu} \approx \frac{\vert V_{\mu N} \vert^4}{\vert V_{\mu N} \vert^2 } = \vert V_{\mu N} \vert^2 = 6\times 10^{-3}
\end{equation}
However, bare results, which are mixing-independent, are presented wherever possible.


\subsection{$N$ Production via the Drell-Yan Process at NNLO}
\begin{figure}[!t]
\begin{center}
\subfigure[]{\includegraphics[scale=1,width=.48\textwidth]{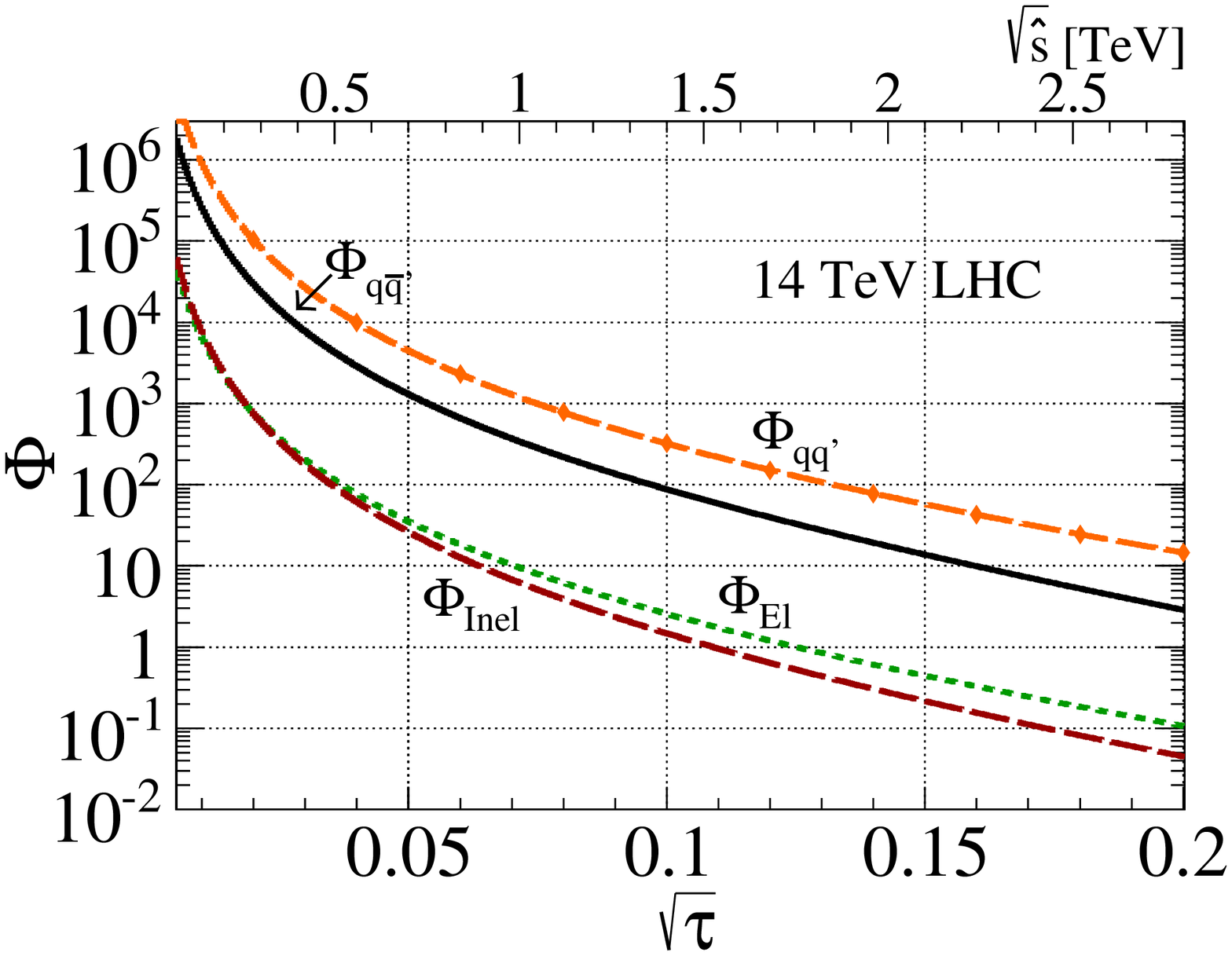}\label{lumilumi.fig}}
\subfigure[]{\includegraphics[scale=1,width=.48\textwidth]{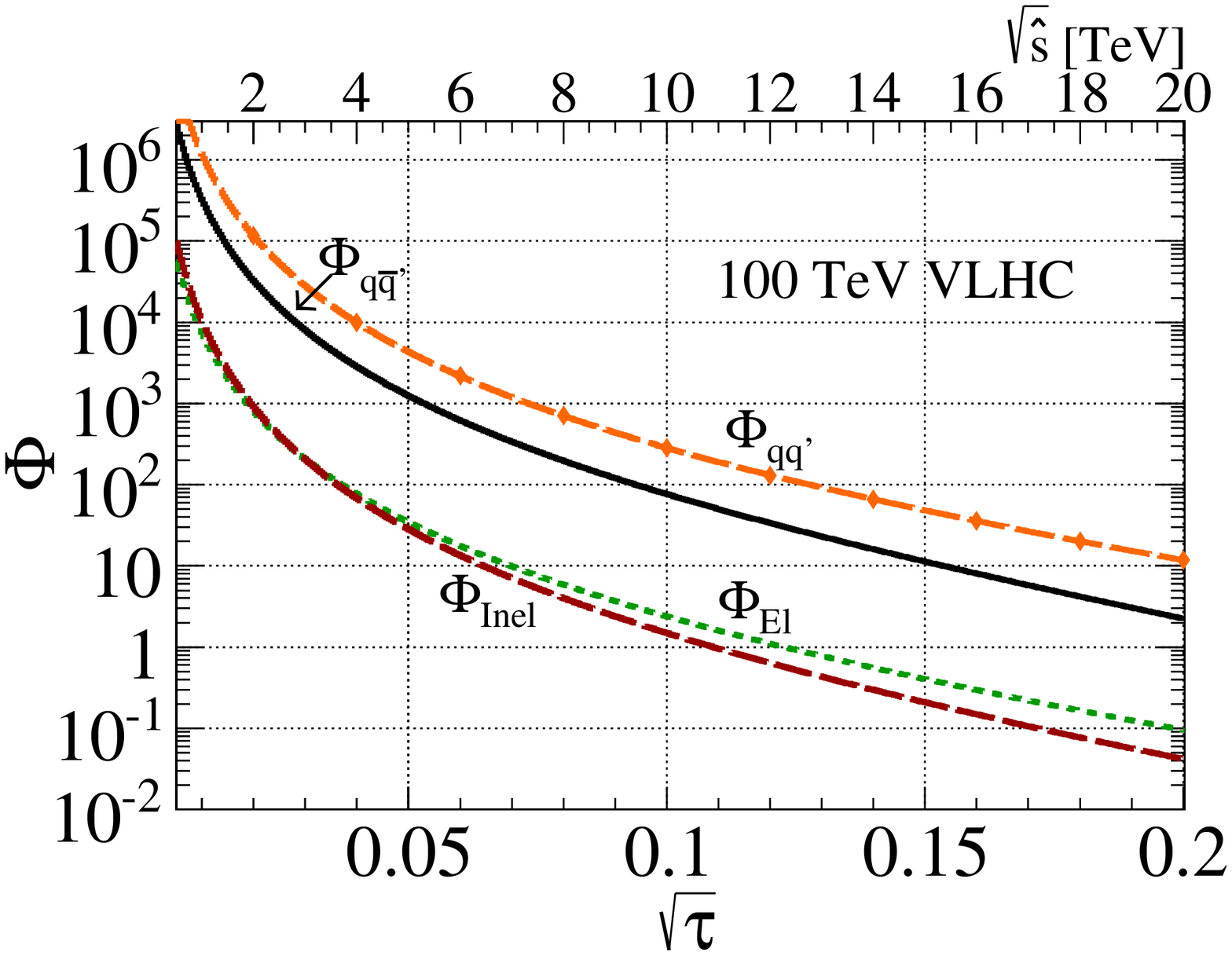}\label{lumilumi100TeV.fig}}
\vspace{.2in}\\
\subfigure[]{\includegraphics[scale=1,width=.48\textwidth]{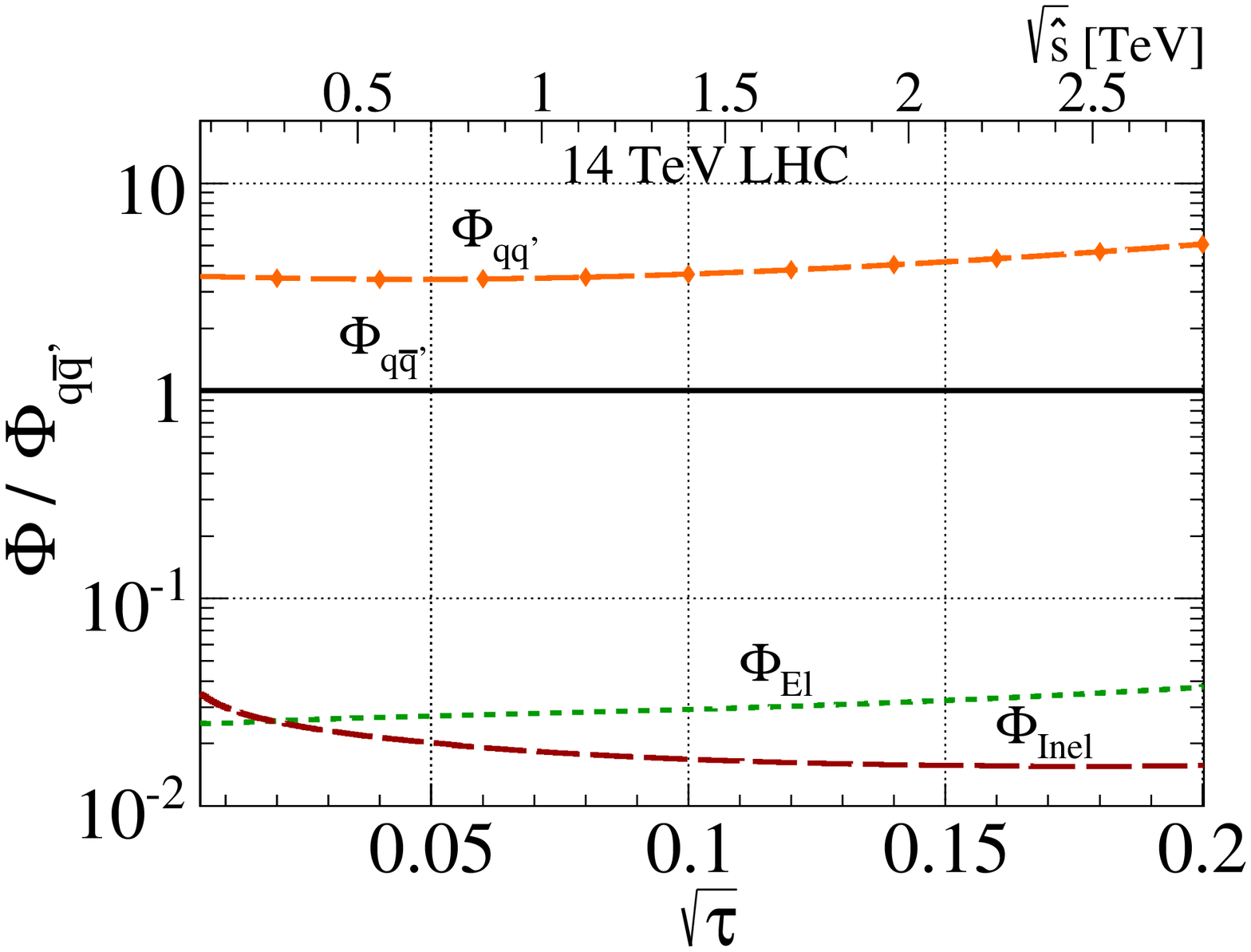}\label{lumiratio.fig}}
\subfigure[]{\includegraphics[scale=1,width=.48\textwidth]{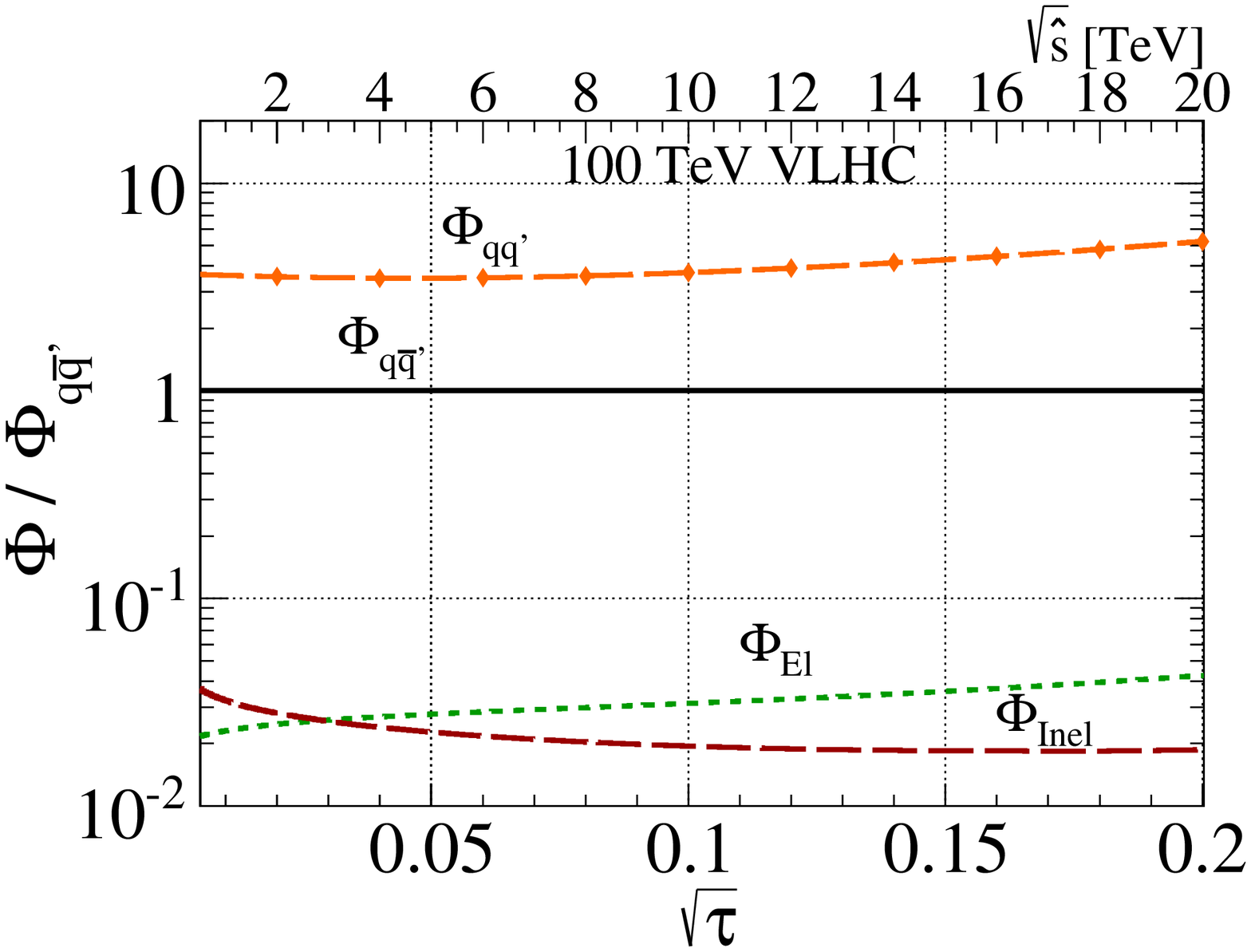}\label{lumiratio100TeV.fig}}
\caption{
Parton luminosities at (a) 14 TeV and (b) 100 TeV 
for the DY (solid), elastic (dot), inelastic (dash), and DIS (dash-diamond) $N\ell X$ processes; 
Ratio of parton luminosities to the DY luminosity in (c) and (d).
}
\label{lumi.fig}
\end{center}
\end{figure}

Before presenting the production cross sections, it is informative to understand the available parton luminosities $(\Phi_{ij})$ as defined in Eq.~(\ref{partonLumi.EQ}).
We show $\Phi_{\rm q\overline{q}'}$ versus $\sqrt \tau$ for $q\overline{q}'$ annihilation summing over light quarks ($u,d,c,s$) by the solid (black) curves in figures~\ref{lumilumi.fig} and \ref{lumilumi100TeV.fig} for the 14 TeV LHC and 100 TeV VLHC, respectively. 
The upper horizontal axis labels the partonic c.m.~energy $\sqrt{\hat{s}}$.
As expected, at a fixed $\sqrt{\hat{s}}$ the DY luminosity at 100 TeV significantly increases over that at 14 TeV. 
At $\sqrt{\hat{s}}\approx500\GeV~(2\TeV)$, the gain is a factor of {$600~(1.8\times10^3)$}, and the discovery potential of heavy Majorana neutrinos is greatly expanded.
Luminosity ratios with respect to $\Phi_{\rm q\overline{q}'}$ are given in figure~\ref{lumiratio.fig} and \ref{lumiratio100TeV.fig}, and will be discussed when appropriate.

Cross sections for resonant $N$ production via the charged current DY process in Eq.~(\ref{dy.EQ}) and shown in figure~\ref{feynDYDecay.fig}
are calculated with the usual helicity amplitudes at the LO $\alpha^2$. Monte Carlo integration is performed using CUBA~\cite{Hahn:2004fe}.
Results are checked by implementing the heavy Majorana neutrino model into FeynRules 2.0.6 \cite{Alloul:2013bka,Christensen:2008py} and MG5\_aMC@NLO 2.1.0 \cite{Alwall:2014hca} (MG5).
For simplicity, percent-level contributions from off-diagonal Cabbibo-Kobayashi-Maskawa (CKM) matrix elements are ignored and the diagonal elements are taken to be unity.
SM inputs $\alpha^{\rm \overline{MS}}(M_{Z})$, $M_{Z}$, and $\sin^{2}_{\rm \overline{MS}}(\theta_{W})$ 
are taken from the 2012 Particle Data Group (PDG)~\cite{Beringer:1900zz}.

\begin{table}[!t]
\caption{LO and NNLO cross sections for $pp\rightarrow W^{*}\rightarrow \mu^\pm \nu$ at 14 and 100 TeV with successive invariant mass cuts
using MSTW2008LO and NNLO PDF Sets.}
 \begin{center}
\begin{tabular}{|c|c|c|c||c|c|c|}
\hline\hline
$\sqrt{\hat{s}^{\min}}$	& 14 TeV LO [pb] & NNLO [pb] & $K$ & 100 TeV LO [pb] & NNLO [pb] & $K$ 
\tabularnewline\hline\hline
100 GeV 	& 152	& 209 	& 1.38	& 1150	& 1420	& 1.23	\tabularnewline\hline
300 GeV 	& 1.54	& 1.90	& 1.23	& 17.0	& 25.6	& 1.50	\tabularnewline\hline
500 GeV 	& 0.248	& 0.304	& 1.22	& 3.56	& 4.97	& 1.40	\tabularnewline\hline
1  TeV 		& 17.0 $\times 10^{-3}$ & 20.5 $\times 10^{-3}$ & 1.20	& 0.380	& 0.485	& 1.28	\tabularnewline\hline
\hline
\end{tabular}
\label{kFactor.TB}
\end{center}
\end{table}

We estimate the 14 and 100 TeV $pp$ NNLO $K$-factor\footnote{The $N^nLO$ $K$-factor is defined as $K = \sigma^{N^nLO}(N\ell) / \sigma^{LO}(N\ell)$, 
where $\sigma^{N^nLO}(N\ell)$ is the N$^n$LO-corrected cross section and $\sigma^{LO}(N\ell)$ is the lowest order $(n=0)$, or  Born, cross section.} 
by using FEWZ 2.1~\cite{Gavin:2010az,Gavin:2012sy} to compute the equivalent quantity for the SM process
\begin{equation}
pp\rightarrow W^{*}\rightarrow \mu^\pm {\nu}, 
\label{smDY.EQ}
\end{equation}
and impose only an minimum invariant mass cut, $\sqrt{\hat{s}^{\min}}$.
Because LO $N\ell$ production and Eq.~(\ref{smDY.EQ}) are identical DY processes (up mass effects) with the same color structure, 
$K$-factors calculated with a fixed $\hat{s}$ are equal. 

\begin{figure}[!t]
\begin{center}
\subfigure[]{\includegraphics[scale=1,width=.47\textwidth]{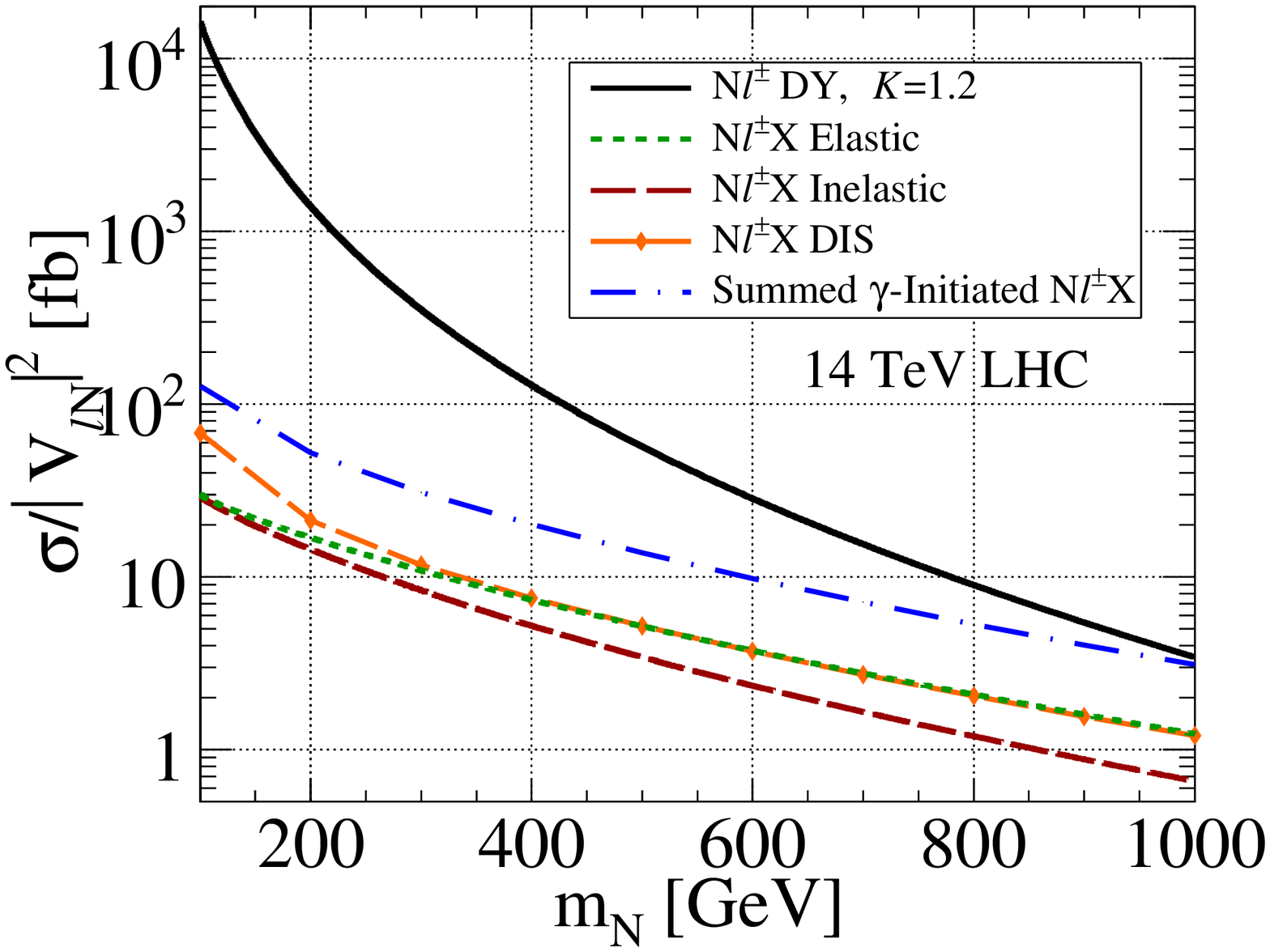}		\label{xsecComb.fig}}
\subfigure[]{\includegraphics[scale=1,width=.48\textwidth]{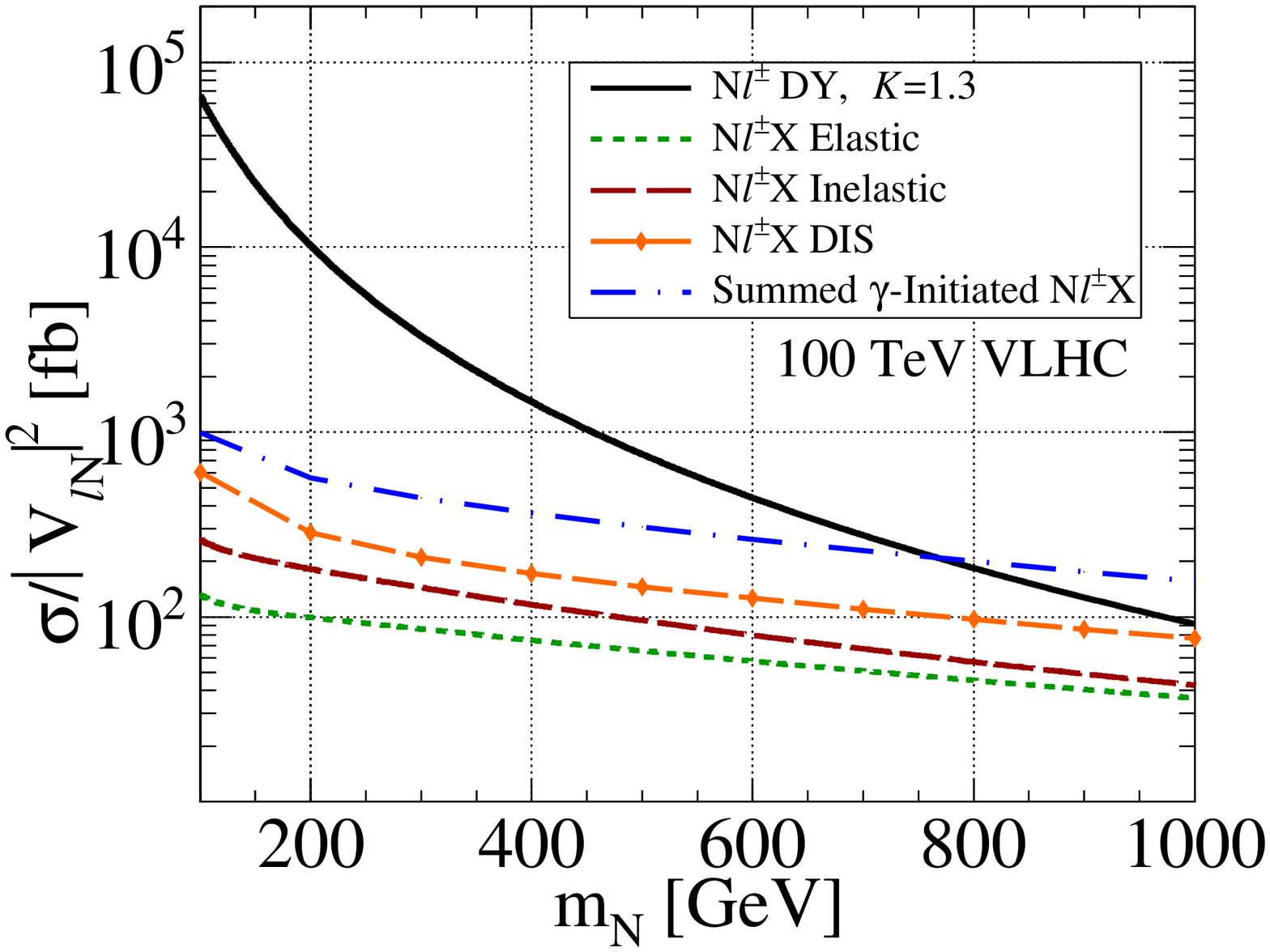}		\label{xsecComb100TeV.fig}}
\vspace{.2in}\\
\subfigure[]{\includegraphics[scale=1,width=.47\textwidth]{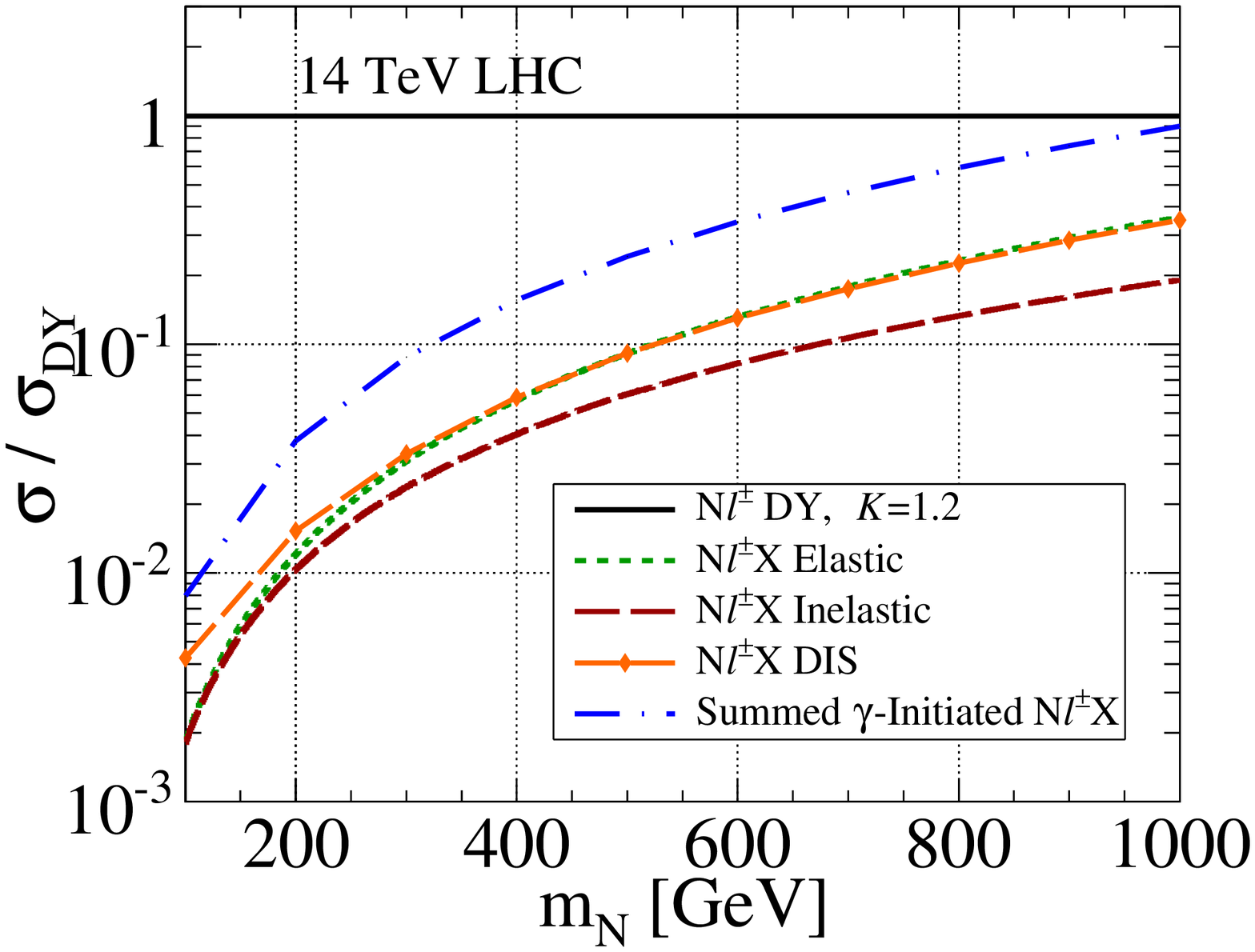}	\label{xsecRatio.fig}}
\subfigure[]{\includegraphics[scale=1,width=.48\textwidth]{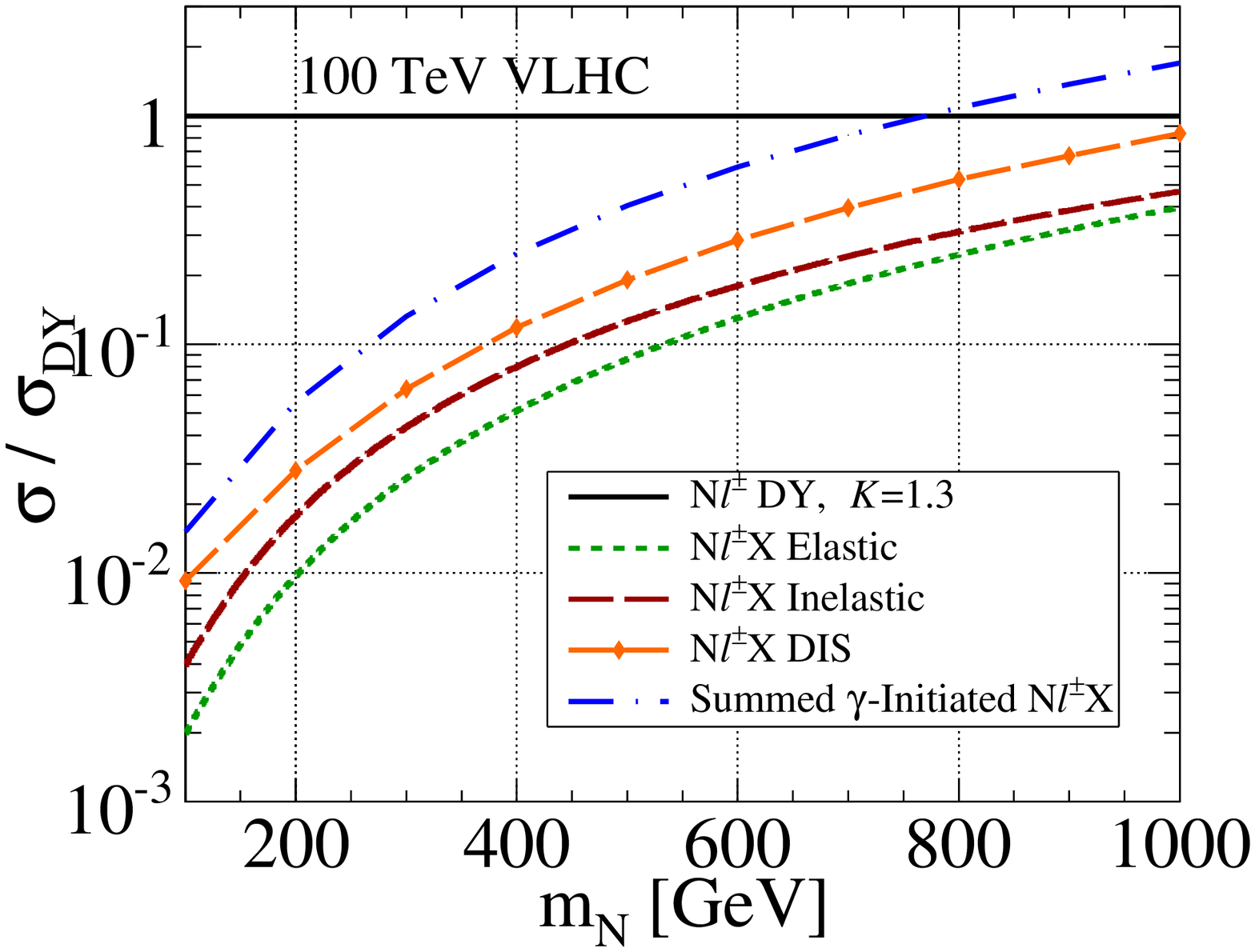}	\label{xsecRatio100TeV.fig}}
\end{center}
\caption{(a) 14 TeV LHC (b) 100 TeV VLHC $N\ell X$ cross section, divided by $\vert V_{\ell N}\vert^2$, as a function of the $N$ mass for 
the NNLO DY (solid), elastic (dot), inelastic (dash), DIS (dash-diamond), and summed $\gamma$-initiated (dash-dot) processes.
(c,d) Ratio of cross sections relative to NNLO DY rate.
} 
\label{xsec.fig}
\end{figure}

Table~\ref{kFactor.TB} 
lists\footnote{As no NNLO CTEQ6L PDF set exists, we have adopted the MSTW2008 series to obtain a self-consistent estimate of the NNLO $K$-factor.} 
the LO and NNLO cross sections as well as the NNLO $K$-factors for several representative values of $\sqrt{\hat{s}^{\min}}$.
At $\sqrt{\hat{s}^{\min}} = 1\TeV$, the QCD-corrected charged current rate can reach tens (several hundreds) of fb at 14 (100) TeV.
Over the range from $\sqrt{\hat{s}^{\min}}=100\GeV - 1\TeV,$ 
\begin{eqnarray}
 K &=& 1.20-1.38 \ \ {\rm at\ 14\ TeV},  \label{K.EQ}\\
   &=& 1.23 -1.50 \ \ {\rm at\ 100\ TeV}.
\end{eqnarray}
This agrees with calculations for similar DY processes~\cite{Nemevsek:2011hz,Chatrchyan:2012meb}.
We see that the higher order QCD corrections to the DY channel are quite stable, which will be important for our discussions in section~\ref{sec:isPhoton}.
Throughout the study, independent of neutrino mass, we apply to the DY-process a $K$-factor of
\begin{equation}
 K = 1.2~(1.3) \quad\text{for}\quad 14~(100)~\TeV.
\end{equation}
Including the QCD K-factor, 
we show the NNLO total cross sections [called the ``bare cross section $\sigma_0$'' by factorizing out $\vert V_{\ell N}\vert^2$ as defined in  Eq.~(\ref{bareProd.EQ})] as a function of heavy neutrino mass in figures~\ref{xsecComb.fig} and \ref{xsecComb100TeV.fig} for the 14-TeV LHC and 100-TeV VLHC, respectively. The curves are denoted by the (black) solid lines.
Here and henceforth, we impose the following basic acceptance cuts on the transverse momentum and pseudorapidity of the charged leptons for 14 (100) TeV, 
\begin{equation}
p_{T}^{\ell} > 10~(30)\GeV,\quad	\vert \eta^{\ell}\vert < 2.4~(2.5).
\label{regCutsLep.EQ}
\end{equation}
The motive to include these cuts is two-fold. 
First,  they are consistent with the detector acceptance for our future simulations and the definition of ``fiducial'' cross section.
Second, they serve as kinematical regulators for potential collinear singularities, to be discussed next.
The $p_T$ and $\eta$ criteria at 100 TeV follow the 2013 Snowmass benchmarks~\cite{Avetisyan:2013onh}.


\subsection{Photon-Initiated Processes}
\label{sec:isPhoton}

\begin{figure}[!t]
\begin{center}
\includegraphics[scale=1,width=.96\textwidth]{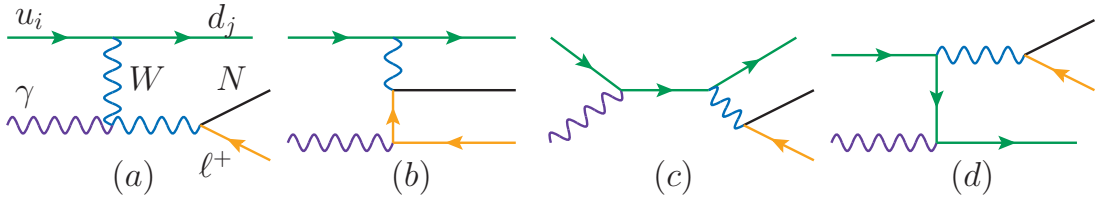}
\end{center}
\caption{Feynman diagrams for photon-initiated process $q\gamma \to N\ell^\pm q'$.} 
\label{feynQA.fig}
\end{figure}

After the dominant DY channel, VBF via $W\gamma$ fusion, as introduced in Eq.~(\ref{eq:fuse}), presents a promising additional contribution to the heavy $N$ production. We do not make any approximation for the initial state $W$ and treat its radiation off the light quarks with exact matrix element calculations.
In fact, we consistently treat the full set of diagrams, shown in figure~\ref{feynQA.fig}, for the photon-initiated process at order $\alpha^3$
\begin{equation}
 q ~ \gamma \rightarrow  N ~\ell^\pm ~q'.
 \label{pgammaDef.EQ}
\end{equation}
Obviously, diagrams  figure~\ref{feynQA.fig}(c) and (d) do not add to $W\gamma$ fusion and are just small QED corrections.\footnote{
Diagram \ref{feynQA.fig}(d) involves a collinear singularity from massless quark splitting.
It is unimportant for our current consideration since its contribution is simply a QED correction to the quark PDF.
For consistency and with little change to our results, $\lamDIS = 15\GeV$ [defined in Eq.~(\ref{qDISMinCut.EQ})] is applied as a regulator.}
Diagram figure~\ref{feynQA.fig}(b) involves a massless $t$-channel charged lepton. 
The collinear pole is regularized by the basic acceptance cuts in Eq.~(\ref{regCutsLep.EQ}). 
What is non-trivial, however, is how to properly treat initial-state photons across the different sources depicted in figure~\ref{pascatt.FIG}.
We now discuss the individual channels in detail. 

\subsubsection{Elastic Scattering: Intact Final-State Nucleons}
\label{sec:semi}
Here and henceforth, the virtuality for the incoming photon in $W\gamma$ fusion is denoted as $Q_\gamma>0$.
In the collinear limit that results in momentum transfers on the order of the proton mass or less, $Q^{2}_\gamma \lesssim m_{p}^{2}$,
initial-state photons are appropriately described as massless radiation by an elastic proton, 
i.e., does not break apart and remains as an on-shell nucleon, as indicated in figure~\ref{gamFromP.fig}.
To model this, we use the ``Improved'' Weizs\"acker-Williams approximation~\cite{Budnev:1974de}
and factorize the photon's collinear behavior into a structure function of the proton to obtain the elastic photon PDF $f_{\gamma/p}^{\rm El }$. 
In Eq.~(\ref{factTheorem.EQ}), this entails replacing one $f_{i/p}$ with $f_{\gamma/p}^{\rm El }$:
\begin{equation}
 f_{i/p}(\xi,Q_f^2) \rightarrow f_{\gamma/p}^{\rm El }(\xi).
 \label{elPDF.EQ}
\end{equation}
The expression for $f_{\gamma/p}^{\rm El }$, given in Appendix~\ref{sec:AppEl}, is dependent on a cutoff scale $\Lambda_{\gamma}^{\rm El }$, 
above which the description of elastic $p\rightarrow \gamma$ emission starts to break down.
Typically, the scale is taken to be $\mathcal{O}(m_{p}-2\GeV)$ 
~\cite{Budnev:1974de,deFavereaudeJeneret:2009db,d'Enterria:2013yra,Chapon:2009hh,Sahin:2010zr,Gupta:2011be,Sahin:2012zm,Sahin:2012mz}
but should be insensitive to small variations if an appropriate scale is chosen.
Based on analysis of $ep$ scattering at low $Q_\gamma$~\cite{Alwall:2004wk}, we take
\begin{equation}
 \Lambda_\gamma^{\rm El } = \sqrt{1.5 \GeV^{2}} \approx 1.22\GeV.
\label{lamEl.EQ}
\end{equation}
The scale dependence associated with $\Lambda_\gamma^{\rm El }$ is discussed in section~\ref{sec:scale}.

In figure~\ref{lumi.fig}, the elastic luminosity spectrum $(\Phi_{\rm El })$ is denoted by the (green) dot line.
For the range studied, $\Phi_{\rm El }$ is roughly {$2-4\%$} of the $q\bar q'$ DY luminosity at 14 and 100 TeV.

We calculate the matrix element for the diagrams in figure~\ref{feynQA.fig} in the same manner as the DY channel.
The results are checked with MG5 using the elastic, asymmetric $p\gamma$ beam mode.
In figures~\ref{xsecComb.fig} and \ref{xsecComb100TeV.fig}, 
we plot the bare cross section for the elastic process, denoted by a (green) dot line, as a function of neutrino mass.
The rate varies between {$1-30 ~(40-100)$} fb at 14 (100) TeV for $m_N = 100$ GeV$-$1 TeV.
As seen in figures~\ref{xsecRatio.fig} and \ref{xsecRatio100TeV.fig}, where the cross sections are normalized to the DY rate,
it reaches about {$30~(40)\%$} of the DY rate for large $m_N$.


\subsubsection{Inelastic Scattering: Collinear Photons From Quarks}
For momentum transfers above the proton mass, the parton model is valid.
When this configuration coincides with the collinear radiation limit, 
initial-state photons are appropriately described as being radiated by quark partons.
To model a quark splitting to a photon, we follow the methodology of Ref.~\cite{Drees:1994zx}
and use the (original) Weizs\"acker-Williams approximation~\cite{Williams:1934ad,vonWeizsacker:1934sx} 
to obtain the inelastic photon PDF $f_{\gamma/p}^{\rm Inel}$. 
Unlike the elastic case, factorization requires us to convolve about a splitting function.
The inelastic $N\ell^{\pm}X$ cross section is obtained by making the replacement in Eq.~(\ref{factTheorem.EQ})
\begin{eqnarray}
 f_{i/p}(\xi,Q_f^2) &\rightarrow & f_{\gamma/p}^{\rm Inel}(\xi,Q_\gamma^2,Q_f^2), 
 \\
 f_{\gamma/p}^{\rm Inel}(\xi,Q_\gamma^2,Q_f^2) &=& 
 \sum_{j}
 \int^{1}_{\xi} \frac{dz}{z} ~ f_{\gamma/j}(z,Q_\gamma^{2}) ~ f_{j/p}\left(\frac{\xi}{z},Q_{f}^2\right) ,
   \label{inelPDF.EQ}
\end{eqnarray}
where $f_{\gamma/j}$ is the Weizs\"acker-Williams $j\rightarrow \gamma$ distribution function,
with $Q_\gamma$ and $Q_f$ being the factorization scales for the photon and quark distributions, respectively. 
The summation is over all charged quarks.
Details regarding Eq.~(\ref{inelPDF.EQ}) can be found in Appendix~\ref{sec:AppIn}. 

Clearly, the scale for the photon momentum transfer should be above the elastic bound $Q_\gamma \ge \lamEl$. What is not clear, however, is how high we should evolve $Q_\gamma$. If we crudely consider the total inclusive cross section, we could simply choose the kinematical upper limit $Q_\gamma^2 \approx Q_f^2 \approx \hat s /4$ or $\hat{s}/4 - m^2_N$, which is a quite common practice in the literature \cite{Drees:1994zx}. 
However, we do not consider this a satisfactory treatment. Well below the kinematical upper limit,  
the photon virtuality $Q_\gamma$ becomes sufficiently large so that the collinear photon approximation as in figure~\ref{feynQA.fig} breaks down. Consequently, ``deeply inelastic scattering'' (DIS), as in figure~\ref{feynDIS.fig}, becomes the dominant feature. Thus, a more reasonable treatment is to introduce an upper limit for the inelastic process $\lamDIS$, above which a full DIS calculation of figure~\ref{feynDIS.fig} should be applied. We adopt the following scheme
\begin{eqnarray}
Q_\gamma = \lamDIS =
\left\{\begin{matrix}
 15 \GeV & \text{for 14 TeV}\\ 
 25 \GeV & \text{for 100 TeV}
\end{matrix}\right.
 \label{disDef.EQ}
\end{eqnarray}
Sensitivity to variations $\lamDIS$  are discussed in section~\ref{sec:scale}.

Consistent with $\Phi_{ij}(\tau)$ in Eq.~(\ref{partonLumi.EQ}),  we define the inelastic $\gamma q$ parton luminosity $\Phi_{\rm Inel}$ to be
\begin{eqnarray}
 \Phi_{\rm Inel}(\tau) =
 \int^{1}_{\tau} \frac{d\xi}{\xi}  \int^{1}_{\tau/\xi}\frac{dz}{z} 
  ~ 
  \sum_{q,q'} 
  \left[
  f_{q/p}(\xi)f_{\gamma/q'}(z)f_{q'/p}\left(\frac{\tau}{\xi z}\right) +    f_{q/p}\left(\frac{\tau}{\xi z}\right)f_{\gamma/q'}(z)f_{q'/p}(\xi) 
 \right].
   \label{inelLumi.EQ}
\end{eqnarray}

In figure~\ref{lumi.fig}, we give the $\Phi_{\rm Inel}$ spectrum as a function of $\sqrt \tau$, denoted by the (red) dash curve, for 14 and 100 TeV.
For the range investigated, $\Phi_{\rm Inel}$ ranges between {$2-4\%$} of the DY luminosity.
Compared to its elastic counterpart, the smallness of the inelastic luminosity is attributed the limited $Q_\gamma^2$ evolution.

The inelastic matrix element is identical to the elastic case. 
In figures~\ref{xsecComb.fig} and \ref{xsecComb100TeV.fig}, 
we show the bare cross section for the inelastic process, denoted by the (red) dash line, as a function of the neutrino mass.
The rate varies between {$0.7-30 ~(40-260)$} fb at 14 (100) TeV for $m_N = 100\GeV-1\TeV$.
As seen in figures~\ref{xsecRatio.fig} and \ref{xsecRatio100TeV.fig}, where the cross sections are normalized to the DY rate,
it reaches about {$10~(50)\%$} of the DY rate at large $m_N$.


\subsubsection{Deeply Inelastic Scattering: High $p_{T}$ Quark Jet}
\begin{figure}[!t]
\begin{center}
\includegraphics[scale=1,width=.96\textwidth]{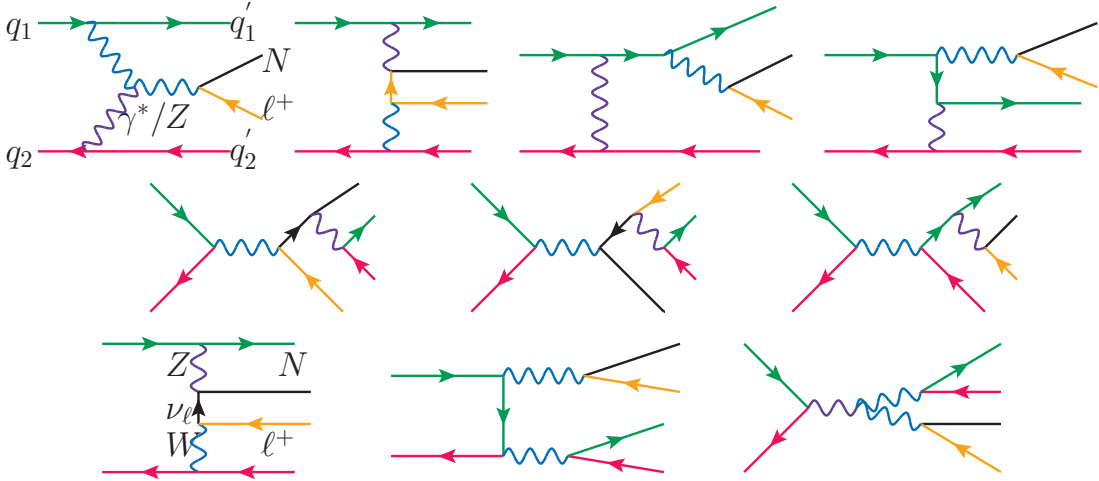}
\end{center}
\caption{Feynman diagrams for the DIS process $q_1 q_2 \rightarrow  N \ell^\pm q'_1 q'_2.$} 
\label{feynDIS.fig}
\end{figure}

As discussed in the previous section, at a sufficiently large momentum transfer
the collinear photon description breaks down and the associated final-state quark emerges as an observable jet. 
The electroweak process at $\alpha^4$
\begin{equation}
  q_1 ~ q_2 \rightarrow  N ~\ell^\pm ~q'_1 ~q'_2.
 \label{dis.EQ}
\end{equation}
becomes DIS, as shown by the Feynman diagrams in figure~\ref{feynDIS.fig}.
The top row of figure~\ref{feynDIS.fig} can be identified as the DIS analog of those diagrams in figure~\ref{feynQA.fig}.
Again, the first two diagrams represent the $W\gamma$ fusion with collinear log-enhancement from $t$-channel $W$ exchange. 
At these momentum transfers, the $WZ$ fusion channel~\cite{Datta:1993nm} turns on but is numerically smaller; 
see figure~\ref{feynDIS.fig}, bottom row, first diagram. 
The center row and two bottom-rightmost diagrams in figure~\ref{feynDIS.fig} represent on-shell $W/Z$ production at $\alpha^3$
with subsequent $W/Z\rightarrow q\overline{q}'$ decay.
Those processes, however, scale as $1/\hat s$ and are not log-enhanced.
A subset of these last diagrams also represent higher-order QED corrections to the DY process.

To model DIS, we use MG5 and simulate Eq.~(\ref{dis.EQ}) at order $\alpha^4$.
We impose\footnote{For consistency, we also require the lepton cuts given in Eq.~(\ref{regCutsLep.EQ})
and a jet separation $\Delta R_{jj} > 0.4$ to regularize irrelevant $\gamma^{*} \to q\overline{q}$ diagrams,
where $\Delta R \equiv \sqrt{\Delta\phi^2 + \Delta\eta^2}$ with $y = \eta \equiv -\log[\tan(\theta/2)]$ in the massless limit.}  
at the generator level a minimum on momentum transfers between initial-state and final-state quarks 
\begin{equation}
\underset{i,j=1,2}{\min}\sqrt{\vert (q_{i} - q'_{j})^2 \vert } > \lamDIS .
 \label{qDISMinCut.EQ}
\end{equation}
This requirement serves to separate the elastic and inelastic channels from  DIS.
Sensitivity to this cutoff is addressed in section~\ref{sec:scale}.

In figure~\ref{lumi.fig}, we show the quark-quark parton luminosity spectrum $\Phi_{\rm qq'}$, 
the source of the DIS processes, and represented by the (orange) dash-diamond  curves.
Though possessing the largest parton luminosity, the channel must overcome its larger coupling and phase space suppression.
At 14 and 100 TeV, $\Phi_{\rm qq'}$ ranges {$3-5$} times larger than $\Phi_{\rm q\overline{q}'}$. 
The difference in size between $\Phi_{\rm qq'}$ and $\Phi_{\rm El~(Inel)}$ is due to the additional coupling $\alpha_{\rm EM}$ in $f_{\gamma/p}^{\rm El~(Inel)}$.

In figures~\ref{xsecComb.fig} and \ref{xsecComb100TeV.fig}, we plot bare cross section as in Eq.~(\ref{bareProd.EQ}), denoted by the (orange) dash-diamond curve.
In figures~\ref{xsecRatio.fig} and \ref{xsecRatio100TeV.fig}, the same curves are normalized to the DY rate.
At 14 (100) TeV, the cross section ranges from {$1-60$ ($80-500$)} fb, reaching about {35\% (80\%)} of the DY rate.

To compare channels, we observe that the DIS (elastic) process increases greatest (least) with increasing collider energies.
This is due to the increase likelihood for larger momentum transfers in more energetic collisions.
A similar conclusion was found for elastic and inelastic $\gamma\gamma$ scattering at the Tevatron and LHC~\cite{Han:2007bk}.

\begin{table}[!t]
\caption{Total cross sections of various $pp\rightarrow N\ell^{\pm}X$ channels for representative values of $m_N$.
Minimal acceptance cuts as in Eqs.~(\ref{regCutsLep.EQ}) have been applied.}
 \begin{center}
\begin{tabular}{|c|c|c|c|}
\hline 
$\sigma_{\rm14\TeV~LHC}/\vert V_{\ell N}\vert^{2}$ [fb] & $m_{N}=300\GeV$ 	&  $m_{N}=500\GeV$ & $m_{N}=1\TeV$ \tabularnewline\hline\hline 
$pp\rightarrow N\ell^{\pm}$ LO DY $[K=1.2]$	&293 (352) 	&47.3 (56.8) 	&2.87 (3.44) \tabularnewline\hline
$pp\rightarrow N\ell^{\pm}X$ Elastic			&10.8971	&5.16756	&1.23693	\tabularnewline\hline
$pp\rightarrow N\ell^{\pm}X$ Inelastic			&8.32241	&3.44245	&0.65728	\tabularnewline\hline
$pp\rightarrow N\ell^{\pm}X$ DIS			&11.7 		&5.19 		&1.21 \tabularnewline\hline
$\sigma_{\gamma{\rm-Initiated}}$/$\sigma_{\rm DY}^{K=1.2}$ & 0.09  & 0.24   & 0.90 \tabularnewline\hline
\hline
\hline 
$\sigma_{\rm 100\TeV~VLHC}/\vert V_{\ell N}\vert^{2}$ [fb] & $m_{N}=300\GeV$ 	&  $m_{N}=500\GeV$ & $m_{N}=1\TeV$ \tabularnewline\hline\hline 
$pp\rightarrow N\ell^{\pm}$ LO DY  $[K=1.3]$	& 2540 (3300) 	& 583 (758) 	& 70.5 (91.6)\tabularnewline\hline
$pp\rightarrow N\ell^{\pm}X$ Elastic			& 85.8 		& 65.5 		& 36.4 \tabularnewline\hline
$pp\rightarrow N\ell^{\pm}X$ Inelastic 			& 144 		& 96.0		& 42.7 \tabularnewline\hline
$pp\rightarrow N\ell^{\pm}X$ DIS			& 210 		& 145 		& 76.7 \tabularnewline\hline
$\sigma_{\gamma{\rm-Initiated}}$/$\sigma_{\rm DY}^{K=1.3}$ & 0.13 	& 0.40  	& 1.7 \tabularnewline\hline
\hline
\end{tabular}
\label{xsec.TB}
\end{center}
\end{table}
\subsubsection{Total Neutrino Production from $\gamma$-Initiated Processes}
\label{sec:totIsPhoton}
The total heavy neutrino production cross section from $\gamma$-initiated processes may be obtained by summing the elastic, inelastic, and DIS channels \cite{Drees:1994zx,Han:2007bk}:
\begin{eqnarray}
 \sigma_{\rm \gamma-Initiated}(N\ell^\pm X) = \sigma_{\rm El }(N\ell^\pm X) + \sigma_{\rm Inel}(N\ell^\pm X) + \sigma_{\rm DIS}(N\ell^\pm X),
\label{gammaSum.EQ}
\end{eqnarray}
We plot Eq.~(\ref{gammaSum.EQ}) as a function of $m_N$ in figures~\ref{xsecComb.fig} and \ref{xsecComb100TeV.fig} at 14 and 100 TeV, denoted by the 
(blue) dash-dot curve. In figures~\ref{xsecRatio.fig} and \ref{xsecRatio100TeV.fig}, the same curves are normalized to the DY rate.
For $m_N = 100\GeV-1\TeV$, the total rate spans {$3-100$~($150-1000$) fb} at 14 (100) TeV,
reaching about {$90~(110)\%$} of the DY rate at large $m_N$.
We find that the $W\gamma$ fusion represents the largest heavy neutrino production mechanism for {$m_N>1\TeV~(770)\GeV$} at 14 (100) TeV.
We expect for increasing collider energy this crossover will occur earlier at lighter neutrino masses.
Cross sections for representative values of $m_N$ for all channels at 14 and 100 TeV are given in Table \ref{xsec.TB}.

\begin{figure}[!t]
\centering
\subfigure[]{\includegraphics[width=.48\textwidth]{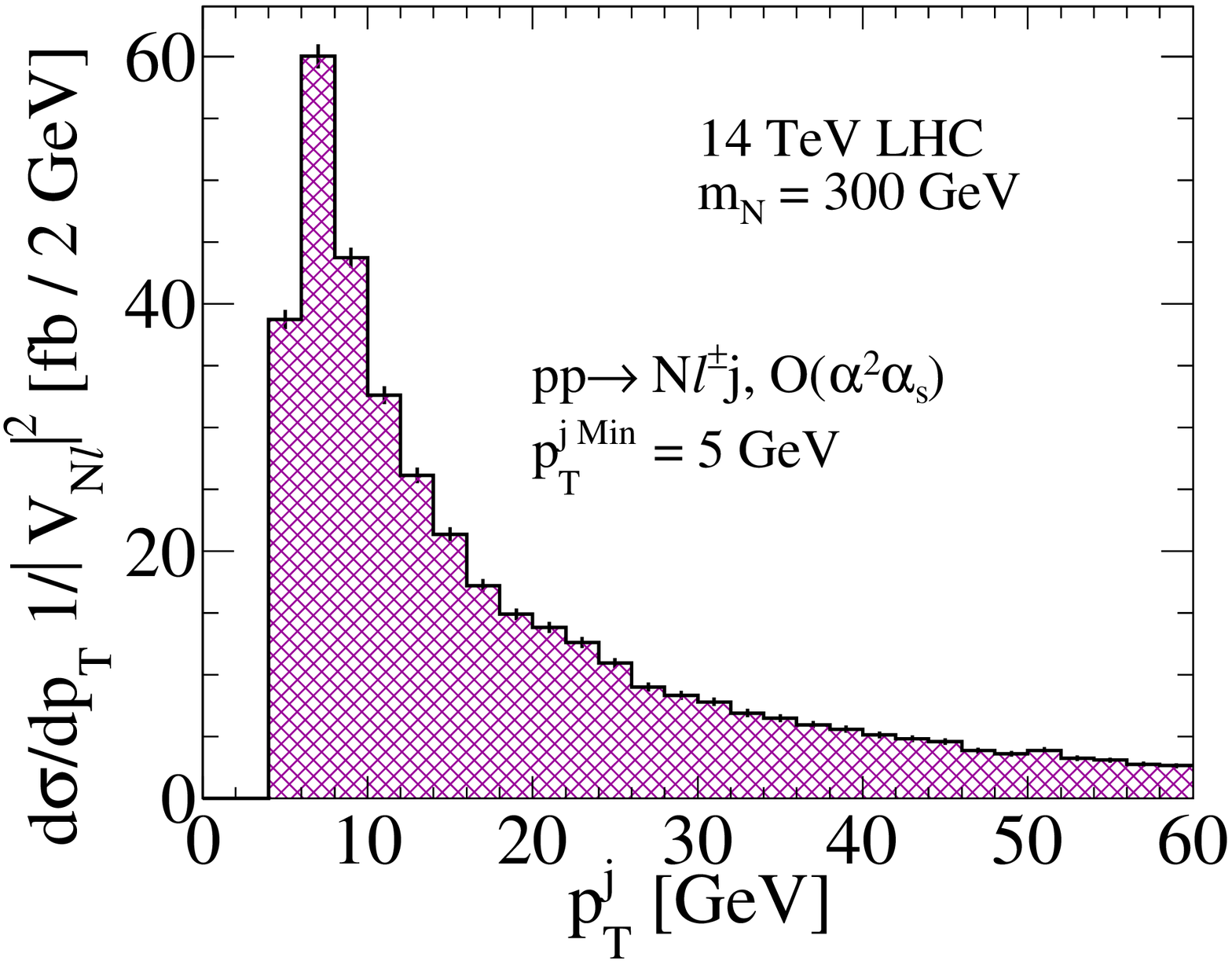}\label{ppNljQCD_dXSec_vs_pt.FIG}}
\subfigure[]{\includegraphics[width=.48\textwidth]{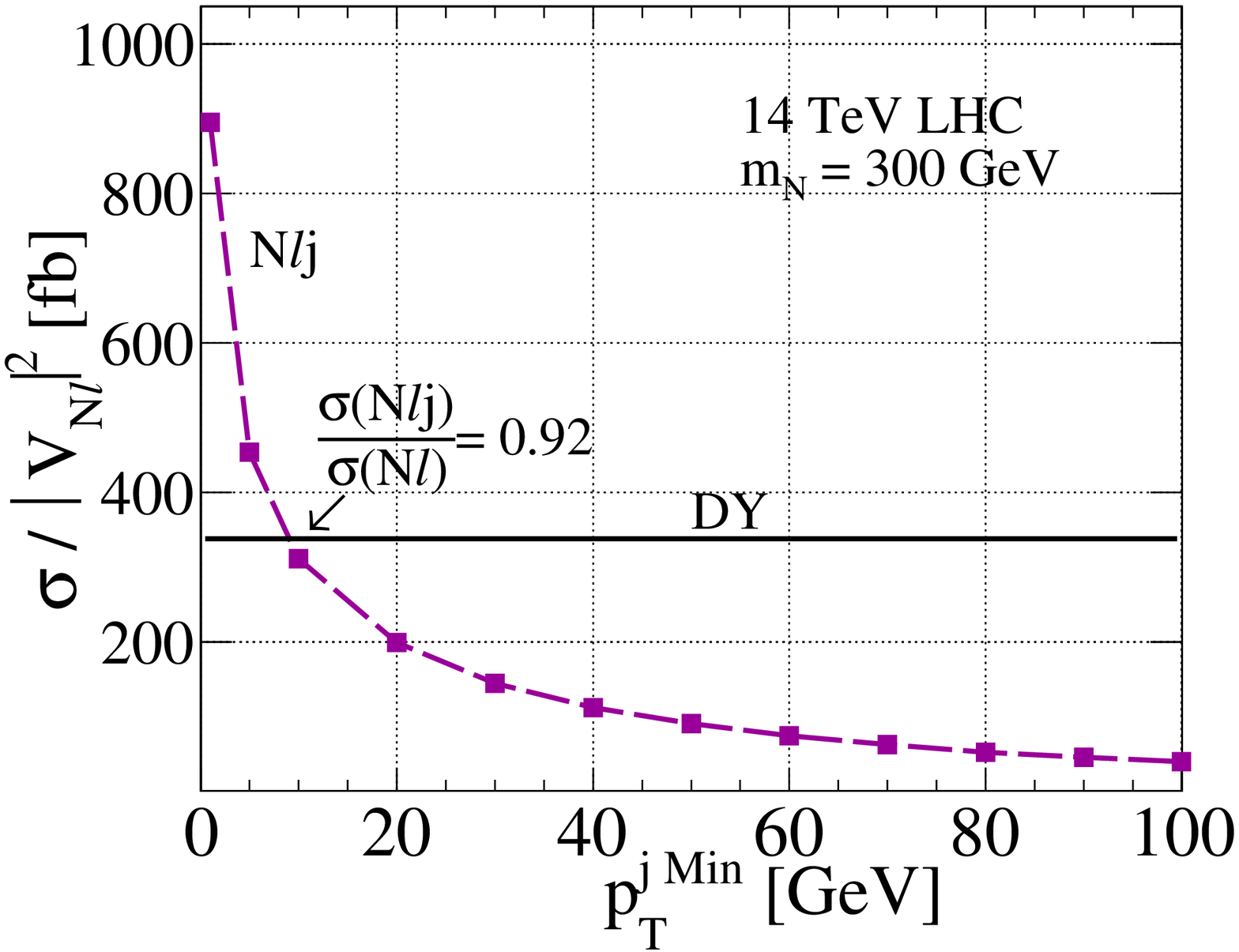}\label{ppNljQCD_XSec_vs_pTMinCut.FIG}}
\caption{
(a) The tree-level differential cross section for $N\ell^\pm j$ at $\alpha^2\alpha_{\rm s}$ with respect to $p_T^j$;
(b) Integrated cross section $\sigma(N\ell^\pm j)$ versus the minimum $p_T^{j}$ cutoff. The solid line denotes the LO DY rate.
} 
\label{ppNl1jQCD.FIG}
\end{figure}

Before closing the discussion for the heavy $N$ production at hadron colliders, an important remark is in order. 
We have taken into account the {\it inclusive} QCD correction at NNLO as a $K$-factor.
In contrast, Ref.~\cite{Dev:2013wba} included only the tree-level process at order $\alpha^2\alpha^2_{s}$ and  $\alpha^4$
\begin{equation}
 p p \rightarrow N \ell^\pm j j .
\end{equation}
When calculating the exclusive $N\ell^\pm jj$ cross section,  
kinematical cuts of $p_{Tj} > 10$ GeV and $\Delta R_{jj}>0.4$ were applied to regularize the cross section. 
For $m_N = 300\GeV$, the exclusive cross section was found to exceed the LO DY channel at 14 TeV, 
whereas we find that the NNLO correction to the inclusive cross section is only $20\%$ with DIS contributing $3\%$.
More recently~\cite{Das:2014jxa}, the tree-level rate for $N\ell j$ with $p_T^j>30\GeV$  was calculated to be $80\%$ of the LO DY rate at $m_N=500\GeV$; 
at NNLO, we find the inclusive correction to be only $20\%$.
We attribute these discrepancies to their too low a $p_T^j$ cut that overestimate the contribution of initial-state radiation based on a tree-level calculation.

To make the point concrete, we consider the tree-level QCD correction to the DY process at order $\alpha^2\alpha_{\rm s}$
\begin{equation}
 p ~ p \rightarrow N ~\ell^\pm ~j,
 \label{qcd.EQ}
\end{equation}
where the final-state jet originates from an initial-state quark or gluon.
MG5 is used to simulated Eq.~(\ref{qcd.EQ}).
In figure~\ref{ppNljQCD_dXSec_vs_pt.FIG}, the differential cross section of $p_T^j$ is shown for a minimal $p_T$ at $5\GeV$. 
The singularity at the origin is apparent. 
In figure~\ref{ppNljQCD_XSec_vs_pTMinCut.FIG}, the 14 TeV LHC cross section as a function of minimum $p_T$ cut on the jet is presented.
A representative neutrino mass of $m_N = 300\GeV$ is used; no additional cut has been imposed. At $p_{T}^{j \min} = 10\GeV$, as adopted in
Ref.~\cite{Dev:2013wba}, the $N\ell j$ rate is nearly equal to the DY rate, well above the NNLO prediction for the inclusive cross section \cite{Hamberg:1990np}.


\subsection{Kinematic Features of $N$ Production with Jets at 14 TeV}
\label{sec:kine}

To explore the kinematic distributions of the inclusive neutrino production, we fix $\sqrt{s} = 14\TeV$ and $m_N = 500\GeV$. 
At 100 TeV, we observe little change in the kinematical features and our conclusions remain the same. The most notable difference, however, is a broadening of rapidity distributions.
This is due an increase in longitudinal momentum carried by the final states,
which follows from the increase in average momentum carried by initial-state partons.
For $m_N \geq 100\GeV$, we observe little difference from the 500 GeV case we present.
Throughout this study, jets are ranked by $p_T$, namely, 
the jet with the largest (smallest) $p_T$ is referred to as hardest (softest).

\begin{figure}[!t]
\begin{center}
\subfigure[]{\includegraphics[scale=1,width=.48\textwidth]{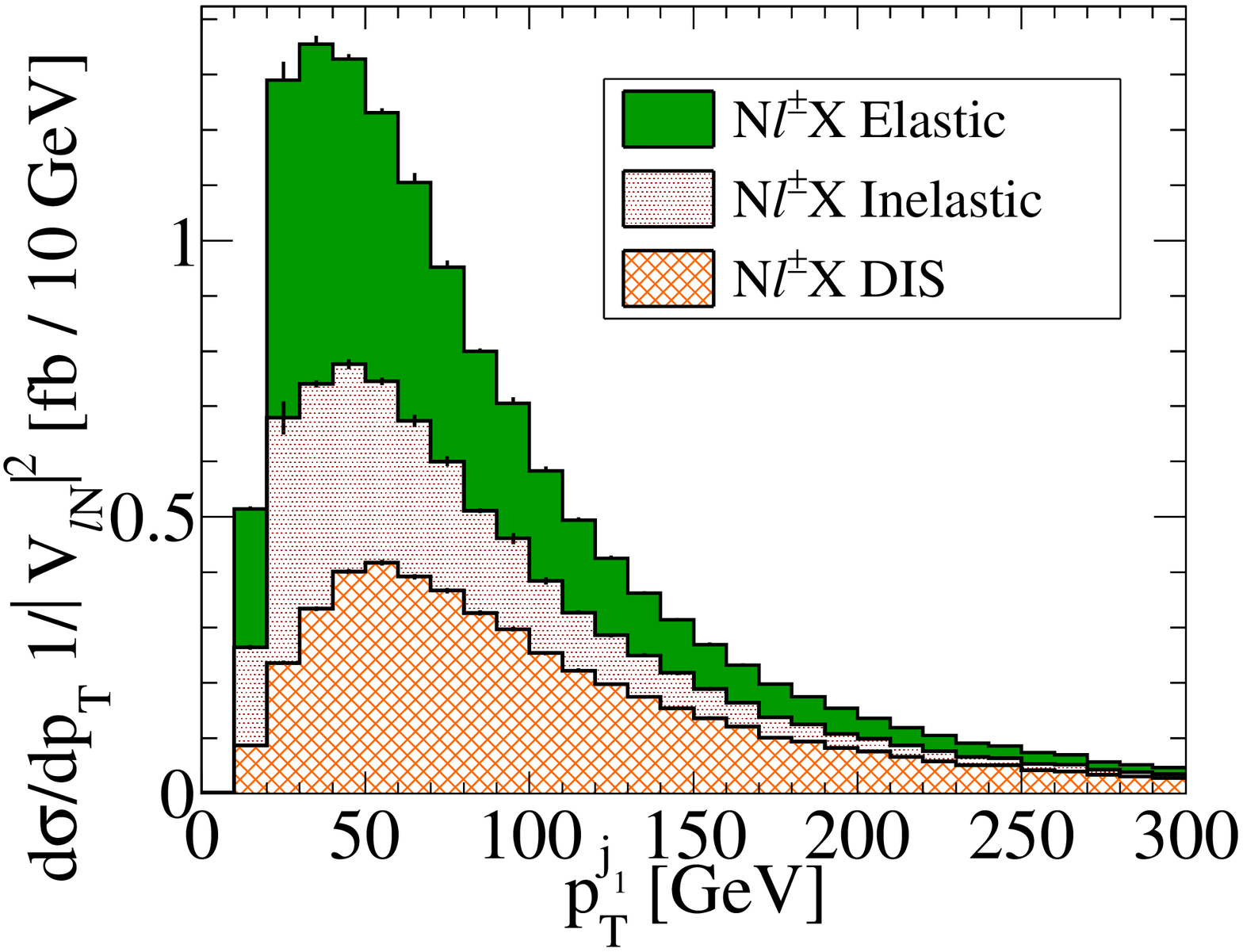}	\label{assocPTj1.FIG} }
\subfigure[]{\includegraphics[scale=1,width=.48\textwidth]{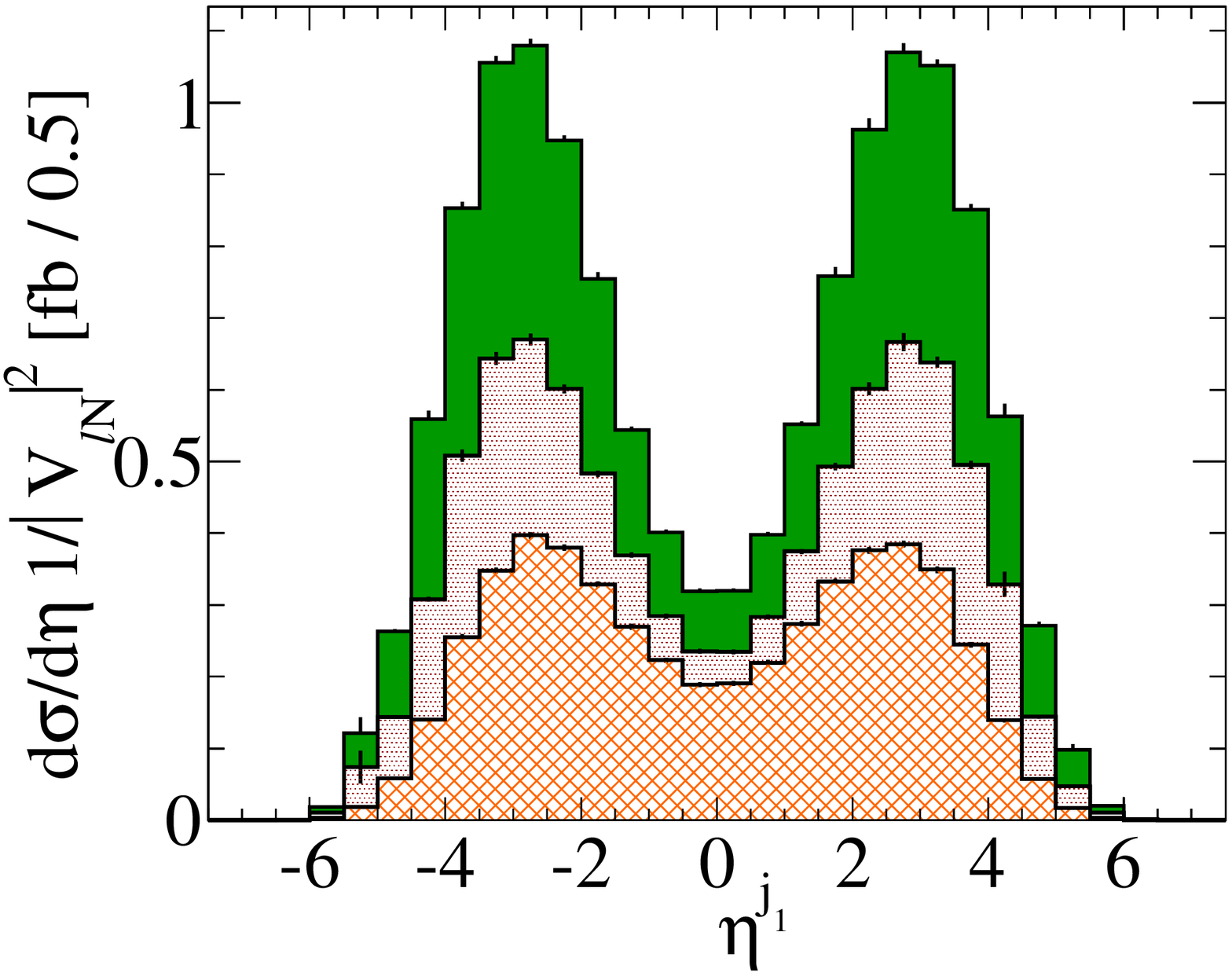}	\label{assocEtaj1.FIG} }
\vspace{.2in}\\
\subfigure[]{\includegraphics[scale=1,width=.48\textwidth]{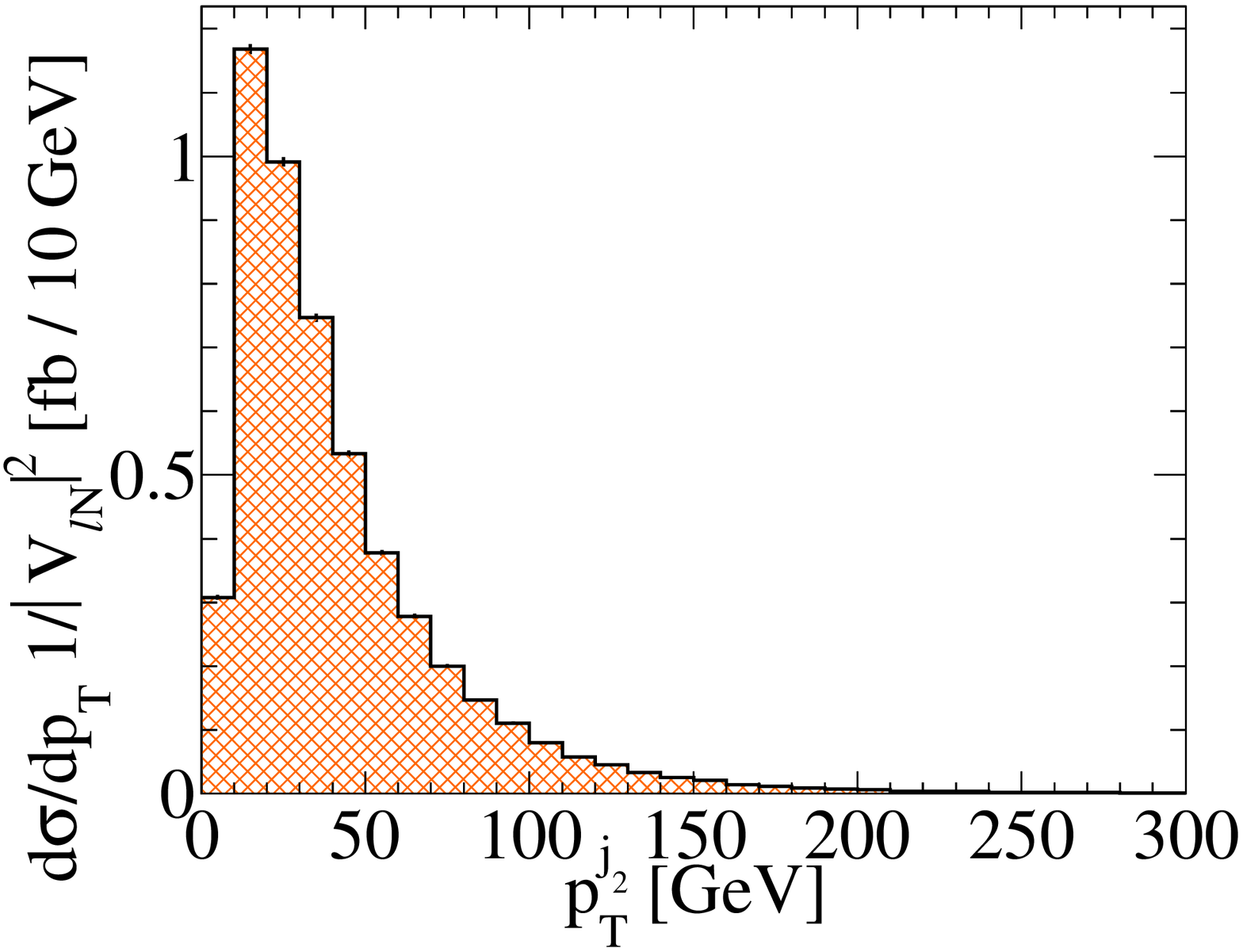}	\label{assocPTj2.FIG} }
\subfigure[]{\includegraphics[scale=1,width=.48\textwidth]{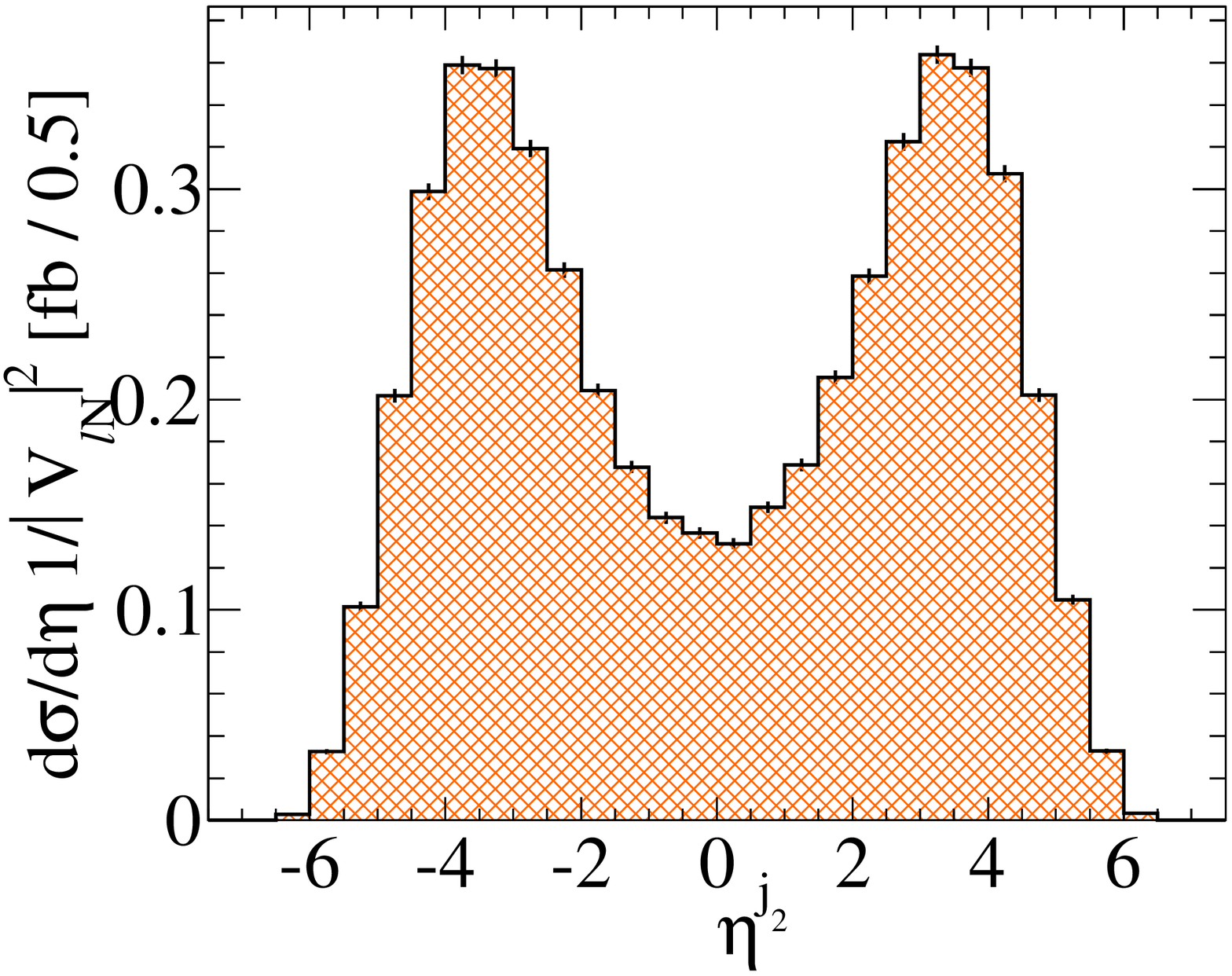}	\label{assocEtal2.FIG} }
\end{center}
\caption{Stacked (a) $p_{T}$ and (b) $\eta$ differential distributions, divided by $\vert V_{\ell N}\vert^{2}$,  
at 14 TeV LHC of the leading jet in the elastic (solid fill), inelastic (dot fill), and DIS (crosshatch fill) processes. 
(c) $p_{T}$ and (d) $\eta$ of the sub-leading jet in DIS.}
\label{assocJet.FIG}
\end{figure}

In figure~\ref{assocJet.FIG}, we plot the (a) $p_T$ and (b) $\eta$ distributions of the hardest jet in $p_T$ produced in association with $N$ 
for the various $W\gamma$ fusion channels.
Also shown are (c) $p_{T}$ and (d) $\eta$ distributions of the sub-leading jet for the DIS channel.
For the hardest jet, we observe a plateau at $p_T \sim M_W / 2$ and a rapidity concentrated at $\vert\eta\vert\sim 3.5$,
suggesting dominance of $t$-channel $W$ boson emission.
For the soft jet, we observe a rise in cross section at low $p_T$ and a rapidity also concentrated at $\vert\eta\vert\sim 3.5$,
indicating $t$-channel emission of a massless vector boson.
We conclude that VBF is the driving contribution $\gamma$-initiated heavy neutrino production.

\begin{figure}[!t]
\begin{center}
\subfigure[]{\includegraphics[scale=1,width=.48\textwidth]{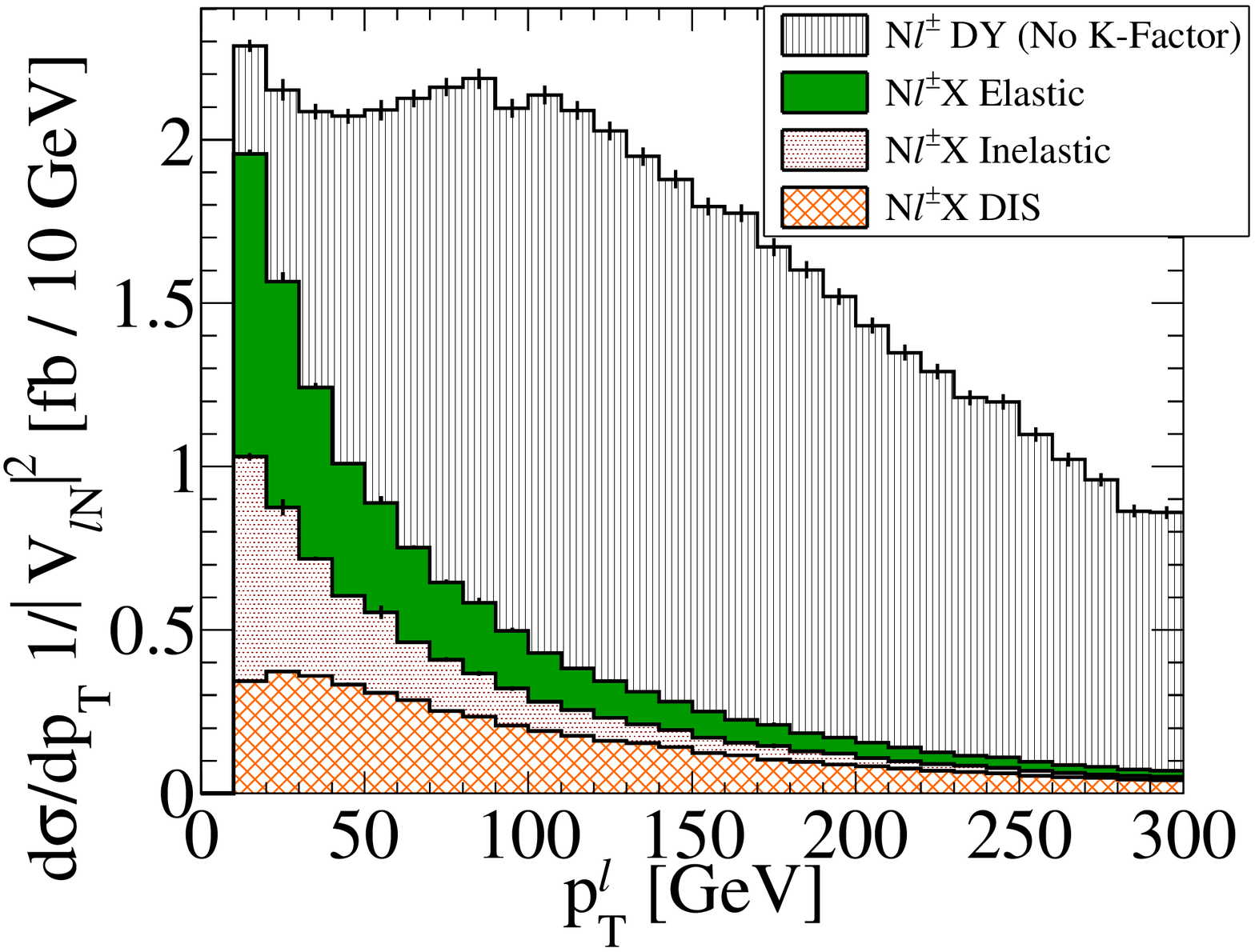}	\label{assocPTl.FIG} }
\subfigure[]{\includegraphics[scale=1,width=.48\textwidth]{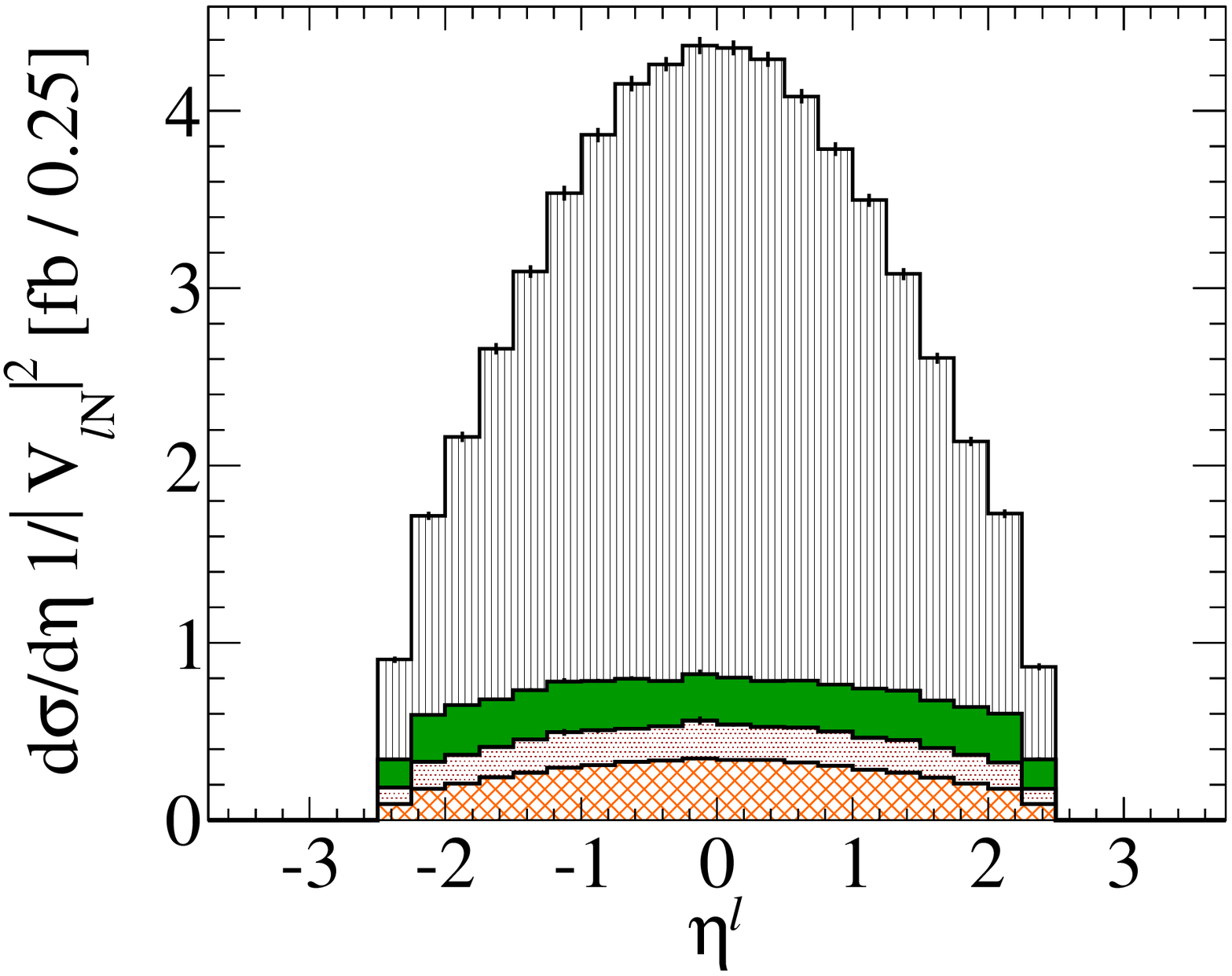}	\label{assocEtal.FIG} }
\vspace{.2in}\\
\subfigure[]{\includegraphics[scale=1,width=.48\textwidth]{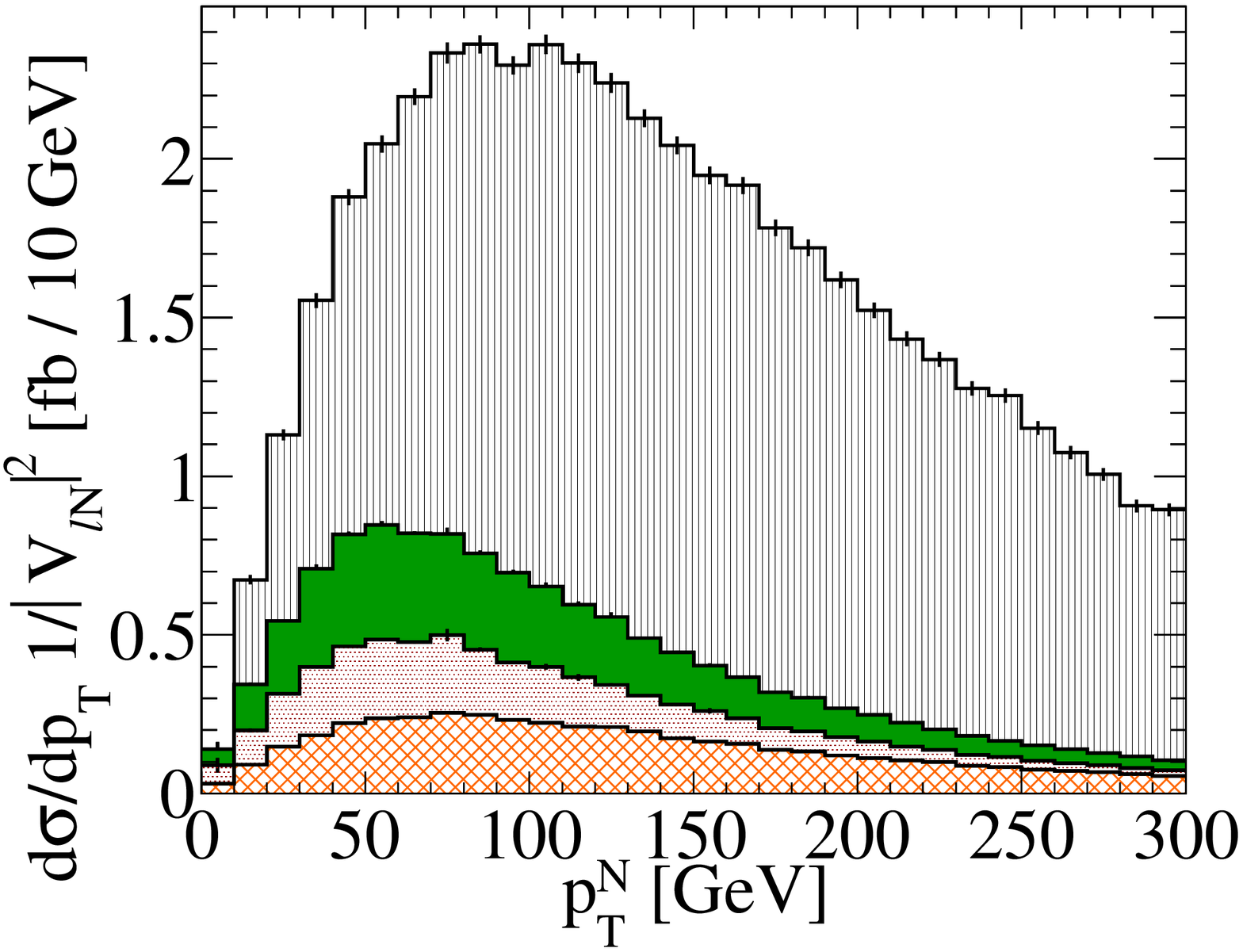}	\label{ptN.FIG} }
\subfigure[]{\includegraphics[scale=1,width=.48\textwidth]{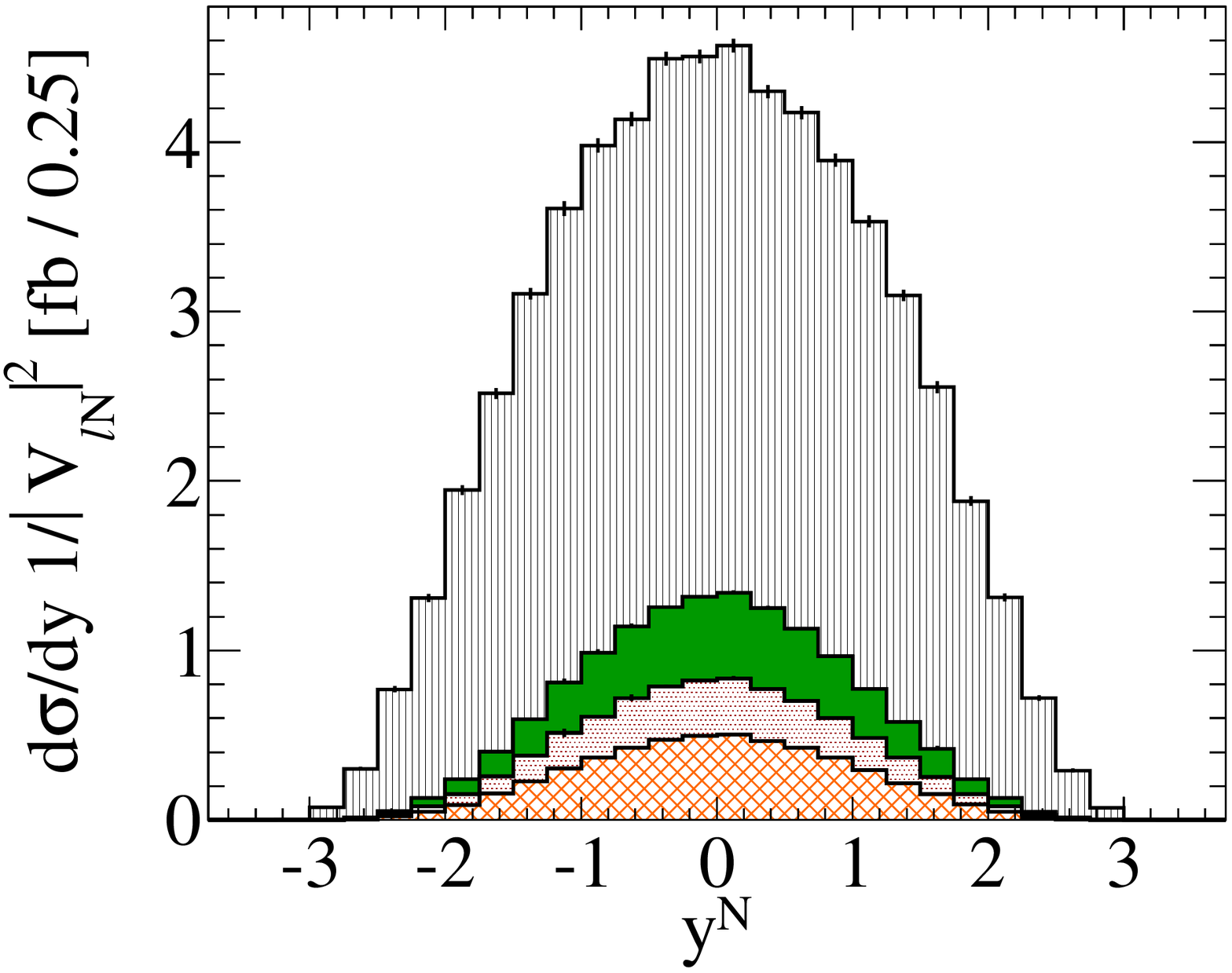}	\label{yN.FIG} }
\end{center}
\caption{Stacked (a) $p_{T}$ and (b) $\eta$ differential distributions at 14 TeV LHC of the charged lepton produced in association with $N$ for the
DY (line fill), elastic, inelastic and DIS processes. (c) $p_{T}$ and (d) $y$ of $N$ for the same processes.
Fill style and normalization remain unchanged from figure~\ref{assocJet.FIG}.}
\label{leptonKin.FIG}
\end{figure}

In figure~\ref{leptonKin.FIG}, we plot the (a) $p_T$ and (b) $\eta$ distributions of the charged lepton produced in association with $N$
for all channels contributing to $N\ell$ production. Also shown are the (c) $p_T$ and (d) $y$ distribution of $N$.
For both leptons, we observed a tendency for softer $p_T$ and broader rapidity distributions in $\gamma$-initiated channels than in the DY channel.
As DY neutrino production proceeds through the $s$-channel, $N$ and $\ell$ possess harder $p_T$ than the $\gamma$-initiated states, 
which proceed through $t$-channel production and are thus more forward.


\subsection{Scale Dependence}
\label{sec:scale}

\begin{table}[!t]
\caption{Summary of scale dependence in $N\ell^\pm X$ production at 14 TeV and 100 TeV.}
 \begin{center}
\begin{tabular}{|c|c|c|c|c|}
\hline\hline
\multirow{2}{*}{Scale Parameter} & Default at & \multirow{2}{*}{Lower} & \multirow{2}{*}{Upper} & Variation \tabularnewline
				 & 14 (100) TeV &		& 				& at 14 (100) TeV	\tabularnewline\hline\hline
\multirow{2}{*}{$\Lambda_\gamma^{\rm El }$ [Eq.~(\ref{elPDF.EQ})]} & \multirow{2}{*}{ 1.22 GeV} &
		$m_p$	& 2.3 GeV	&	$\mathcal{O}(10\%)\ \  (12\%)$ \tabularnewline
	  & &	$m_p$	& 5 GeV		&	$\mathcal{O}(22\%)\ \  (28\%)$ \tabularnewline\hline
\multirow{2}{*}{$\lamDIS$ [Eq.~(\ref{disDef.EQ})]} & \multirow{2}{*}{15 GeV (25 GeV)} &
	        5 GeV	& 50 GeV	&	$\mathcal{O}(10\%)\ \ (15\%)$ \tabularnewline
	  & &   5 GeV	& 150 GeV	&	$\mathcal{O}(18\%)\ \ (27\%)$ \tabularnewline\hline\hline
$Q_f^{\rm DY}$  [Eq.~(\ref{factTheorem.EQ})]	& $\sqrt{\hat{s}}/2$	& $m_N/2$	& $\sqrt{\hat{s}}$	&$\mathcal{O}(10\%)\ \ (5\%)$\tabularnewline\hline
$Q_f^{\rm DIS}$ [Eq.~(\ref{factTheorem.EQ})]	& $\sqrt{\hat{s}}/2$	& $m_N/2$	& $\sqrt{\hat{s}}$	&$\mathcal{O}(15\%)\ \ (8\%)$\tabularnewline\hline
\hline
\end{tabular}
\label{scale.TB}
\end{center}
\end{table}

For the processes under consideration, namely DY and $W\gamma$ fusion, there are two factorization scales involved: $Q_{f}$ and $Q_{\gamma}$.
They characterize typical momentum transfers of the physical processes. 
For the $\gamma$-initiated channels, we separate the contributions into three regimes using $\lamEl$ and $\lamDIS$. 
Though the quark parton scale $Q_{f}$ is present in all channels, we assume it to be near $m_N^{}$ and set it as in Eq.~(\ref{QfScale.EQ}). 

\begin{figure}[!t]
\begin{center}
\subfigure[]{\includegraphics[scale=1,width=.45\textwidth]{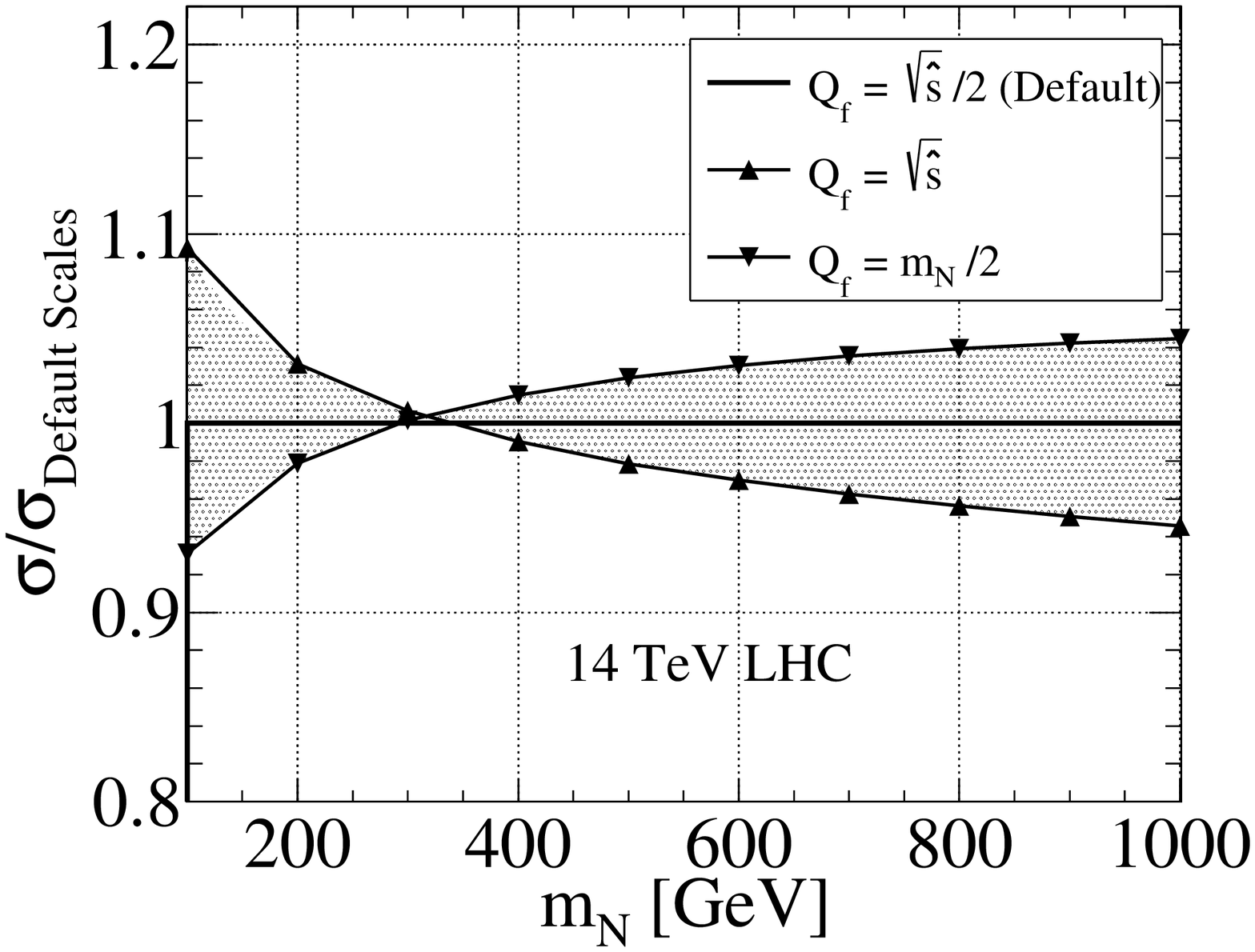}		\label{scale_DY.FIG} }
\subfigure[]{\includegraphics[scale=1,width=.45\textwidth]{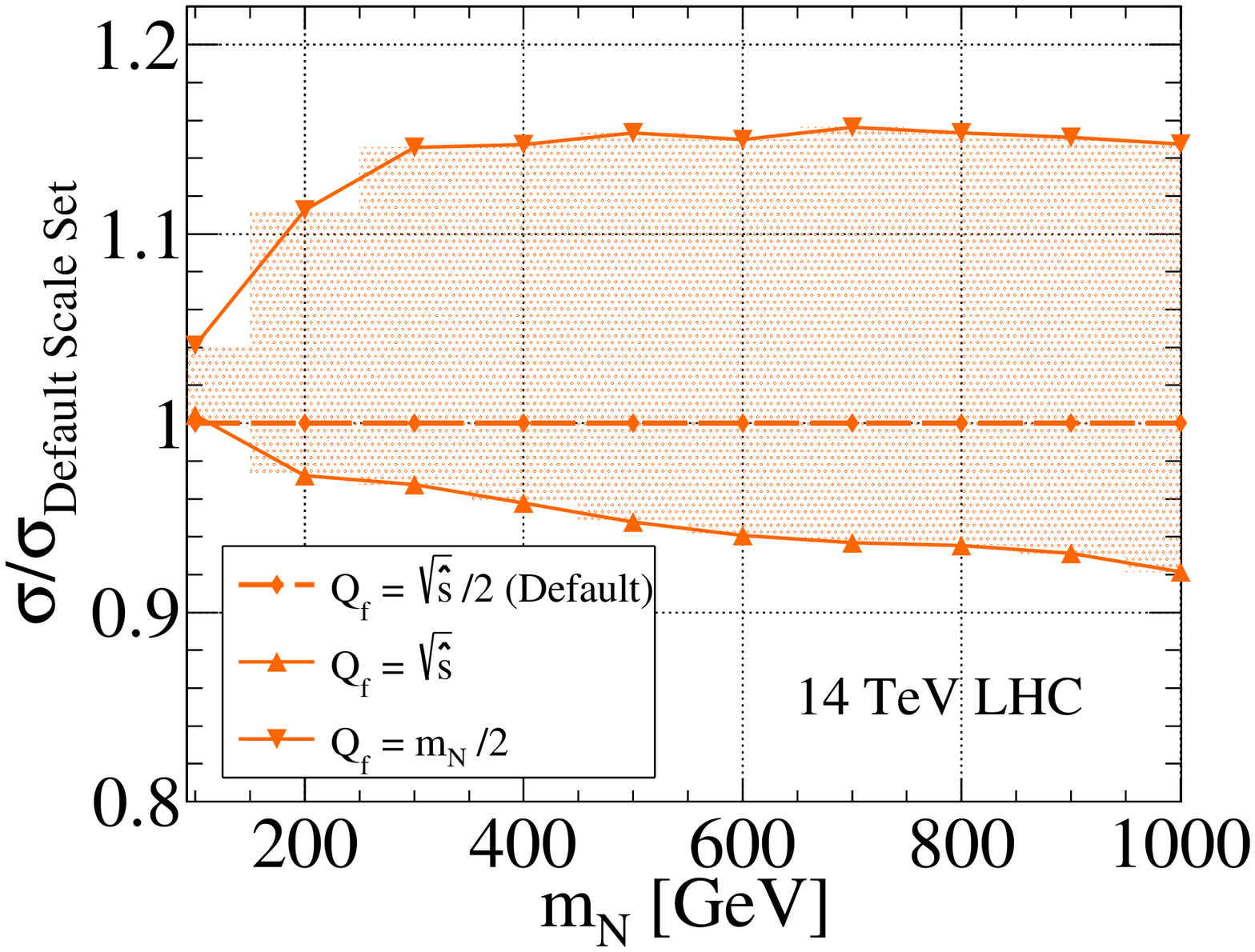}		\label{scale_DIS_qFact.FIG}}
\\
\subfigure[]{\includegraphics[scale=1,width=.45\textwidth]{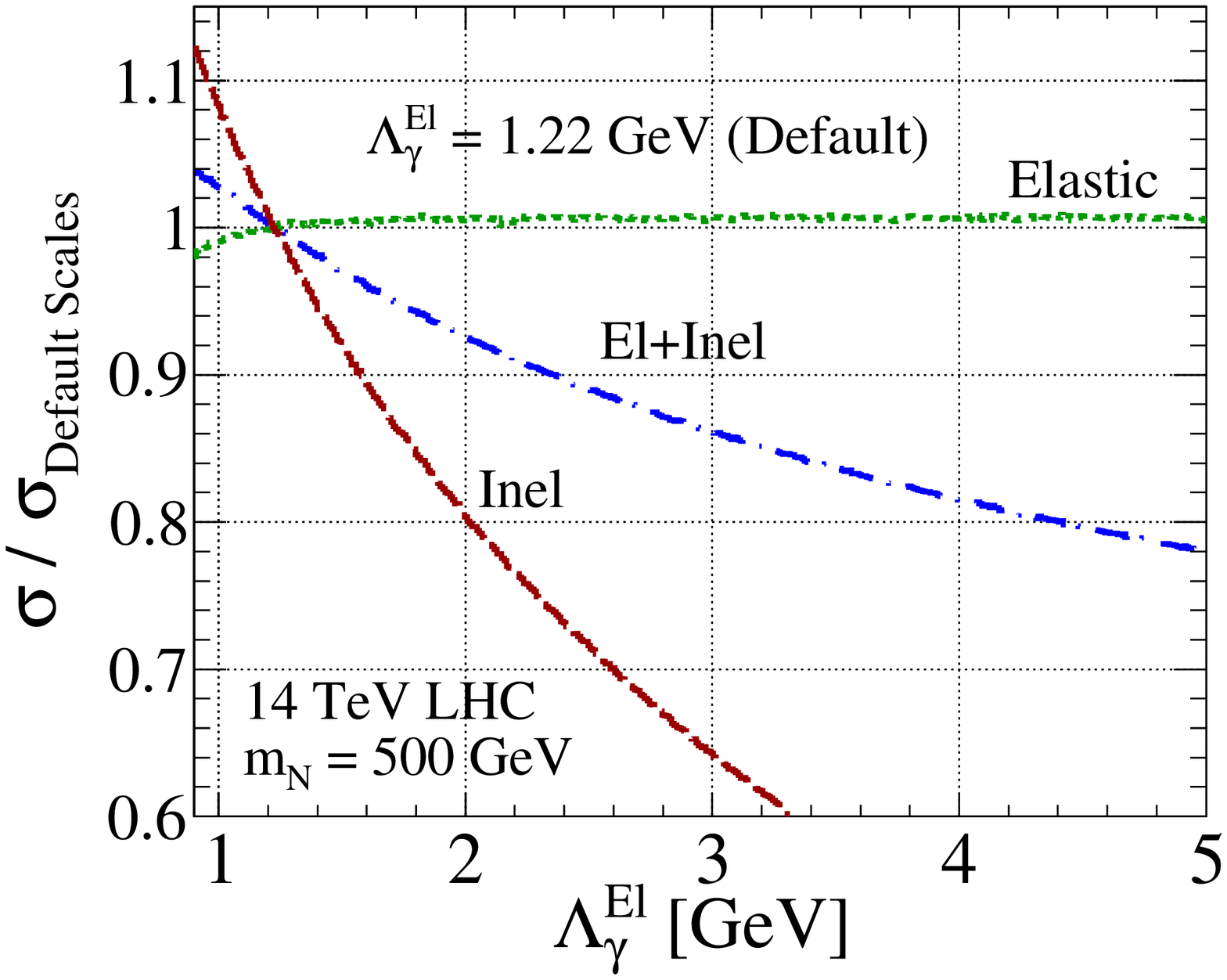}		\label{scale_ElMatch.FIG} }
\subfigure[]{\includegraphics[scale=1,width=.45\textwidth]{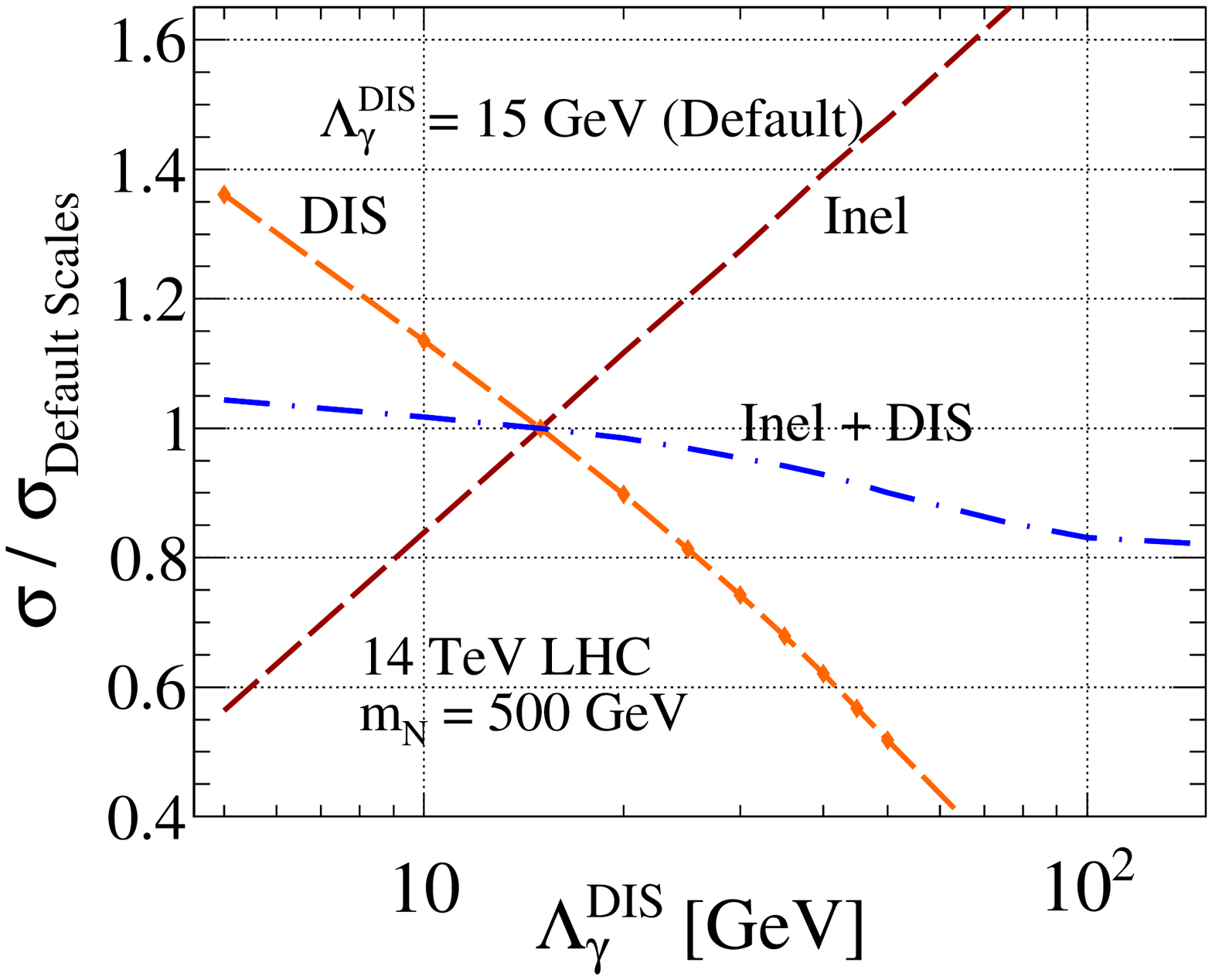}		\label{scale_DISMatch.FIG} }
\end{center}
\caption{
Cross section ratios relative to the default scale scheme, as a function of $m_N$, 
for the high-scale (triangle) and low-scale (upside-down triangle) $Q_{f}$ scheme in (a) DY and (b) DIS.
The same quantity as a function of (c) $\lamEl$ in elastic (dot), inelastic (dash), elastic+inelastic (dash-dot) scattering;
(d) $\lamDIS$ in inelastic (dash), DIS (dash-diamond), and inelastic+DIS (dash-dot).}
\label{scale.fig}
\end{figure}

To quantify the numerical impact of varying these scales, 
each relevant cross section as a function of $m_N$ is computed with one scale varied while all other scales are held at their default values. 
The test ranges are taken as 
\begin{eqnarray}
  m_{p} \leq \Lambda_\gamma^{\rm El } \leq 5\GeV,
  \quad
  5\GeV \leq Q_\gamma = \lamDIS \leq  150\GeV,
  \quad
  \frac{m_N}{2}\leq Q_{f} \leq \sqrt{\hat{s}},
\label{scaleVar.EQ}
\end{eqnarray}
In figure~\ref{scale.fig}, we plot the variation band in each production channel cross section due to the shifting scale.
For a given channel, rates are normalized to the cross section using the default scale choices, as discussed in the previous sections and summarized in the first column of Table \ref{scale.TB}.
High-(low-) scale choices are denoted by a solid line with right-side (upside-down) up triangles.

For the 14 TeV LO DY process, we observe in figure~\ref{scale_DY.FIG} maximally a 9\% upward (7\% downward) variation for the range of $m_N$ investigated.
Below $m_N\approx 300\GeV$, the default scale scheme curve is below (above) the high (low) scale scheme curve.
The trend is reversed for above $m_N\approx 300\GeV$.
At 100 TeV, the crossover point shifts to much higher values of $m_N$.
Numerically, we observe a smaller scale dependence at the {5\%} level.

In figure~\ref{scale_DIS_qFact.FIG}, we plot scale variation associated with the factorization scale $Q_{f}$ for DIS.
Maximally, we observe a 16\% upward (8\% downward) shift.
We observe that the crossover between the high and low scale schemes now occurs at $m_N\lesssim 100\GeV$.
This is to be expected as $\hat{s}$ for the 4-body DIS at a fixed neutrino mass is much larger than that for the 2-body DY channel.
Similarly, as $\sqrt{\hat{s}}$ and $m_N$ are no longer comparable, as in the DY case, an asymmetry between the high- and low-scale scheme curves emerges. 
At 100 TeV, we observe a smaller dependence at the {10\%} level.

In figure~\ref{scale_ElMatch.FIG}, 
we show the dependence on $\lamEl$ in the elastic (dot) and inelastic (dash) channels, as well as the sum of the two channels (dash-dot). 
For the elastic channel we find very small dependence on $\lamEl$ between $m_p$ and $5\GeV$,
with the analytical expression for $f^{\rm El}_{\gamma/p}$ given in appendix~\ref{sec:AppEl}.
For the inelastic channel, on the other hand, we find rather large dependence on $\lamEl$ between $m_p$ and $5\GeV$.
Since $\lamEl$ acts as the regulator of the inelastic channel's collinear logarithm, this large sensitivity is expected;
see appendix~\ref{sec:AppIn} for details regarding $f^{\rm Inel}_{\gamma/p}$.
We find that the summed rate is slightly more stable.
In the region $m_p < \lamEl < 2.3\GeV$, the variation is below the 10\% level.
Over the entire range studied, this grows to {$20\%$}. 
At 100 TeV, similar behavior is observed and the dependence grows to the {$30\%$} level over the whole range.

In figure~\ref{scale_DISMatch.FIG}, for $m_N=500\GeV$, 
we plot the scale dependence on $\lamDIS$ in the inelastic (dash) and DIS (dash-diamond) channels, as well as the sum of the two channels (dash-dot).
Very large sensitivity on the scale is found for individual channels, ranging {$40\%-60\%$} over the entire domain.
However, as the choice of $\lamDIS$ is arbitrary, we expect and observe that their sum is considerably less sensitive to $\lamDIS$.
For $\lamDIS = 5-50~(5-150)$ GeV, we find maximally a {10\% (18\%)} variation.
The stability suggests the channels are well-matched for scales in the range of $5-50\GeV$. 
Results are summarized in Table \ref{scale.TB}.


\section{HEAVY NEUTRINO OBSERVABILITY AT HADRON COLLIDERS}
\label{sec:100TeV}

\subsection{Kinematic Features of Heavy $N$ Decays to Same-Sign Leptons with Jets at 100 TeV}
We consider at a $100$ TeV $pp$ collider charged current production of a heavy Majorana neutrino $N$ in association with $n=0,~1~\text{or}~2$ jets,
and its decay to same-sign leptons and a dijet via the subprocess $N\to \ell W \to \ell jj$:
\begin{equation}
 p~p \rightarrow ~N ~\ell^{\pm} ~+~ nj 	\rightarrow 
   \ell^{\pm} ~\ell^{'\pm} ~+~ (n+2)j, \quad n = 0,~1,~2.
 \label{ppllnj.EQ}
\end{equation}
Event simulation for the DY and DIS channels was handled with MG5.
A NNLO $K$-factor of $K=1.3$ is applied to the LO DY channel; kinematic distributions are not scaled by $K$.
Elastic and inelastic channels were handled by extending neutrino production calculations to include heavy neutrino decay.
The NWA with full spin correlation was applied.
The elastic channel matrix element was again checked with MG5.

Detector response was modeled by applying a Gaussian smearing to jets and leptons.
For jet energy, the energy resolution is parameterized by \cite{Aad:2009wy} 
\begin{equation}
 \frac{\sigma_E}{E} = \frac{a}{\sqrt{E/\GeV}} \oplus b, 
 \label{jetSmear.EQ}
\end{equation}
with $a = 0.6~(0.9)$ and $b= 0.05~(0.07)$ for $\vert\eta\vert\leq3.2 ~(>3.2)$,  
and where the terms are added in quadrature, i.e., $x\oplus y = \sqrt{x^2 + y^2}$.
For muons, the inverse-$p_T$ resolution is parameterized by \cite{Aad:2009wy}
\begin{equation}
 \frac{\sigma_{1/p_T}}{(1/p_T)} = \frac{0.011\GeV}{p_T} \oplus 0.00017.
 \label{muSmear.EQ}
\end{equation}
We will eventually discuss the sensitivity to the $e^\pm\mu^\pm$ final state and thus consider electron $p_T$ smearing.
For electrons,\footnote{
For this group of exotic searches, the dominant lepton uncertainty stems  from $p_T$ mis-measurement.
The energy uncertainty is only 1\% versus a 20\% uncertainty in the electron $p_T$ resolution~\cite{Aad:2009wy}.} 
the $p_T$ resolution is parameterized by \cite{Aad:2009wy}
\begin{equation}
 \frac{\sigma_{p_T}}{p_T} = 0.66 \times \left( \frac{0.10}{\sqrt{p_T/\GeV}} \oplus 0.007 \right). 
 \label{eleSmear.EQ}
\end{equation}
Both the muon $1/p_T$ and electron $p_T$ smearing are translated into an energy smearing, keeping the polar angle unchanged.
We only impose the cuts on the charged leptons as listed in Eq.~(\ref{regCutsLep.EQ}).

\begin{figure}[!t]
\begin{center}
\subfigure[]{\includegraphics[scale=1,width=.48\textwidth]{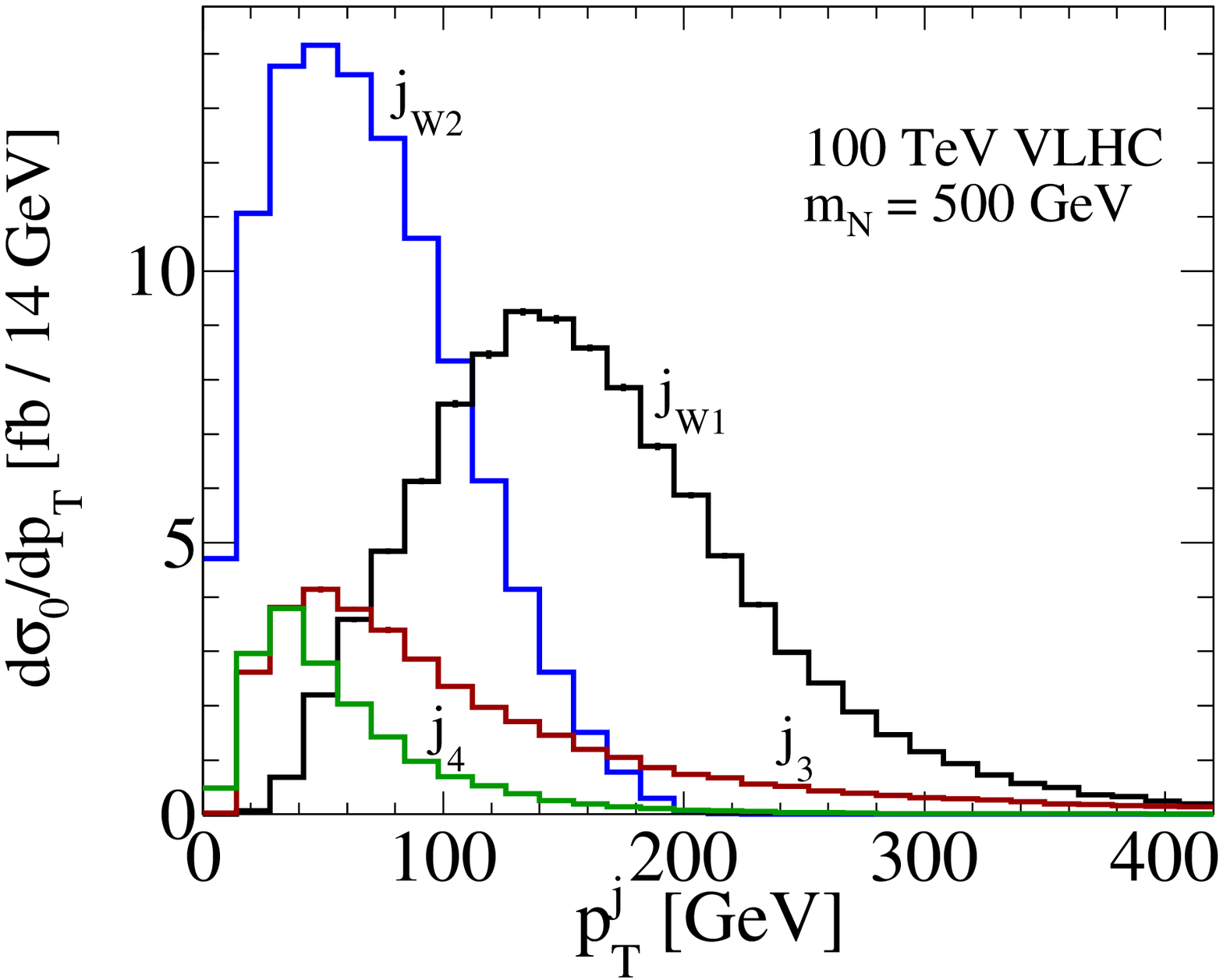}\label{ptj_100TeV_500GeV.fig}}
\subfigure[]{\includegraphics[scale=1,width=.48\textwidth]{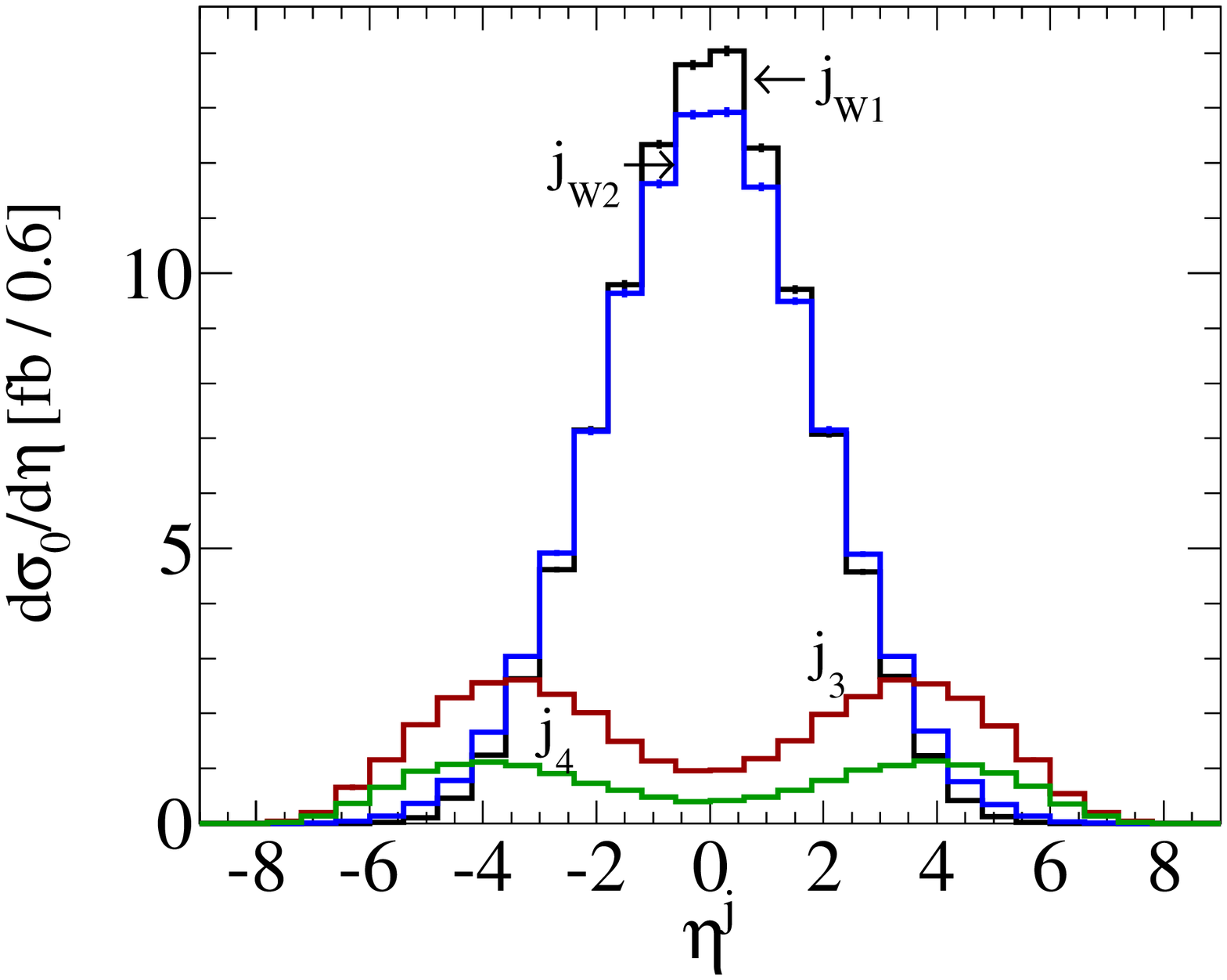}\label{etaj_100TeV_500GeV.fig}}
\\
\subfigure[]{\includegraphics[scale=1,width=.48\textwidth]{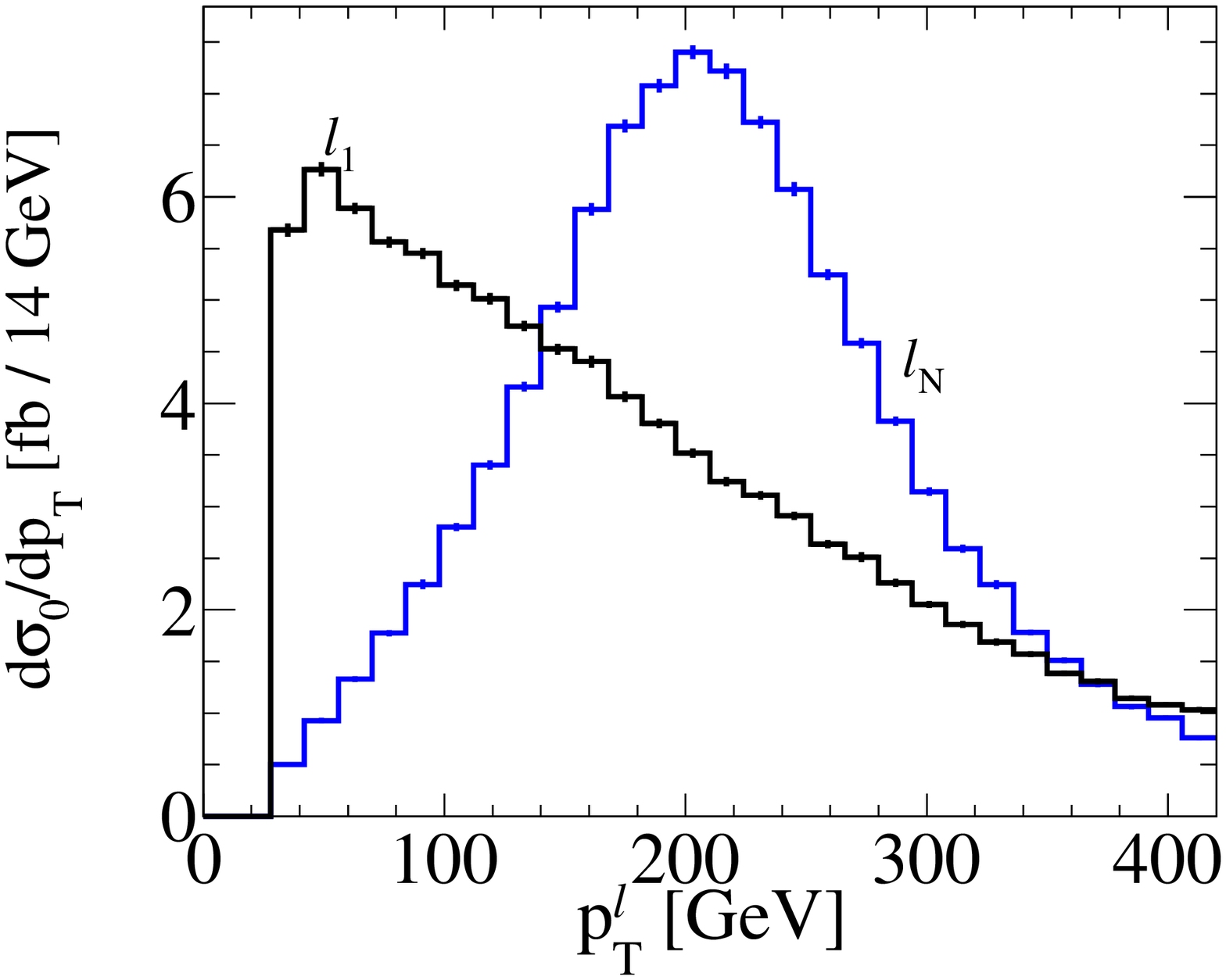}\label{ptl_100TeV_500GeV.fig}}
\subfigure[]{\includegraphics[scale=1,width=.48\textwidth]{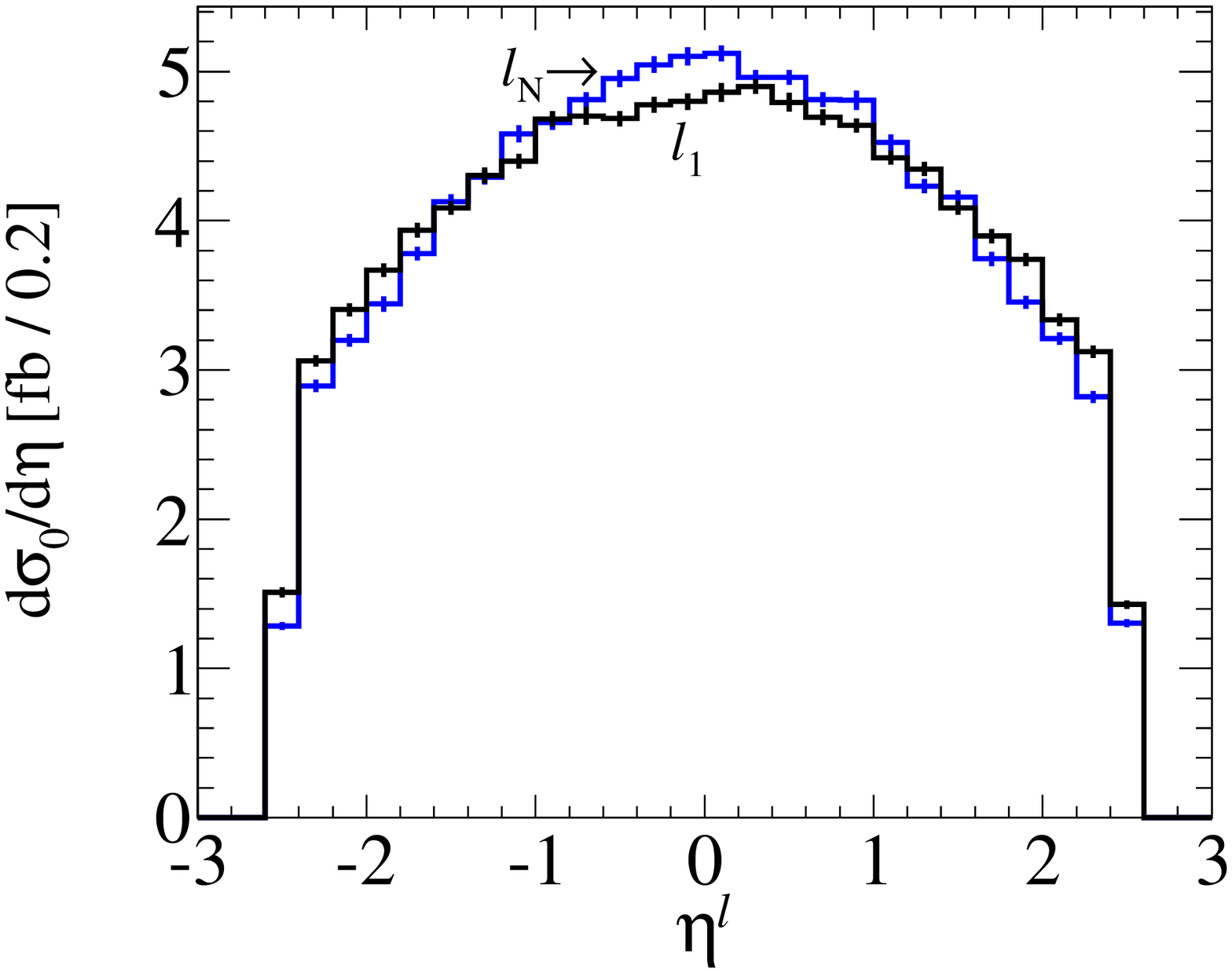}\label{etal_100TeV_500GeV.fig}}
\end{center}
\caption{(a) $p_T$ and (b) $\eta$ differential distributions of the final-state jets for the processes in Eq.~(\ref{ppllnj.EQ}), for $m_N=500\GeV$;
(c,d) the same for final-state same-sign dileptons.}
\label{jets_leps_100TeV.fig}
\end{figure}

In figure~\ref{jets_leps_100TeV.fig}, we show the transverse momentum and pseudorapidity distributions of the final-state jets and same-sign dileptons for the processes in Eq.~(\ref{ppllnj.EQ}), for $m_N = 500$ GeV. 
Jets originating from $N$ decay are denoted by $j_{W_{i}}$, for $i=1,2$, and are ranked by $p_T~(p_{T}^{j_{W_1}}>p_{T}^{j_{W_2}})$. As the three-body $N\rightarrow \ell j j$ decay is preceded by the two-body $N\rightarrow \ell W$ process, $p_T^{j_W}$ scales like $m_N/4$,
as seen in figure~\ref{ptj_100TeV_500GeV.fig}.
The jets produced in association with $N$ are denoted by $j_{3}$ or $j_{4}$, and also ranked by $p_T$.
As VBF drives these channels, we expect $j_3$ (associated with $W^*$) and $j_4$ (associated with $\gamma^*$) to scale 
like $M_W/2$ and $\lamDIS$, respectively.
In figure~\ref{etaj_100TeV_500GeV.fig}, the $\eta$ distributions of all final-state jets are shown.
We see that $j_3$ and $j_4$ are significantly more forward than $j_{W1}$ and $j_{W2}$, consistent with jets participating in VBF.
The high degree of centrality of $j_{W1}$ and $j_{W2}$ follows from the central $W$ decay.

In figures~\ref{ptl_100TeV_500GeV.fig} and~\ref{etal_100TeV_500GeV.fig}, we plot the $p_T$ and $\eta$ distributions of the final-state leptons.
The charged lepton produced in association with $N$ is denoted by $\ell_1$ and the neutrino's child lepton by $\ell_N$. 
As a decay product, $p_T^{\ell_N}$ scales like $(m_N-M_W)/2$, 
whereas $p_{T}^{\ell_1}$ scales as $(\sqrt{\hat{s}}-m_N)/2$.
$\ell_1$ tends to be soft and more forward in the $\gamma$-initiated channels.

\begin{table}[!t]
\caption{Parton-level cuts on 100 TeV $\mu^\pm\mu^\pm jjX$ Analysis.}
 \begin{center}
\begin{tabular}{|c|c|c|}
\hline\hline
Lepton Cuts & Jet Cuts & Other Cuts \tabularnewline\hline\hline
 $\Delta R_{\ell\ell}>0.2$				&$\Delta R_{jj}>0.4$				& $\Delta R_{\ell j}^{\rm Min} > 0.6$	
 \tabularnewline
 $p_T^\ell ~(p_{T}^{\ell ~\rm Max})> 30~(60)\GeV$ 	&$p_T^j ~(p_{T}^{j~\rm Max})> 30~(40)\GeV$ 	& $\not\!\! E_T < 50\GeV$ \tabularnewline
$\vert \eta^\ell\vert<2.5$ 				&$\vert\eta^j\vert < 2.5$	 		& $\vert m_{N}^{\rm Candidate} - m_N \vert < \rm 20\GeV$
\tabularnewline
							&$\vert M_{W}^{\rm Candidate} - M_W \vert < 20\GeV$ 	& \tabularnewline	
							&$\vert m_{jjj} - m_t \vert < 20\GeV$ (Veto) 		&\tabularnewline\hline	
\hline
\end{tabular}
\label{100TeVCuts.TB}
\end{center}
\end{table}
\begin{table}[!t]
\caption{Acceptance rates and percentage efficiencies for the signal $\mu^\pm\mu^\pm jjX$ at 100 TeV VLHC.}
 \begin{center}
\begin{tabular}{|c|c|c|c|}
\hline 
 $\sigma_{0}$ [Eq.~(\ref{bareXSecDef.EQ})] [fb]	$\quad\backslash\quad$ $m_N$ [GeV] & $300$ & $500$ & $1000$ \tabularnewline\hline\hline  
Fiducial + Kin.~ + Smearing	[Eq.~(\ref{fidkinsmCut.EQ})]	&281~(41\%)	&83.9~(45\%)   	& 11.6~(28\%)  	\tabularnewline\hline
Leading $p_{T}$ Minimum			[Eq.~(\ref{leadPTCut.EQ})]	&278~(99\%)	&83.8~($>$99\%)	& 11.6~($>$99\%) \tabularnewline\hline
$\Delta R_{\ell j}$ Separation		[Eq.~(\ref{dRljCut.EQ})]    	&264~(95\%)	&79.3~(95\%)	& 10.7~(92\%)  \tabularnewline\hline
$\not\!\! E_{T}$ Maximum		[Eq.~(\ref{metCut.EQ})]   	&263~($>$99\%)	&78.1~(99\%) 	& 10.1~(95\%) \tabularnewline\hline
$M_{W}$ Reco.~ 				[Eq.~(\ref{mWCut.EQ})]	   	&252~(96\%)	&74.1~(95\%) 	& 9.51~(94\%) \tabularnewline\hline
$m_{t}$ Veto				[Eq.~(\ref{mtCut.EQ})]		&251~(99\%)	&73.5~(99\%) 	& 9.42~(99\%)\tabularnewline\hline
$m_{N}$ Reco.~ 				[Eq.~(\ref{mNCut.EQ})]		&244~(98\%)	&64.7~(88\%) 	 & 7.79~(83\%)  \tabularnewline\hline
\hline
Acceptance $[\mathcal{A}] = \sigma_{0}^{\rm ~All~Cuts} / \sigma_{0}^{\rm Fid.+Kin.+Sm.}$	& 87\%	& 77\%	& 67\%	\tabularnewline\hline
\hline
\end{tabular}
\label{acceptXSec.TB}
\end{center}
\end{table}

\subsection{Signal Definition and Event Selection: Same-Sign Leptons with Jets}

For simplicity,  we restrict our study to electrons and muons.
We design our cut menu based on the same-sign muon channel.
Up to detector smearing effects, the analysis remains unchanged for electrons.
A summary of imposed cuts are listed in Table~\ref{100TeVCuts.TB}.
Jets and leptons are identified by imposing an isolation requirement; we require
\begin{equation}
 \Delta R_{jj} > 0.4,\quad \Delta R_{\ell\ell}>0.2.
\label{regCutsDIS.EQ}
\end{equation}
We define our signal as two muons possessing the same electric charge and at least two jets satisfying the following fiducial and kinematic requirements:
\begin{equation}
 \vert \eta^\ell \vert < 2.5, \quad 
 p_{T}^\ell > 30\GeV,\quad  
 \vert \eta^j \vert < 2.5,\quad
 p_{T}^j > 30\GeV.
 \label{fidkinsmCut.EQ}
\end{equation}
The bare cross sections [defined by factorizing out $S_{\ell\ell}$ as defined in  Eq.~(\ref{bareXSecDef.EQ})]
after cuts listed in Eqs.~(\ref{fidkinsmCut.EQ}) and (\ref{regCutsDIS.EQ}) and smearing are given in the first row of Table~\ref{acceptXSec.TB},
for representative masses $m_N = 300,~500,$ and 1000 GeV.
Events with additional leptons are rejected. 
Events with additional jets are kept; we have not tried to utilize the VBF channel's high-rapidity jets.
About 30-45\% of all $\ell^\pm\ell^{'\pm} jjX$ events survive these cuts.
As learned from figure~\ref{jets_leps_100TeV.fig}, the $\eta$ requirement given in Ref.~\cite{Avetisyan:2013onh} considerably reduces selection efficiency.
Extending the fiducial coverage to $\eta^{\rm Max} = 3$ or larger, though technically difficult, can be very beneficial experimentally.

\begin{figure}[!t]
\begin{center}
\subfigure[]{\includegraphics[scale=1,width=.48\textwidth]{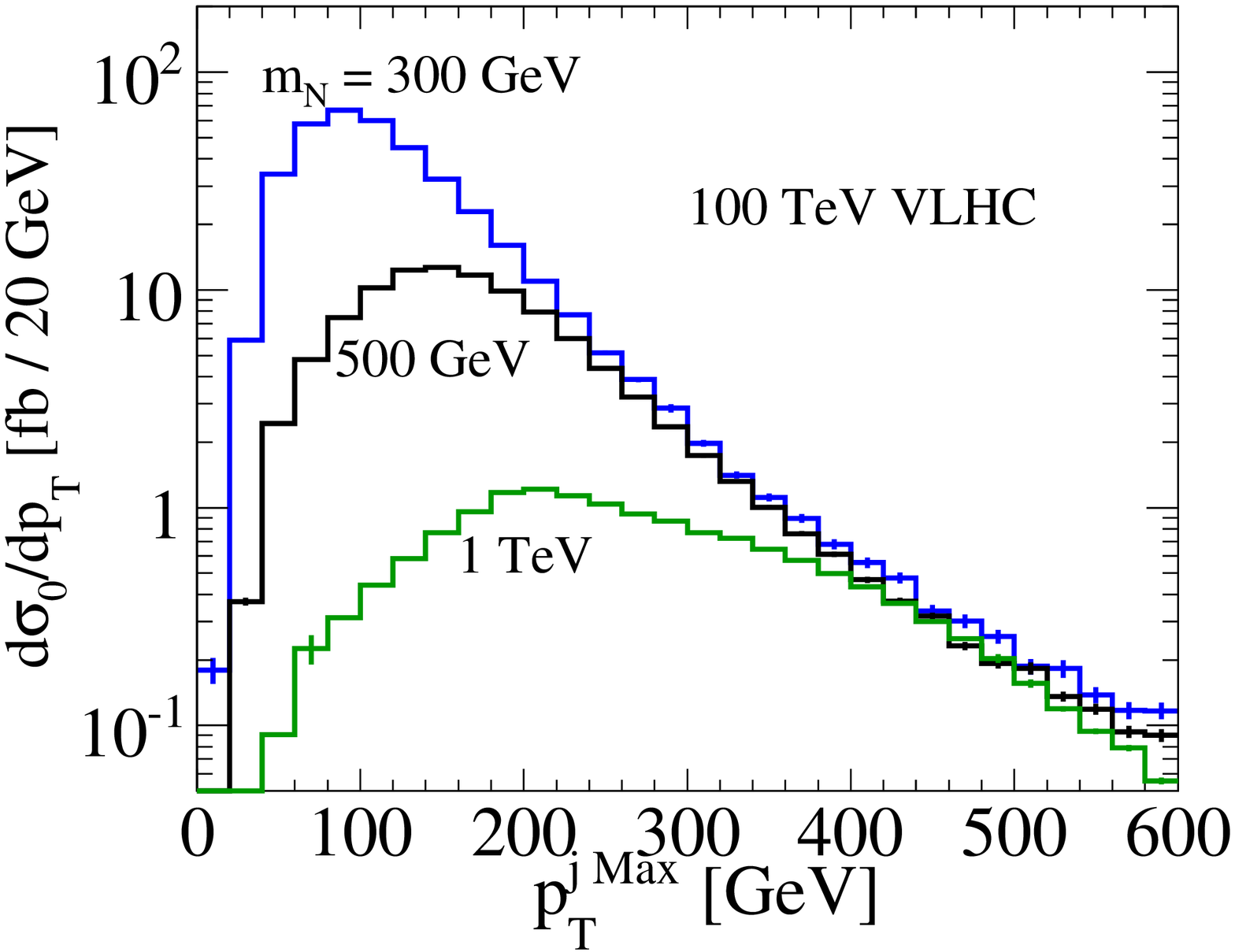}\label{ptjMax_100TeV_MultiGeV.fig}}
\subfigure[]{\includegraphics[scale=1,width=.48\textwidth]{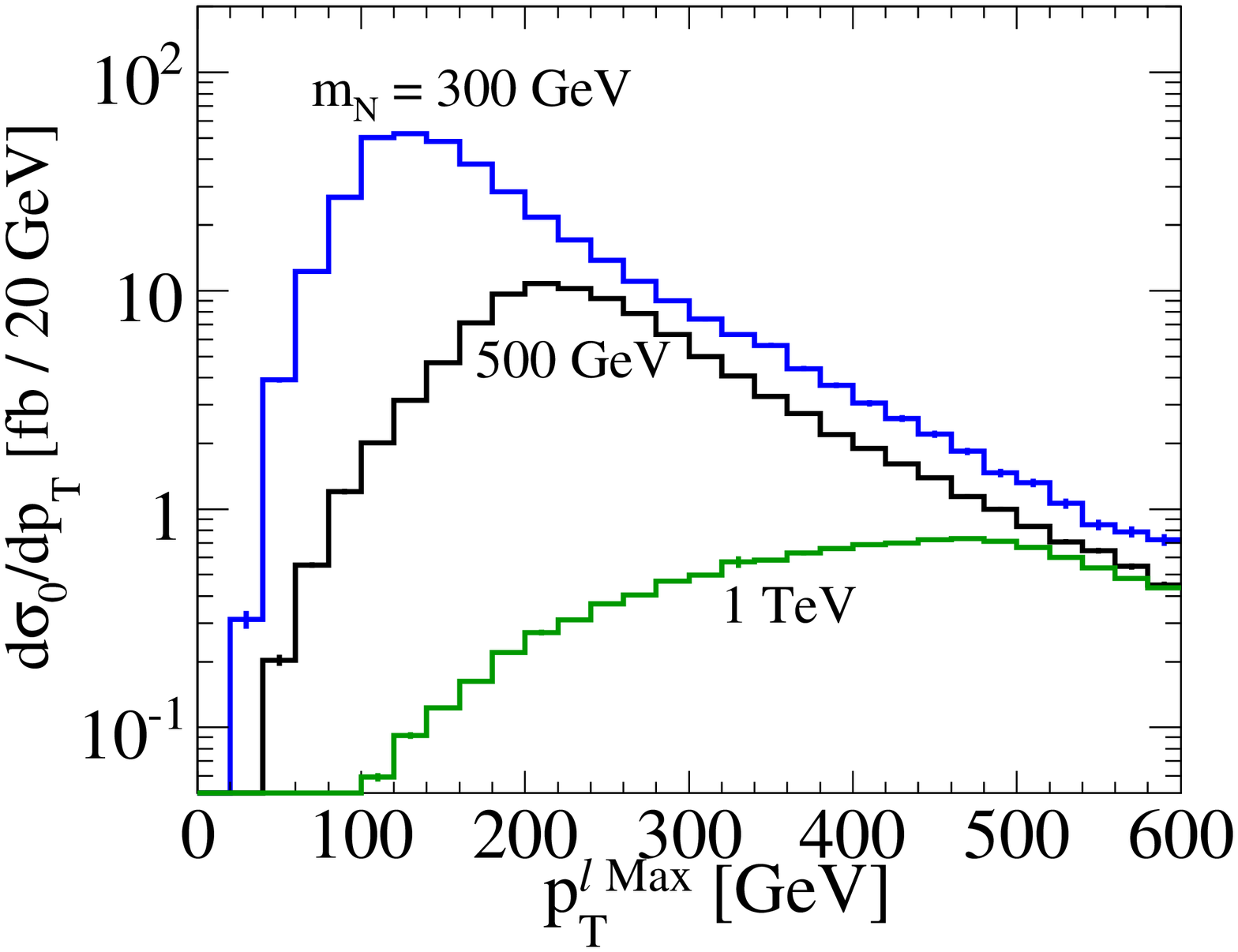}\label{ptlMax_100TeV_MultiGeV.fig}}
\vspace{.2in}\\
\subfigure[]{\includegraphics[scale=1,width=.48\textwidth]{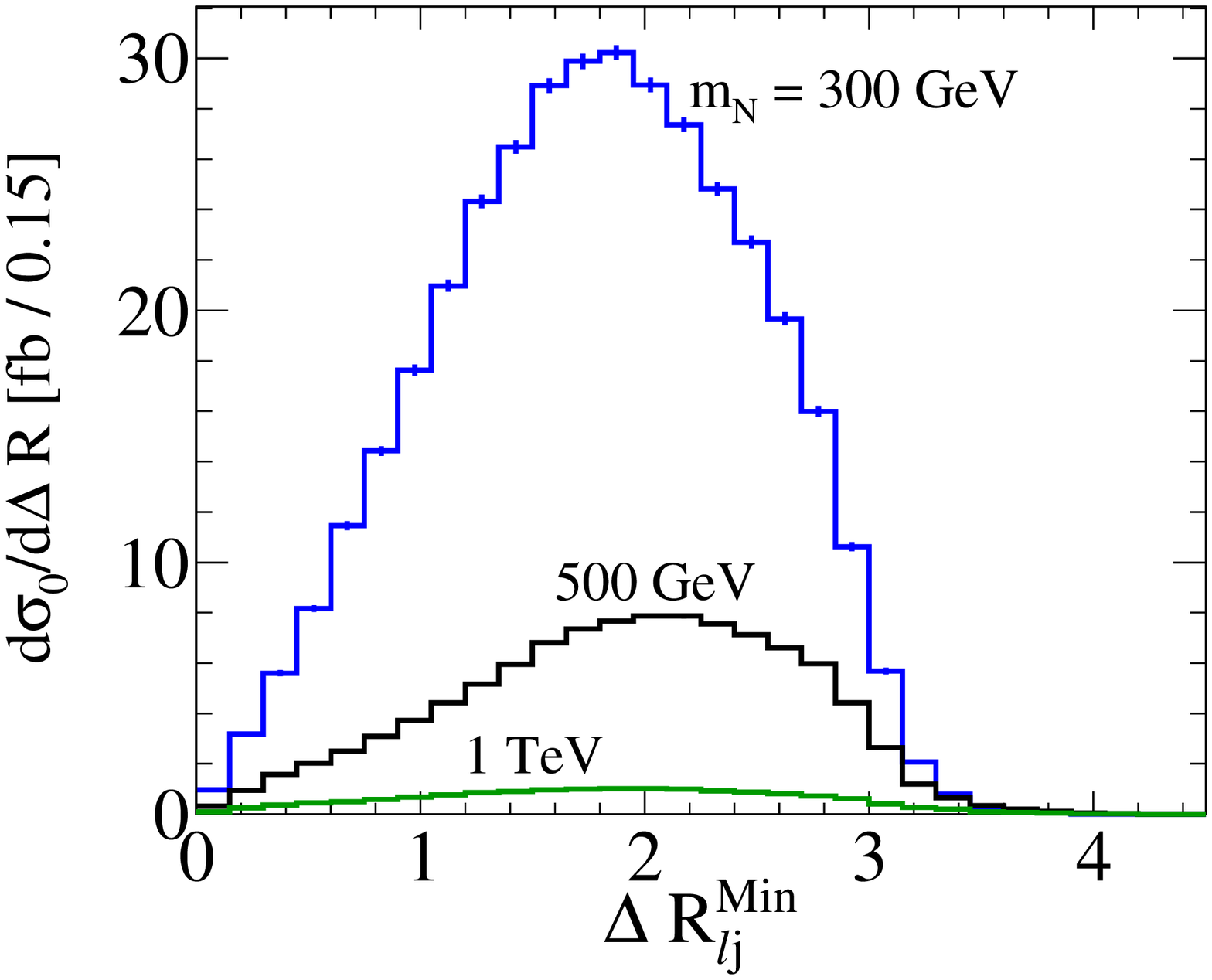}\label{dRljMin_100TeV_MultiGeV.fig}}
\subfigure[]{\includegraphics[scale=1,width=.48\textwidth]{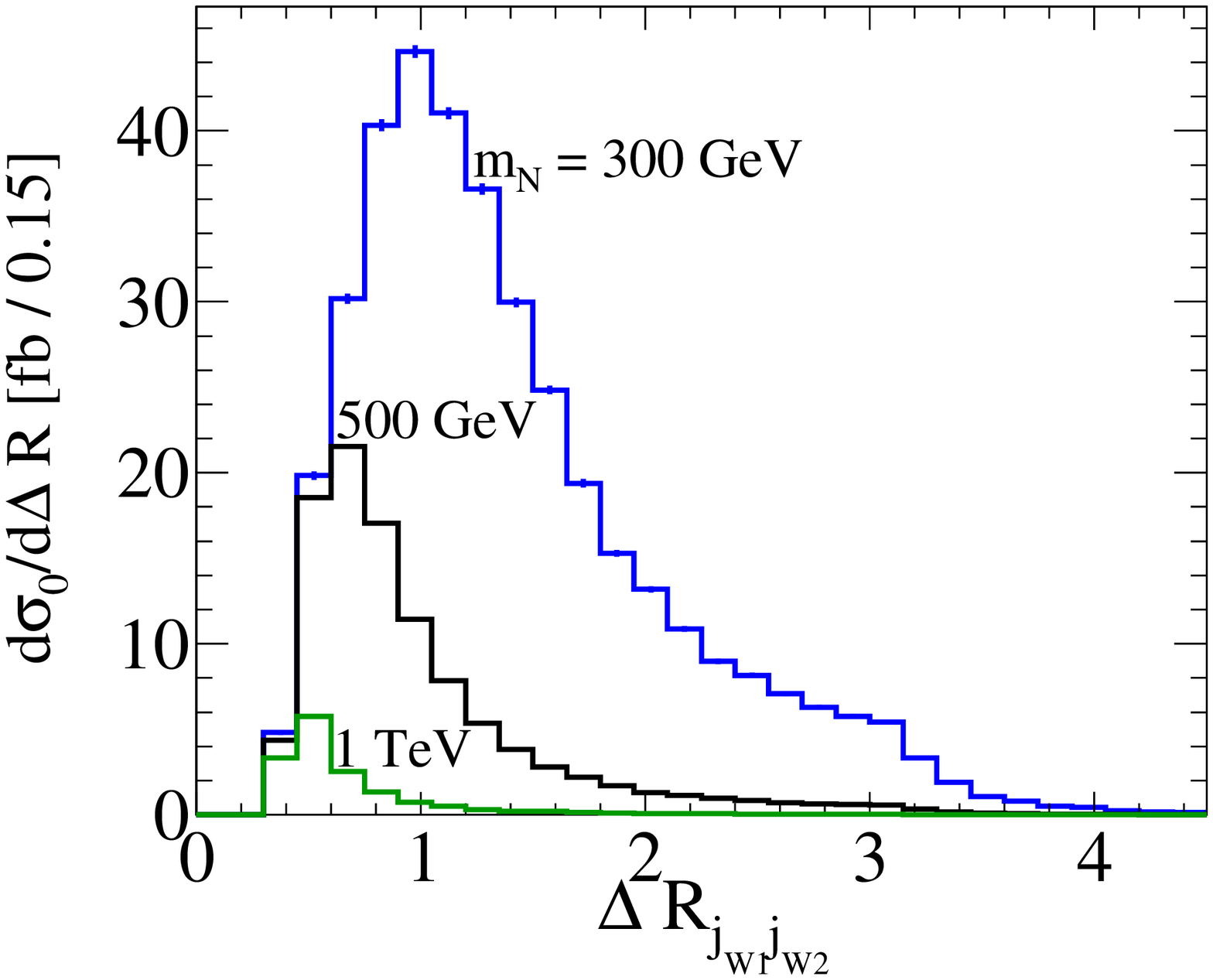}\label{dRjW1jW2_100TeV_MultiGeV.fig}}
\end{center}
\caption{
(a) Maximum jet $p_T$, (b) maximum charged lepton $p_T$, (c) minimum $\Delta R_{\ell j}$, (d) $\Delta R_{j_{W1} j_{W2}}$ distributions for $m_N = 300, \ 500,$ and 1000 GeV.}
\label{maxPt_sep_100TeV.fig}
\end{figure}

We plot the maximum $p_T$ of jets in figure~\ref{ptjMax_100TeV_MultiGeV.fig} and of charged leptons in figure~\ref{ptlMax_100TeV_MultiGeV.fig}, 
for $m_N = 300, ~500,$ and 1000 GeV.
One finds that the $p_T^{j ~\rm Max}$ scale is $m_N/4$ and is set by the $N\rightarrow W \rightarrow jj$ chain.
For the lepton case, $p_T^{\ell ~\rm Max}$ is set by the neutrino decay and scales as $m_N/2$.
In light of these, we apply the following additional selection cuts to reduce background processes:
\begin{equation}
p_{T}^{j~\rm Max} > 40\GeV, \quad p_{T}^{\ell ~\rm Max} > 60\GeV.
\label{leadPTCut.EQ}
\end{equation}
The corresponding rate is given in the second row of Table~\ref{acceptXSec.TB} and we find that virtually all events pass Eq.~(\ref{leadPTCut.EQ}).
As both $p_{T}^{\rm Max}$ are sensitive to $m_N$, searches can be slightly optimized by instead imposing the variable cut 
\begin{equation}
p_{T}^{j~\rm Max} \gtrsim \mathcal{O}\left(\frac{m_N}{4}\right), \quad p_{T}^{\ell ~\rm Max} \gtrsim \mathcal{O}\left(\frac{m_N}{2}\right).
\label{leadPTCutAlt.EQ}
\end{equation}

\begin{figure}[!t]
\begin{center}
\subfigure[]{\includegraphics[scale=1,width=.48\textwidth]{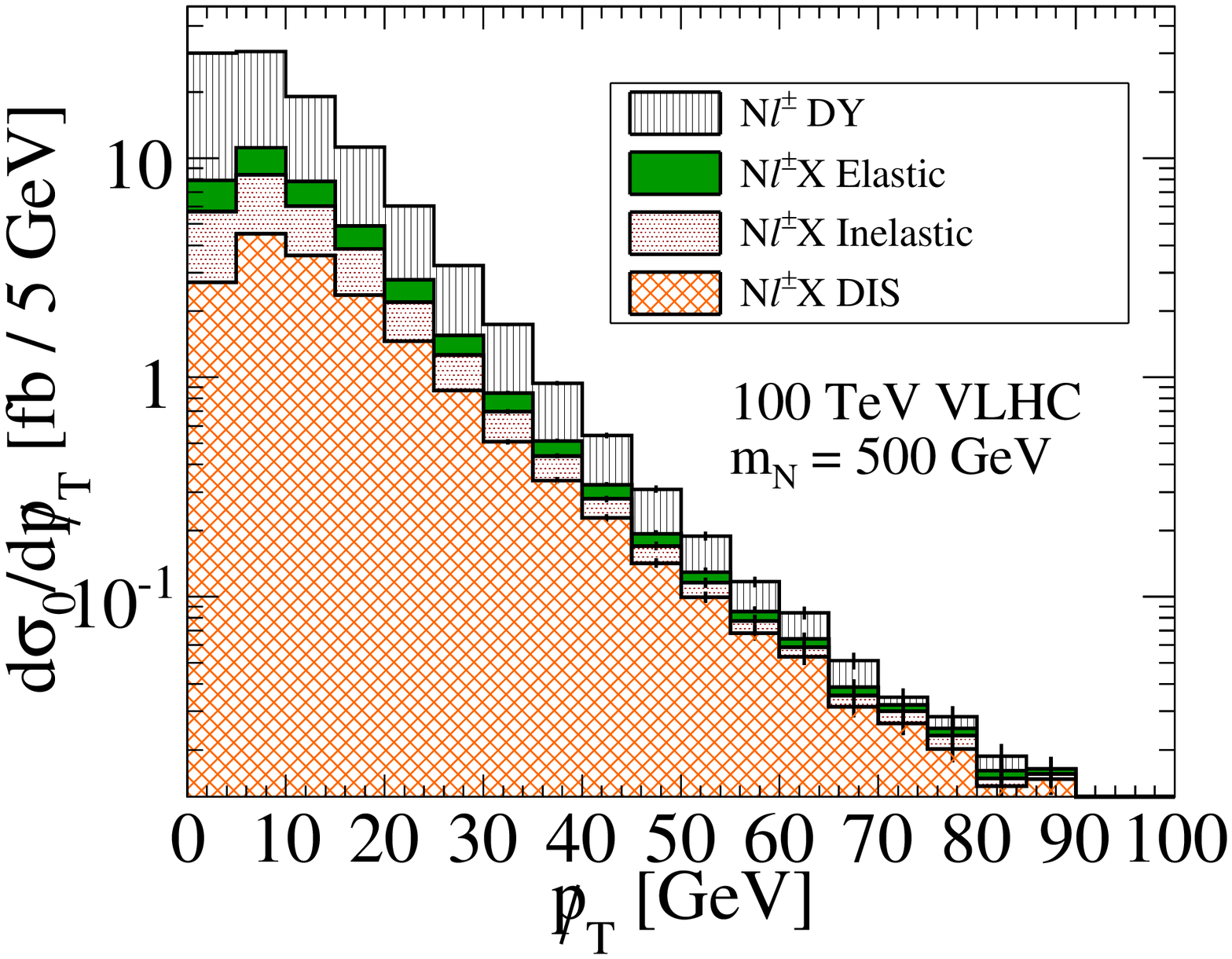}\label{metStacked_100TeV_500GeV.fig}}
\subfigure[]{\includegraphics[scale=1,width=.48\textwidth]{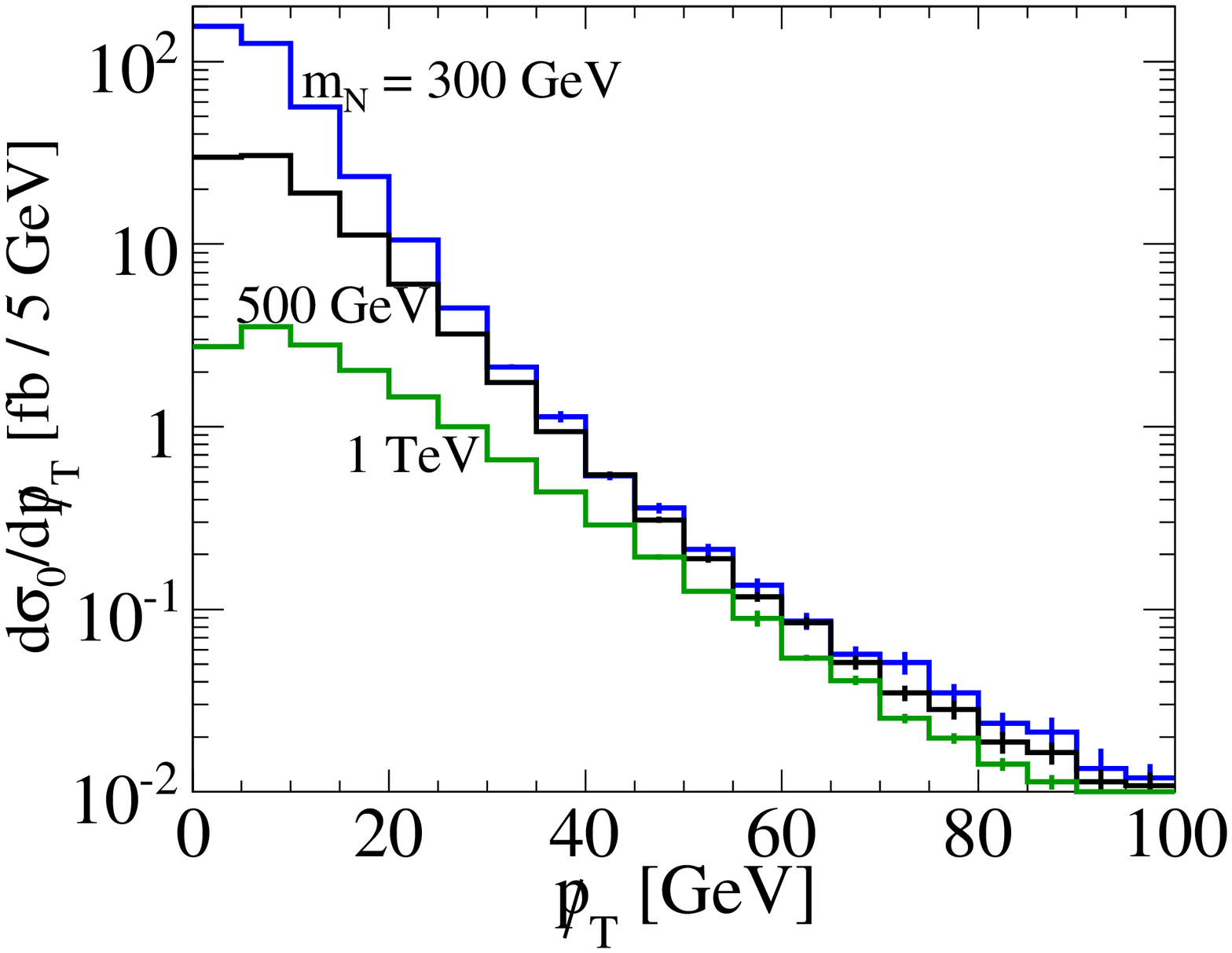}\label{met_100TeV_MultiGeV.fig}}
\end{center}
\caption{
(a) $\slashchar{p_T}$  for individual contributions to $pp\rightarrow \ell^{\pm}\ell^{'\pm}jjX$ at $m_N = 500\GeV$.
(b) Total $\slashchar{p_T}$ 
for same $m_N$ as figure~\ref{maxPt_sep_100TeV.fig}.
}
\label{met_100TeV.fig}
\end{figure}

In each of the several production channels, the final-state charged leptons and jets are widely separated in $\Delta R$; 
see figure~\ref{dRljMin_100TeV_MultiGeV.fig}.
With only a marginal effect on the signal rate, we impose the following cut that greatly reduce heavy quarks backgrounds such as $t\overline{t}$ production~\cite{Han:2006ip}:
\begin{equation}
\Delta R_{\ell j}^{\min} > 0.6.
 \label{dRljCut.EQ}
 \end{equation}
The corresponding rate is given in the third row of Table~\ref{acceptXSec.TB}.
If needed, Eq.~(\ref{dRljCut.EQ}) can be set as high as $1.0$ and still maintain a high signal efficiency. 

In figure~\ref{dRjW1jW2_100TeV_MultiGeV.fig}, the separation between the jets in the $N$ decay is shown.
For increasing $m_N$, the separation decreases.
This is the result of the $W$ boson becoming more boosted at larger $m_N$, resulting in more collimated jets.
For TeV-scale $N$, substructure techniques become necessary for optimize event identification and reconstruction.
We reserve studying the inclusive same-sign leptons with at least one (fat) jet for a future analysis.
 
For the signature studied here, no light neutrinos are present in the final state.
For the heavy neutrino widths listed in Eq.~(\ref{widthParam.EQ}), the decay length $\beta c \tau$ is from $10^{-2}$ $-$ 1 fm, indicating that $N$ is very short-lived.
Thus, there is no source of missing transverse momentum (MET) in the same-sign leptons with $(n+2)j$ aside from detector-level mis-measurements, 
which are parameterized by Eqs.~(\ref{jetSmear.EQ})-(\ref{eleSmear.EQ}).
With this smearing parameterization, forward (large $\eta$) jets are observed with less precision than central (small $\eta$) jets.
Due to the naturally larger energies associated with forward jets participating in VBF at 100 TeV, 
the energy-dependent term in Eq.~(\ref{jetSmear.EQ}) provides a potentially large source of momentum mis-measurements in our analysis.
This channel-dependent behavior can be seen in figure~\ref{metStacked_100TeV_500GeV.fig} for $m_N =$500 GeV.
The increase in MET is found to be modest.
In figure~\ref{met_100TeV_MultiGeV.fig}, we plot the combined MET differential distribution for representative $m_N$.
To maximize the contributions to our signal rate, we impose the loose criterion
\begin{equation}
 \slashchar{p_T} < 50\GeV.
 \label{metCut.EQ}
\end{equation}
The corresponding rate is given in the fourth row of Table~\ref{acceptXSec.TB} and show that most events pass.
Though technically difficult, tightening this cut can greatly enhance the signal-to-noise ratio.

\begin{figure}[!t]
\begin{center}
\subfigure[]{\includegraphics[scale=1,width=.48\textwidth]{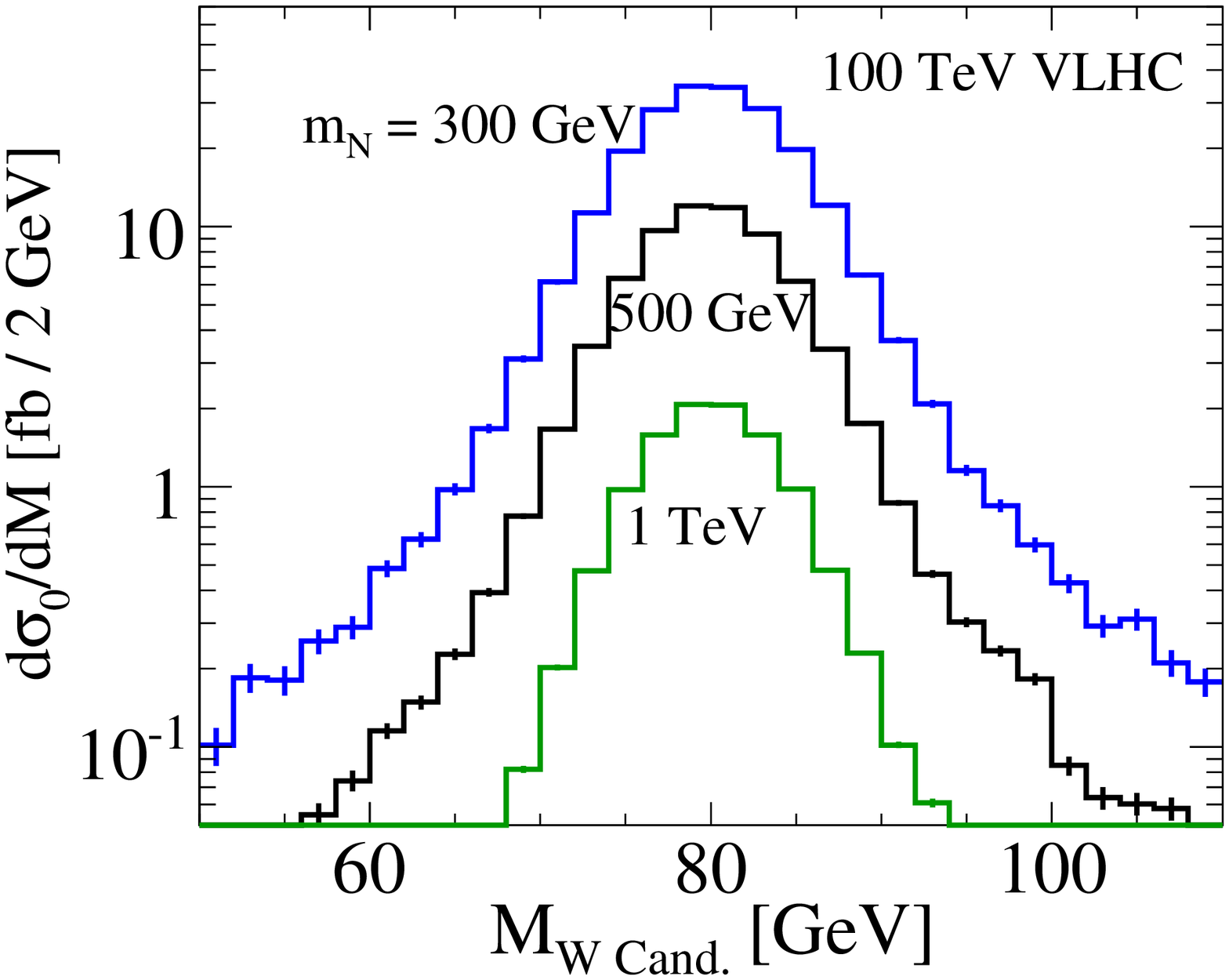}\label{mWReco_100TeV_MultiGeV.fig}}
\subfigure[]{\includegraphics[scale=1,width=.48\textwidth]{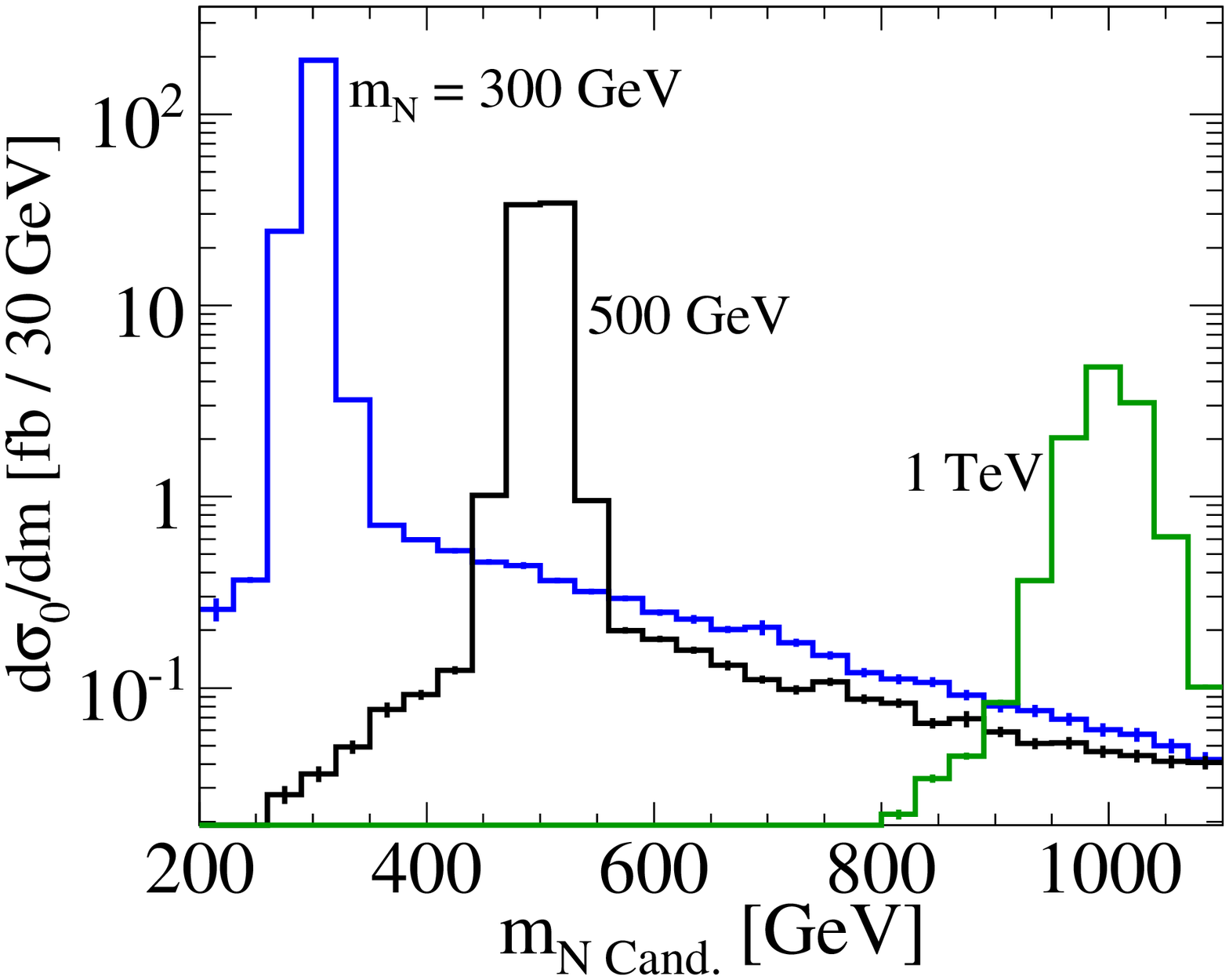}\label{mNReco_100TeV_MultiGeV.fig}}
\end{center}
\caption{Reconstructed invariant mass of the (a) $W$ boson and (b) heavy $N$ candidates for same $m_N$ as figure~\ref{maxPt_sep_100TeV.fig}.}
\label{mN_100TeV.fig}
\end{figure}

To identify the heavy neutrino resonance in the complicated $\ell^\pm\ell^\pm+(n+2)j$ topology, 
we exploit that the $N\rightarrow \ell^\pm jj$ decay results in two very energetic jets that remain very central and possess a resonant invariant mass.
In the $4j$ final-state channel, (rare) contributions from $N\ell^\pm W^\mp$ can lead to the existence of a second $W$ boson in our signal.
To avoid identifying a second $W$ (or a continuum distribution) as the $W$ boson from our heavy neutrino decay, we employ the following algorithm: 
(i) First consider all jets satisfying Eq.~(\ref{fidkinsmCut.EQ}) and require that at least one pair possesses an invariant mass close to $M_W$, i.e.,
\begin{equation}
\vert m_{j_m j_n} - M_W \vert < 20\GeV.
 \label{mWCut.EQ}
\end{equation}
(ii) If no such pair has an invariant mass within 20 GeV of $M_W$, then the event is rejected.
(iii) If more than one pair satisfies Eq.~(\ref{mWCut.EQ}), 
including the situation where one jet can satisfy Eq.~(\ref{mWCut.EQ}) with multiple partners, 
we identify the $jj$-system with the highest $p_T$ as the child $W$ boson from the heavy neutrino decay.
This last step is motived by the fact that the $p_{T}$ of neutrino's decay products scale like $p_T \sim m_N/2$, 
and thus at larger values of $m_N$ the $W$ boson will become more boosted.
This is contrary to $N\ell^\pm W^\mp$ and continuum events, 
in which all states are mostly produced close to threshold.
In figure~\ref{mWReco_100TeV_MultiGeV.fig}, we plot the reconstructed invariant mass of the dijet system satisfying this procedure and observe a very clear resonance at $M_W$.
The corresponding rate is given in the fifth row of Table~\ref{acceptXSec.TB} and show most events pass.

To remove background events from $t\overline{t}W$ production, events containing four or more jets with any three jets satisfying 
\begin{equation}
\vert m_{jjj} - m_t \vert < 20\GeV
 \label{mtCut.EQ}
\end{equation}
are rejected. 
As this is a non-resonant distribution in the $N\ell+nj$ channels, its impact on the signal rate is minimal.
The corresponding rate is given in the sixth row of Table~\ref{acceptXSec.TB} and show that nearly all events pass.
A top quark-veto can be further optimized by introducing high-purity anti-$b$-tagging, e.g., Ref.~\cite{Chatrchyan:2012jua}.

We identify $N$ by imposing the $m_N$-dependent requirement on the two $(\ell_i,W_{\rm Cand.})$ pairs 
and choose whichever system possesses an invariant mass closer to $m_N$.
In figure~\ref{mNReco_100TeV_MultiGeV.fig}, we plot the reconstructed invariant mass of this system observing very clear peaks at $m_N$.
It is important to take into account that the width of the heavy neutrino grows like $m_N^3$, and reaches the 10 GeV-level at $m_N = 1\TeV$.
Therefore, we apply the following width-sensitive cut:
\begin{equation}
 \vert m_{N\rm ~Cand.} - m_N \vert < \rm Max(20\GeV,~3\Gamma_N).
 \label{mNCut.EQ}
\end{equation}
The corresponding rate is given in the seventh row of Table~\ref{acceptXSec.TB} and show most events pass.

\begin{table}[!t]
\caption{Expected $\mu^\pm\mu^{\pm}jjX$ (bare) signal and SM background rates at 100 TeV VLHC after cuts.
 Number of background events and required signal events for $2\sigma$ sensitivity after 100$\invfb$.}
 \begin{center}
\begin{tabular}{|c|c|c|c|c|c|c|}
\hline\hline  
  $m_N$	[GeV] & $100$	& $200$	& $300$	& $400$	& $500$	&$600$   \tabularnewline\hline\hline
 $\sigma_{0}^{\rm ~All~Cuts}$ [fb]		&205	&588	&244	&118	&64.7	&48.1	\tabularnewline\hline
 $\sigma_{\rm Tot}^{\rm SM}$  [ab]		&16.3	&115	&53.2	&22.2	&11.4	&6.01	\tabularnewline\hline
 $n^{b+\delta_{\rm Sys}}_{2\sigma}(100~\invfb)$	&4	&18	&9	&5	&3	&2	\tabularnewline\hline
 $n^{s}_{2\sigma}(100~\invfb)$			&8	&16	&11	&9	&7	&6	\tabularnewline\hline\hline
  $m_N$	[GeV]	& $700$	& $800$	& $900$	& $1000$	& $1100$	&$1200$   \tabularnewline\hline\hline
 $\sigma_{0}^{\rm ~All~Cuts}$ [fb]		&23.4	&14.4	&10.5	&7.79	&4.61	&4.01	\tabularnewline\hline
 $\sigma_{\rm Tot}^{\rm SM}$  [ab]		&3.47	&1.94	&1.57	&1.25	&0.795	&0.649	\tabularnewline\hline
 $n^{b+\delta_{\rm Sys}}_{2\sigma}(100~\invfb)$	&2	&1	&1	&1	&1	&1	\tabularnewline\hline
 $n^{s}_{2\sigma}(100~\invfb)$			&7	&5	&5	&5	&5	&5	\tabularnewline\hline
 \hline
\end{tabular}
\label{100TeVAnaMuMu.TB}
\end{center}
\end{table}

\begin{table}[!t]
\caption{Same as Table \ref{100TeVAnaMuMu.TB} for $e^\pm\mu^{\pm}jjX$.}
 \begin{center}
\begin{tabular}{|c|c|c|c|c|c|c|}
\hline\hline  
  $m_N$ [GeV]	& $100$	& $200$	& $300$	& $400$	& $500$	&$600$   \tabularnewline\hline\hline
 $\sigma_{0}^{\rm ~All~Cuts}$ [fb]		&408	&1160	&480	&230	&125	&93.2	\tabularnewline\hline
 $\sigma_{\rm Tot}^{\rm SM}$  [ab]		&196	&4000	&578	&82.2	&17.7	&8.20	\tabularnewline\hline
 $n^{b+\delta_{\rm Sys}}_{2\sigma}(100~\invfb)$	&27	&434	&71	&13	&4	&3	\tabularnewline\hline
 $n^{s}_{2\sigma}(100~\invfb)$			&18	&71	&30	&13	&8	&8	\tabularnewline\hline\hline
  $m_N$ [GeV]	& $700$	& $800$	& $900$	& $1000$	& $1100$	&$1200$   \tabularnewline\hline\hline
 $\sigma_{0}^{\rm ~All~Cuts}$ [fb]		&44.9	&27.7	&20.3	&15.1	&8.98	&7.86	\tabularnewline\hline
 $\sigma_{\rm Tot}^{\rm SM}$  [ab]		&4.79	&2.68	&2.07	&1.87	&1.29	&0.932	\tabularnewline\hline
 $n^{b+\delta_{\rm Sys}}_{2\sigma}(100~\invfb)$	&2	&1	&1	&1	&1	&1	\tabularnewline\hline
 $n^{s}_{2\sigma}(100~\invfb)$			&6	&5	&5	&5	&5	&5	\tabularnewline\hline
 \hline
\end{tabular}
\label{100TeVAnaEMu.TB}
\end{center}
\end{table}

The acceptance $\mathcal{A}$ of our signal rate, defined as 
\begin{equation}
 \mathcal{A} = \sigma_{\rm All~Cuts} ~/~ \sigma_{\rm Fidcuial~Cuts + Kinematic~Cuts + Smearing},
\end{equation}
is given in the last row of Table~\ref{acceptXSec.TB}.
The total bare rate for the $\mu\mu$ and $\mu e$ channels at representative values of $m_N$ are given, respectively, 
in the Tables ~\ref{100TeVAnaMuMu.TB} and ~\ref{100TeVAnaEMu.TB}.


\subsection{Background}
\label{sec:background}
Although there are no lepton-number violating processes in the SM, 
there exist rare processes with final-state, same-sign leptons as well as  ``faked'' backgrounds from detector mis-measurement.
Here we describe our estimate of the leading backgrounds to the final-state
\begin{equation}
 p p \rightarrow \ell^\pm \ell^{'\pm} + n\geq2 j + X
 \label{finalState.EQ}
\end{equation}
for the $\mu\mu$ and $e\mu$ channels.
The principle SM processes are $t\overline{t}X$, $W^\pm W^\pm X$, and electron charge misidentification.
We model the parton-level matrix elements of these processes using MG5\_aMC@NLO~\cite{Alwall:2014hca} and the CTEQ6L PDFs~\cite{Pumplin:2002vw} with
factorization and renormalization scales $Q = \sqrt{\hat{s}}/2.$ We perform the background analysis in the same manner as for the signal-analysis.

\begin{table}[!t]
\caption{Acceptance rates for SM $t\overline{t}$ at 100 TeV $pp$ collider.}
 \begin{center}
\begin{tabular}{|c|c|c|}
\hline\hline
 $\sigma(t\overline{t}W)$ [fb]	& $e\mu $ & $\mu\mu$ \tabularnewline\hline
Fiducial + Kinematics + Smearing [$K = 1.2$]	[Eq.~(\ref{fidkinsmCut.EQ})]	  
									  &20.5 	&10.3~(26\%) \tabularnewline\hline
Leading $p_{T}$ Minimum			[Eq.~(\ref{leadPTCut.EQ})]	  &16.5 	&8.23~(80\%) \tabularnewline\hline
$\Delta R_{\ell j}$ Separation		[Eq.~(\ref{dRljCut.EQ})]    	  &11.8 	&5.91~(72\%) \tabularnewline\hline
$\not\!\! E_{T}$ Maximum		[Eq.~(\ref{metCut.EQ})]   	  &3.58 	&1.78~(30\%) \tabularnewline\hline
$M_{W}$ Reconstruction			[Eq.~(\ref{mWCut.EQ})]	   	  &2.54 	&1.27~(72\%) \tabularnewline\hline
$m_{t}$ Veto				[Eq.~(\ref{mtCut.EQ})] 		  &0.0452 	&0.0213~(2\%) \tabularnewline\hline
\hline
 $\sigma(t\overline{t})$ (Electron Charge Mis-ID) [fb]	&  \multicolumn{2}{|c|}{$e\mu$}  \tabularnewline\hline
Fiducial + Kinematics + Smearing [Eq.~(\ref{fidkinsmCut.EQ})] [$K = 0.96$]
								& \multicolumn{2}{|c|}{94.5 $\times10^{3}$~(21\%)}  	\tabularnewline\hline
Leading $p_{T}$ Minimum		[Eq.~(\ref{leadPTCut.EQ})]	& \multicolumn{2}{|c|}{67.0 $\times10^{3}$~(71\%)} 	\tabularnewline\hline
$\Delta R_{\ell j}$ Separation	[Eq.~(\ref{dRljCut.EQ})]    	& \multicolumn{2}{|c|}{55.2 $\times10^{3}$~(82\%)} 	\tabularnewline\hline
$\not\!\! E_{T}$ Maximum	[Eq.~(\ref{metCut.EQ})]   	& \multicolumn{2}{|c|}{21.4 $\times10^{3}$~(39\%)} 	\tabularnewline\hline
$M_{W}$ Reconstruction		[Eq.~(\ref{mWCut.EQ})]	   	& \multicolumn{2}{|c|}{3.12 $\times10^{3}$~(15\%)} 	\tabularnewline\hline
$m_{t}$ Veto			[Eq.~(\ref{mtCut.EQ})] 		& \multicolumn{2}{|c|}{3.12 $\times10^{3}$~(100\%)}	\tabularnewline\hline
Charge Mis-ID [$\epsilon_{e~ \rm Mis-ID}$]	[Eq.~(\ref{misID.EQ})] 	& \multicolumn{2}{|c|}{10.9 (0.4\%)}     		\tabularnewline\hline
\hline
\end{tabular}
\label{ttBarBkg.TB}
\end{center}
\end{table}

\subsubsection{$t\overline{t}$}
At 100 TeV, radiative EW processes at $\alpha_{\rm s}^{2}\alpha$ such as 
\begin{eqnarray}
 p~p~ \rightarrow ~t ~\overline{t} ~W^\pm 	
    \rightarrow ~b ~\overline{b} ~W^+ ~W^- ~W^\pm \rightarrow ~\ell^\pm ~\ell^{'\pm} ~b ~\overline{b} ~j ~j ~
    \nu_\ell ~\nu_{\ell'},
      \label{ttW.EQ}
\end{eqnarray}
possess non-negligible cross sections. 
At LO, $\sigma(t\overline{t}W \rightarrow \mu^\pm\mu^\pm b\overline{b}jj\nu_\mu\nu_\mu) \approx 40$ fb, and threatens discovery potential.
At 14 TeV, $t\overline{t}W$ possesses a NLO $K$-factor of $K=1.2$~\cite{Campbell:2012dh}. As an estimate, this is applied at 100 TeV.
As shown in Table \ref{ttBarBkg.TB}, the tight acceptance cuts reduce the rate by roughly 75\%.
Unlike the signal process, $t\overline{t}W$ produces two light neutrinos, an inherent source of  MET.
After the MET cut, the background rate is reduced to the 2 fb level. 
Lastly, the decay chain
\begin{equation}
 t ~\rightarrow ~b ~ W~ \rightarrow ~b~j~j
\end{equation}
can be reconstructed into a top quark.
Rejecting any event with a three-jet invariant mass near the top quark mass, i.e., Eq.~(\ref{mtCut.EQ}), 
dramatically reduces this background to the tens of ab level. 
At this point, approximately {0.2\%} of events passing initial selection criteria survive.

At 100 TeV, the NLO $t\overline{t}$ cross section is estimated to be $\sigma(t\overline{t})\approx 1.8\times 10^7$ fb~\cite{Avetisyan:2013onh}.
Hence, rare top quark decays have the potential to spoil our sensitivity, e.g.,
\begin{eqnarray}
 pp ~ \rightarrow ~t ~\overline{t} ~ 
      \rightarrow ~b ~\overline{b} ~ W^+ ~W^- 
      \rightarrow ~b ~\overline{c} ~\ell^{+} ~\ell^{+'} ~\nu_\ell ~\nu_{\ell'} ~W^{-}  + {\rm c.c.},\label{bToc.EQ}
\end{eqnarray}
where a $b$-quark hadronizes into a $B$-meson that then decays semi-leptonically through the $b\rightarrow c\ell \nu_\ell$ subprocess, 
which is proportional to the small mixing $\vert V_{cb}\vert^{2}$.
The MET and $\Delta R_{\ell j}$ cuts render the  rate negligible~\cite{Atre:2009rg}.
Usage of high-purity anti-$b$ tagging techniques~\cite{Chatrchyan:2012jua} can further suppresses this process.
The $b\rightarrow u$ transition offers a similar background but is $|V_{ub}/V_{cb}|^2\sim (0.1)^2$ smaller~\cite{Beringer:1900zz}.


\subsubsection{Electron Charge Misidentification}
An important source of background for the $e^\pm\mu^\pm$ channel is from electron charge misidentification in fully leptonic decays of top quark pairs:
\begin{equation}
p ~p \rightarrow t ~\overline{t} \rightarrow b ~\overline{b} ~W^{+} ~W^{-}\rightarrow b ~\overline{b} ~ e^{\pm} ~\ell^{\mp} ~\nu_e \nu_\ell, \quad \ell = e, ~\mu.
\label{ttBarOS.EQ}
\end{equation}
Such misidentification occurs when an electron undergoes bremsstrahlung in the tracker volume and the associated 
photon converts into an $e^+e^-$ pair. 
If the electron of opposite charge carries a large fraction of the original electron's energy, 
then the oppositely charged electron may be misidentified as the primary electron.
For muons, this effect is negligible due the near absence of photons converting to muons~\cite{Aad:2011vj,Chatrchyan:2011wba}.
At the CMS detector, the electron charge misidentification rate, $\epsilon_{\rm e~ Mis-ID}$, 
has been determined as a function of generator-level $\eta$~\cite{Chatrchyan:2011wba}.
We assume a conservative, uniform rate of 
\begin{equation}
 \epsilon_{e ~\rm Mis-ID} = 3.5\times 10^{-3}.
 \label{misID.EQ}
\end{equation}

To estimate the effect of electron charge mis-ID at 100 TeV, we consider Eq.~(\ref{ttBarOS.EQ}), normalized to NLO.
Other charge mis-ID channels, including $Z+nj$, are coupling/phase space suppressed compared to $t\overline{t}$.
The $t\overline{t}$ rate after selection cuts is recorded  in Table~\ref{ttBarBkg.TB}, and exists at the 100 pb level.
We find that the electron charge mis-ID rate for Eq.~(\ref{ttBarOS.EQ}) can be as large as 11 fb before the $m_{N~\rm Cand}$ cut is applied.
As either electron in the $e^\pm e^\pm$ channel can be tagged, the mis-ID background is the same size as the $e^\pm\mu^\pm$ channel.
Applying the  $m_{N~\rm Cand}$ cut we observe that the background quickly falls off for $m_N\gtrsim 200\GeV$.
As with other backgrounds possessing final-state bottoms, high purity anti-$b$-tagging offers improvements.
We conclude that the effects of charge misidentification are the dominant background in electron-based final states.

\begin{table}[!t]
\caption{Acceptance rates for SM $W^{\pm}W^{\pm}$ at 100 TeV $pp$ collider.}
 \begin{center}
\begin{tabular}{|c|c|c|}
\hline\hline 
 $\sigma(W^{\pm}W^{\pm}+2j)$ [fb]	 & $e\mu $ & $\mu\mu$ \tabularnewline\hline
Fiducial + Kinematics + Smearing 	 [Eq.~(\ref{fidkinsmCut.EQ})]		&11.6	&5.78~(11\%)     \tabularnewline\hline
Leading $p_{T}$ Minimum			[Eq.~(\ref{leadPTCut.EQ})]		&9.45	&4.72~(82\%) \tabularnewline\hline
$\Delta R_{\ell j}$ Separation		[Eq.~(\ref{dRljCut.EQ})]    		&7.46	&3.63~(77\%)  \tabularnewline\hline
$\not\!\! E_{T}$ Maximum		[Eq.~(\ref{metCut.EQ})]   		&2.56	&1.28~(35\%)  \tabularnewline\hline
$M_{W}$ Reconstruction			[Eq.~(\ref{mWCut.EQ})]	  		&0.132	&0.0664~(5\%)	\tabularnewline\hline
$m_{t}$ Veto				[Eq.~(\ref{mtCut.EQ})]			&0.132	&0.0664~(100\%)	\tabularnewline\hline
\hline 
 $\sigma(W^{\pm}W^{\pm}W^{\mp})$ [fb]	&  $e\mu $ & $\mu\mu$ \tabularnewline\hline
Fiducial + Kinematics + Smearing [$K = 1.8$]	[Eq.~(\ref{fidkinsmCut.EQ})]	&3.35 	&1.68~(13\%)   \tabularnewline\hline
Leading $p_{T}$ Minimum			[Eq.~(\ref{leadPTCut.EQ})]			&2.53	&1.26~(75\%) \tabularnewline\hline
$\Delta R_{\ell j}$ Separation		[Eq.~(\ref{dRljCut.EQ})]    			&2.31	&1.11~(87\%)  \tabularnewline\hline
$\not\!\! E_{T}$ Maximum		[Eq.~(\ref{metCut.EQ})]   			&0.754	&0.375~(34\%)  \tabularnewline\hline
$M_{W}$ Reconstruction			[Eq.~(\ref{mWCut.EQ})]	  			&0.735	&0.368~(98\%)	\tabularnewline\hline
$m_{t}$ Veto				[Eq.~(\ref{mtCut.EQ})]				&0.735	&0.368~(100\%)	\tabularnewline\hline
\hline
\end{tabular}
\label{wpwpBkg.TB}
\end{center}
\end{table}


\subsubsection{$W^{\pm}W^{\pm}$}
The QCD and EW processes at orders $\alpha_{\rm s}^2 \alpha^{2}$ and $\alpha^{3}$ , respectively,
\begin{eqnarray}
  p ~p~ &\rightarrow& ~W^\pm   ~W^\pm ~j ~j 	\label{WW2j}\\
  p ~p~ &\rightarrow& ~W^{\pm} ~W^{\pm} ~W^{\mp}	\label{WWW.EQ}
\end{eqnarray}
present a challenging background due to their sizable rates and kinematic similarity to the signal process.
The triboson production rate at NLO in QCD for 14 TeV LHC has been calculated~\cite{Binoth:2008kt}.
As an estimate, we apply the 14 TeV $K$-factor of $K=1.8$ to the 100 TeV LO $W^\pm W^\pm W^\mp$ channel.
After requiring the signal definition criteria, we find the $W^\pm W^\pm$ backgrounds are present at the several fb-level.
Like $t\overline{t}$, the $W^\pm W^\pm X$ final states possess light neutrinos and non-negligible MET. 
Imposing a maximum on the allowed MET further reduces the background by about {35\%}.
As no $W\rightarrow jj$ decay exists in the QCD process, the reconstructed $M_W$ requirement drops the rate considerably.
After the $m_t$ veto, the SM $W^\pm W^\pm X$ rate is {0.4~(0.9) fb} for the $\mu\mu$ ($e\mu$) channel.

\begin{figure}[!t]
\begin{center}
\subfigure[]{\includegraphics[scale=1,width=.45\textwidth]{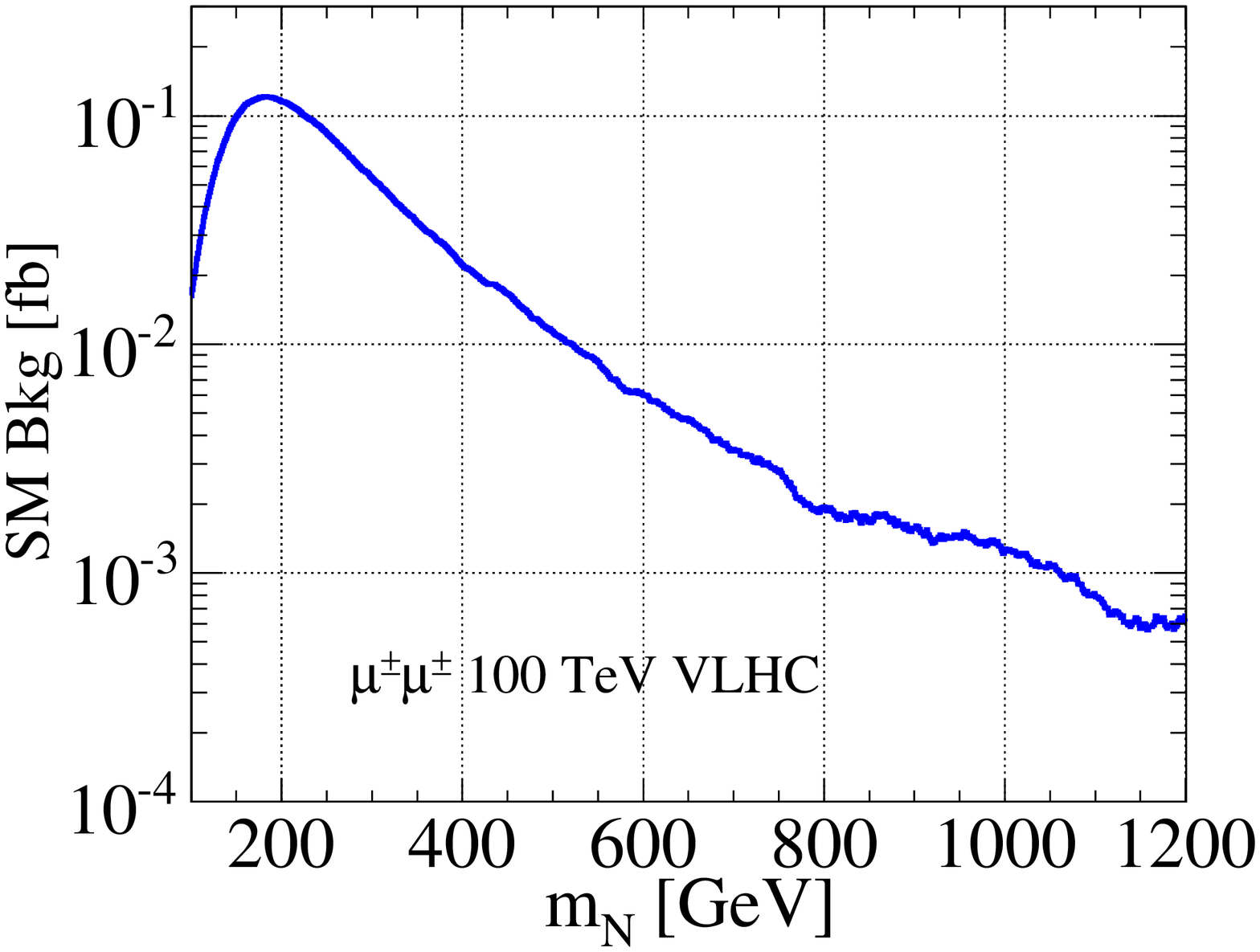}	\label{smBkgMuMu.fig}}
\subfigure[]{\includegraphics[scale=1,width=.45\textwidth]{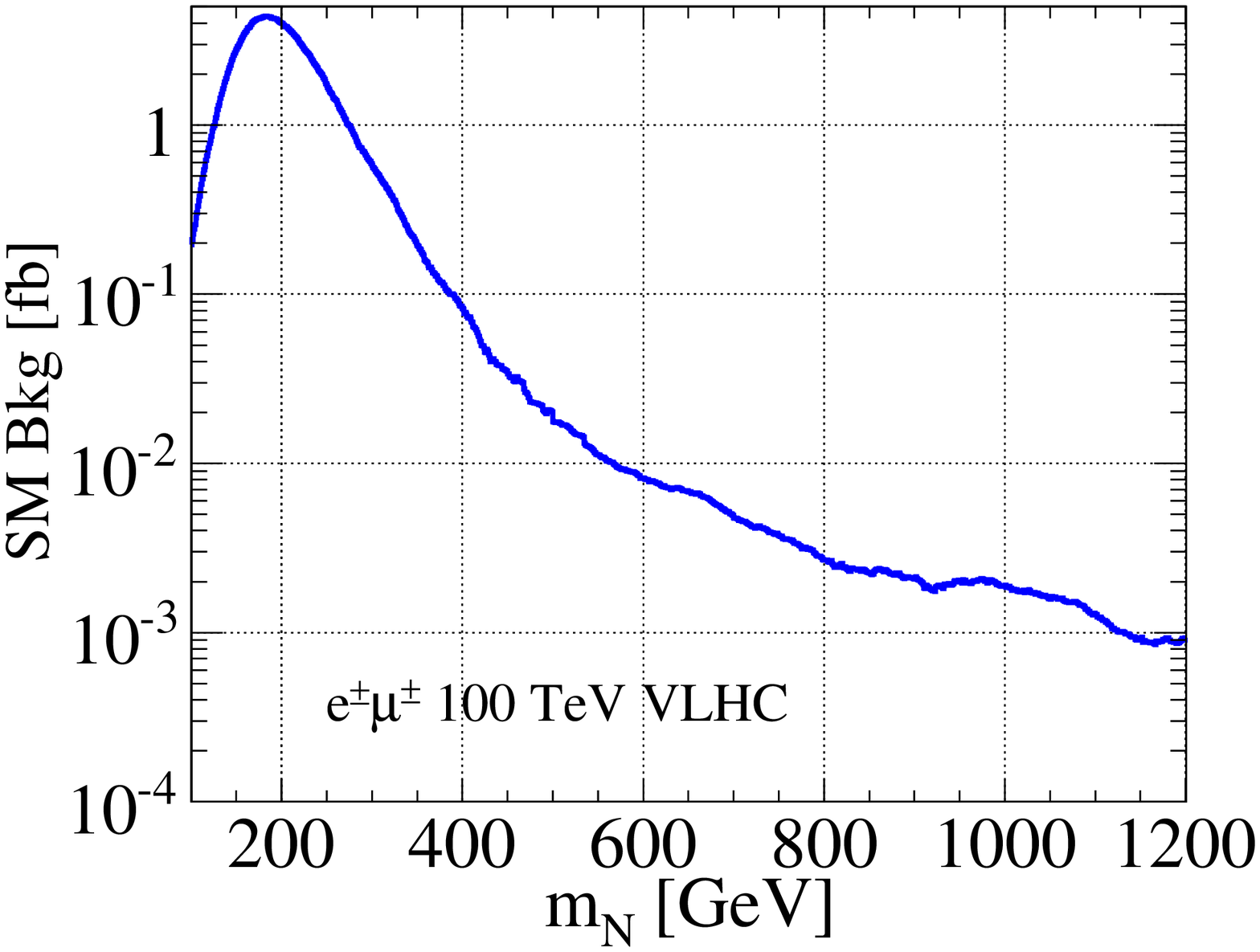}	\label{smBkgEMu.fig}}
\end{center}
\caption{Total SM background versus $m_N$ for (a) $\mu^{\pm}\mu^{\pm}$ and (b) $e^{\pm}\mu^{\pm}$ channels at 100 TeV.}
\label{smBkg.fig}
\end{figure}

For all background channels, 
we apply the $m_N$-dependent cut given in Eq.~(\ref{mNCut.EQ}) on the invariant mass of the reconstructed $W$ candidate with either charged lepton.
The total expected SM background after all selection cuts as a function of $m_N$ are given
for the $\mu\mu$ channel in figure~\ref{smBkgMuMu.fig}, and the $e\mu$ channel in figure~\ref{smBkgEMu.fig}.
The total expected SM background for representative values of $m_N$ are given in Tables~\ref{100TeVAnaMuMu.TB} and ~\ref{100TeVAnaEMu.TB}, respectively.
For these channels, we find a SM background of {$1-115$ ab and $1-4000$ ab} for the neutrino masses considered.
For both channels, the background is greatest for $m_N\lesssim 400\GeV$ and become comparable for $m_N\gtrsim 600\GeV$.

\subsection{Discovery Potential at 100 TeV}
\label{sec:discovery}
\begin{figure}[!t]
\begin{center}
\subfigure[]{\includegraphics[scale=1,width=.48\textwidth]{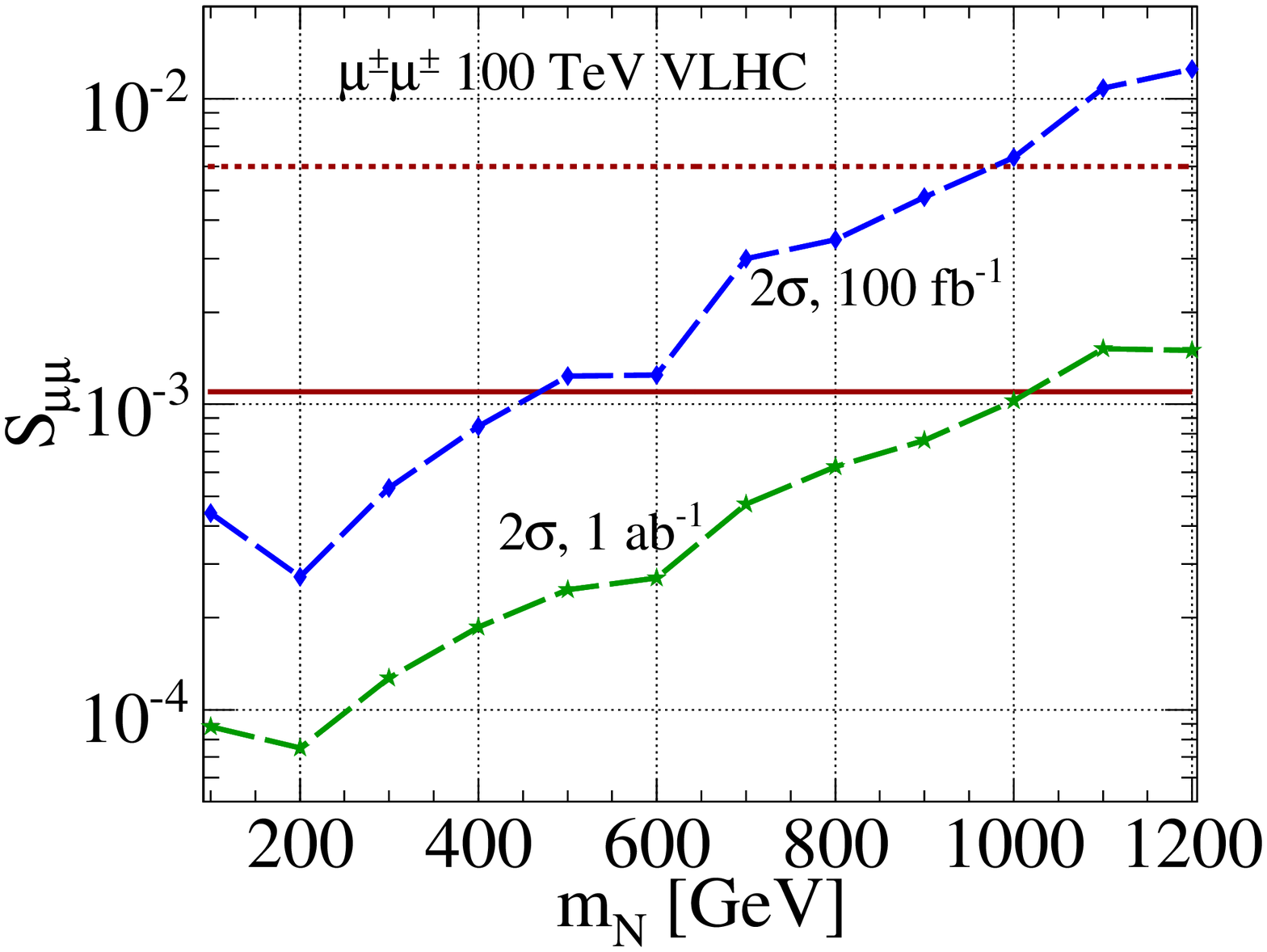}	\label{sMuMuVsMN.fig}}
\subfigure[]{\includegraphics[scale=1,width=.48\textwidth]{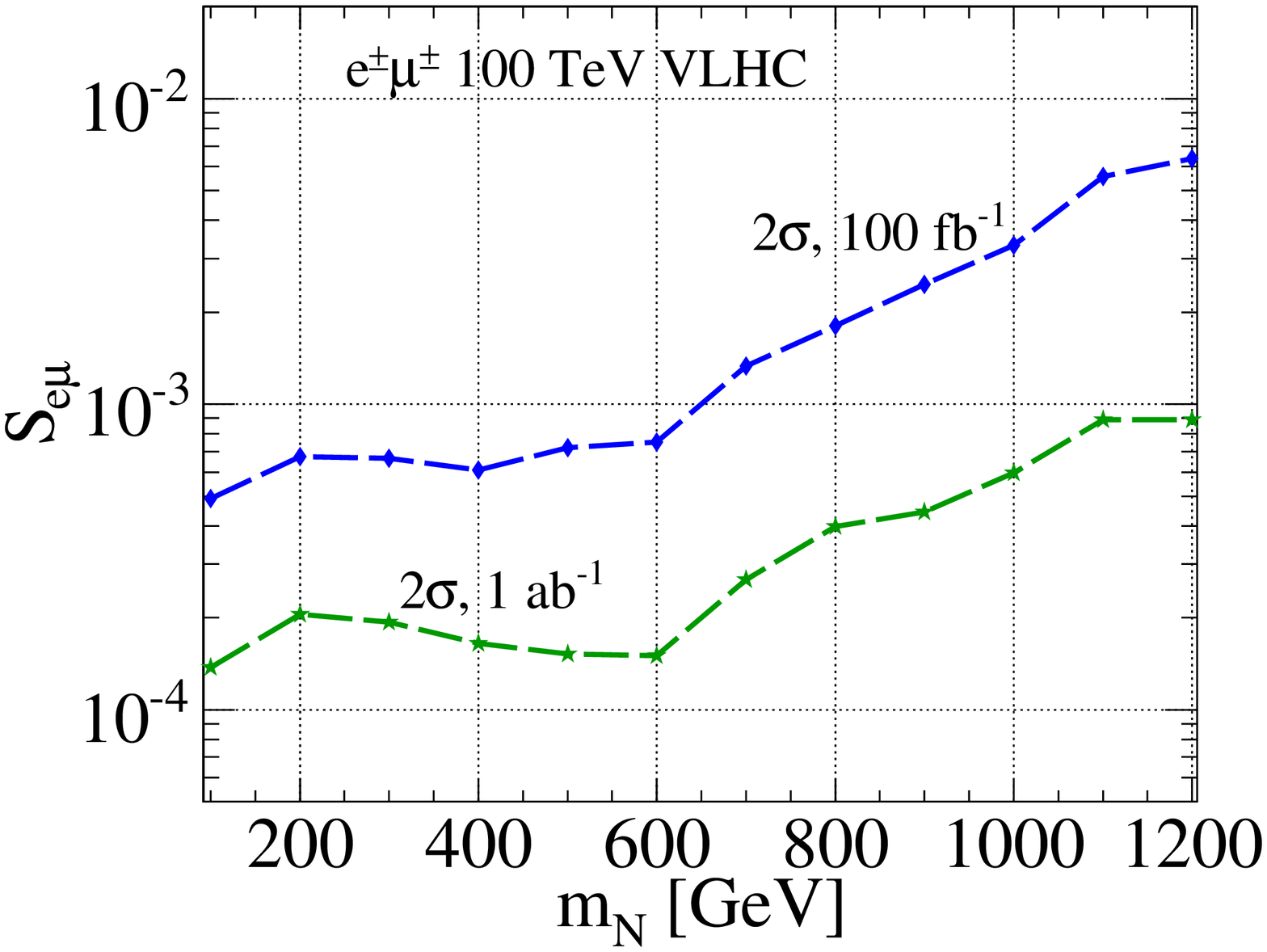}	\label{sEMuVsMN.fig}}
\vspace{.2in}\\
\subfigure[]{\includegraphics[scale=1,width=.48\textwidth]{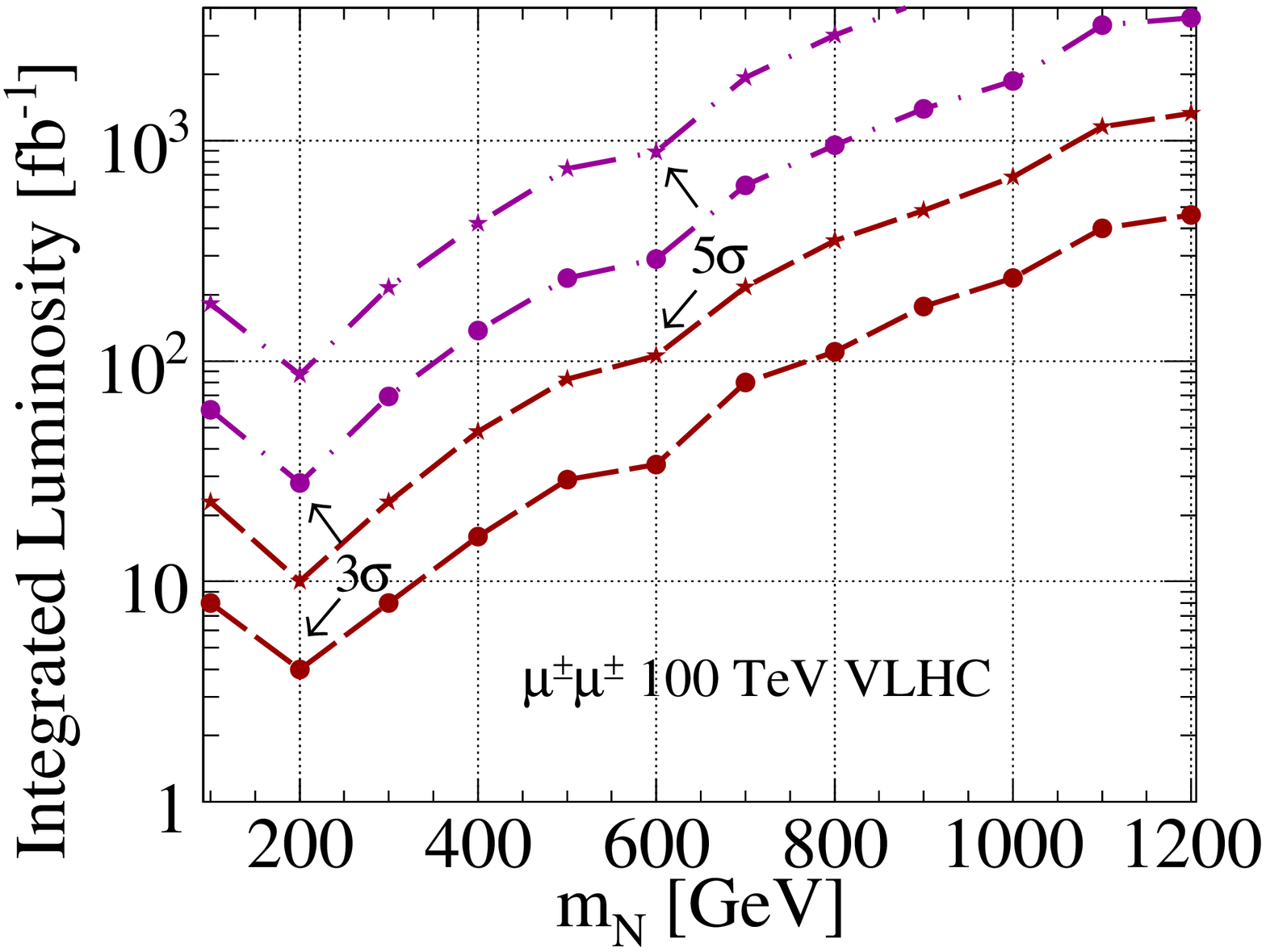}	\label{lumiVsMNMuMu.fig}}
\end{center}
\caption{
At 100 TeV and as a function of $m_N$, the $2\sigma$ sensitivity to $S_{\ell\ell^{'}}$ after $100\ \invfb$ (dash-diamond) and 1 ab$^{-1}$ (dash-star) 
for the (a) $\mu^{\pm}\mu^{\pm}$ and (b) $e^{\pm}\mu^{\pm}$ channels.
The optimistic (pessimistic) bound is given by the solid (short-dash) horizontal line.
(c) The required luminosity for a 3 (dash-circle) and 5$\sigma$ (dash-star) discovery in the $\mu^{\pm}\mu^{\pm}$ channel}
\label{100TeVdiscovery.fig}
\end{figure}

We now estimate the discovery potential at the 100 TeV VLHC of $L$-violation via same-sign leptons and jets. We quantify this using Poisson statistics. 
Details of our treatment can be found in Appendix~\ref{sec:stats}.
The total neutrino cross section is related to the total bare cross section by the expression 
\begin{equation}
 \sigma(\ell^\pm\ell^{'\pm}jj+X) = S_{\ell \ell'} ~\times~ \sigma_{0}(\ell^\pm\ell^{'\pm}jj+X).
\end{equation}
We consider two scenarios for $S_{\mu\mu}$, one used by Ref.~\cite{Atre:2009rg}, dubbed the ``optimistic'' scenario,
\begin{equation}
 S_{\mu\mu} = 6\times 10^{-3},
\end{equation}
and the more stringent value obtained in Eq.~(\ref{smumu.EQ}), dubbed the ``pessimistic'' scenario,
\begin{equation}
 S_{\mu\mu} = 1.1\times 10^{-3}.
\end{equation}
For $S_{e\mu}$, we use the $m_N$-dependent quantity obtained in Eq.~(\ref{semu.EQ}), i.e., $10^{-5}-10^{-6}$.
We introduce a $20\%$ systematic uncertainty by making the following scaling to the SM background cross section
\begin{equation}
 \sigma_{\rm SM} \rightarrow \delta_{\rm Sys} \times \sigma_{\rm SM }, \quad  \delta_{\rm Sys}=1.2.
\end{equation}
For the $\mu\mu$ and $e\mu $ channels, respectively, the maximum number of background events
and requisite number of signal events at a 2$\sigma$ significance after 100$\invfb$ are given in Tables ~\ref{100TeVAnaMuMu.TB} and ~\ref{100TeVAnaEMu.TB}.
For the $\mu\mu$ channel, these span $1-18$ background and $5-16$ signal events; for $e\mu$, $1-434$ and $5-71$ events.

We translate this into sensitivity to the mixing parameter $S_{\ell\ell'}$ and
plot the $2\sigma$ contours in $S_{\ell\ell'}-m_N$ space assuming 100 fb$^{-1}$ (dash-diamond) and 1 ab$^{-1}$ (dash-star) for the 
$\mu\mu$ [figure~\ref{sMuMuVsMN.fig}] and $e\mu$ [figure~\ref{sEMuVsMN.fig}] channels.
In the $\mu\mu$ scenario and $m_N = 500\GeV$, a mixing at the level of 
{$S_{\mu\mu}=1.2\times10^{-3}~(2.5\times10^{-4})$} with 100$^{-1}$ (1 ab$^{-1}$) can be probed.
The optimistic (pessimistic) bound is given by the solid (short-dash) horizontal line.
In the $e\mu$ scenario and the same mass, we find sensitivity to {$S_{e\mu}=7.2~(1.5)\times 10^{-4}$}.
For the $e\mu$ channel, the EW+$0\nu\beta\beta$ bound is at the $10^{-6}-10^{-5}$ level.
Sensitivity to $S_{\ell\ell'}$ at 100 TeV is summarized in Table~\ref{mixingReach.TB}.

Comparatively, we observe a slight ``dip'' (broad ``bump'') in the $\mu\mu~(e\mu)$ curve around 200 GeV.
For the $\mu\mu$ channel, this is due to the low signal acceptance rates for Majorana neutrinos very close to the $W$ threshold;
the search methodology for $m_N$ near or below the $M_W$ has been studied elsewhere~\cite{Han:2006ip,Atre:2009rg}.
For $m_N\geq 200\GeV$, the signal acceptance rate grows rapidly, greatly increasing sensitivity.
In the $e\mu$ channel, the electron charge mis-ID background is greatest in the region around 200 GeV and quickly dwindles for larger $m_N$.
In the low-mass regime, we find greater sensitivity in the $\mu\mu$ channel.
However, due to flavor multiplicity and comparable background rates, the $e\mu$ channel has greater sensitivity in the large-$m_N$ regime.

In figure~\ref{lumiVsMNMuMu.fig}, 
we plot as a function of $m_N$ the required luminosity for a $3\sigma$ (circle) and $5\sigma$ (star) discovery in the $\mu\mu$ channel
for the optimistic (red, dash) and pessimistic (purple, dash-dot) mixing scenarios.
With $100\invfb (1\invab)$ and in the optimistic scenario, a Majorana neutrino with {$m_N=580~(1070)\GeV$} can be discovered at $5\sigma$ significance;
with the same integrated luminosity but in the pessimistic scenario, the reach is {$m_N=215~(615)\GeV$}.
In the optimistic (pessimistic) scenario, for a 375 GeV Majorana neutrino, a benchmark used by Ref.~\cite{Atre:2009rg}, 
a $5\sigma$ discovery can be achieved with {$40~(350)\invfb$}; for 500 GeV, this is {$80~(750)\invfb$}.
Sensitivity to $m_N$ at 100 TeV is summarized in Table~\ref{massReach.TB}.

\begin{table}[!t]
\caption{Sensitivity to the mixing parameter $S_{\ell\ell'}$ at the 14 TeV LHC and 100 TeV VLHC}
 \begin{center}
\begin{tabular}{|c|c|c|c|c|}
\hline\hline 
		& $\mathcal{L}$	& $S_{e\mu}(100\TeV) $ & $S_{\mu\mu}(100\TeV) $  	& $S_{\mu\mu}(14\TeV) $ \tabularnewline\hline
\multirow{2}{*}{$2\sigma$} 	& $100\invfb$	& $4.9 \times10^{-4}$	& $2.7 \times10^{-4}$	& $1.4 \times10^{-4}$	\tabularnewline
		& $1\invab$	& $1.4 \times10^{-4}$	& $7.5 \times10^{-5}$	& $3.1 \times10^{-5}$	\tabularnewline\hline
\multirow{2}{*}{$375\GeV$} 	& $100\invfb$	& $6 \times10^{-4}$	& $7.5 \times10^{-4}$	& $3   \times10^{-3}$	\tabularnewline
		& $1\invab$	& $1.7 \times10^{-4}$	& $1.8 \times10^{-4}$	& $5.5 \times10^{-4}$	\tabularnewline\hline
\multirow{2}{*}{$500\GeV$} 	& $100\invfb$	& $7.2 \times10^{-4}$	& $1.2 \times10^{-3}$	& $8   \times10^{-3}$	\tabularnewline
		& $1\invab$	& $1.5 \times10^{-4}$	& $2.5 \times10^{-4}$	& $1.1 \times10^{-3}$	\tabularnewline\hline
\hline
\end{tabular}
\label{mixingReach.TB}
\end{center}
\end{table}

\begin{table}[!t]
\caption{Sensitivity to heavy neutrino production in the $\mu\mu$ channel at 14 and 100 TeV.}
 \begin{center}
\begin{tabular}{|c|c|c|c|c|c|c|}
\hline\hline 
$100\TeV$ & $2\sigma(100\invfb)$  & $5\sigma(100\invfb)$ & $5\sigma(1\invab)$ & $\mathcal{L}_{5\sigma}(375\GeV)$ & $\mathcal{L}_{5\sigma}(500\GeV)$
   \tabularnewline\hline
 Optimistic	& $980 \GeV$		& $580 \GeV$ 		& $1070 \GeV$		& 40$\invfb$	& 80$\invfb$ \tabularnewline\hline
 Pessimistic	& $470 \GeV$		& $215 \GeV$		& $615 \GeV$		& 380$\invfb$	& 750$\invfb$ \tabularnewline\hline
 \hline
$14\TeV$ & $2\sigma(100\invfb)$ & $5\sigma(100\invfb)$ & $5\sigma(1\invab)$ & $\mathcal{L}_{5\sigma}(375\GeV)$ & $\mathcal{L}_{5\sigma}(500\GeV)$
   \tabularnewline\hline
 Optimistic	& $465 \GeV$ 		& $270\GeV$ 		& $530\GeV$		& 300$\invfb$	& 810$\invfb$  \tabularnewline\hline
 Pessimistic	& $255 \GeV$		& $135\GeV$		& $280\GeV$		& 2.6$\invab$	& 6.9$\invab$ \tabularnewline\hline
\hline
\end{tabular}
\label{massReach.TB}
\end{center}
\end{table}

\subsection{Updated Discovery Potential at 14 TeV LHC}
\label{sec:14TeVLHC}
\begin{table}[!t]
\caption{Parton-level cuts on 14 TeV $\mu^\pm\mu^\pm jjX$ Analysis}
 \begin{center}
\begin{tabular}{|c|c|c|}
\hline\hline
Lepton Cuts & Jet Cuts & Other Cuts \tabularnewline\hline\hline
 $\Delta R_{\ell\ell}>0.2$	&$\Delta R_{jj}>0.4$	& $\Delta R_{\ell j}^{\rm Min} > 0.5$	\tabularnewline
 $p_T^\ell ~(p_{T}^{\ell ~\rm Max})> 10~(30)\GeV$ 	& $p_T^j ~(p_{T}^{j~\rm Max})> 15~(40)\GeV$ 	& $\not\!\! E_T < 35\GeV$ \tabularnewline
$\vert \eta^\ell\vert<2.4$ 	& $\vert\eta^j\vert < 2.4$	 	& $\vert m_{N}^{\rm Candidate} - m_N \vert < \rm 20\GeV$ \tabularnewline
				&$\vert M_{W}^{\rm Candidate} - M_W \vert < 20\GeV$ & \tabularnewline	
				&$\vert m_{jjj} - m_t \vert < 20\GeV$ (Veto) &\tabularnewline\hline	
\hline
\end{tabular}
\label{14TeVCuts.TB}
\end{center}
\end{table}

\begin{table}[!t]
\caption{Same as Table \ref{100TeVAnaMuMu.TB} for 14 TeV LHC.}
 \begin{center}
\begin{tabular}{|c|c|c|c|c|c|c|c|}
\hline\hline  
 $\sigma$	$\backslash$ $m_N$ [GeV] & $100$	& $200$	& $300$	& $400$	& $500$	&$600$ &$700$  \tabularnewline\hline\hline
 $\sigma_{0}^{\rm ~All~Cuts}$ [fb]	&576	&132	&36.0	&14.0	&6.28	&3.05  &1.55 \tabularnewline\hline
 $\sigma_{\rm Tot}^{\rm SM}$  [ab]	&14.1	&18.6	&5.62	&2.05	&0.837	&0.393	&0.195 \tabularnewline\hline
 $n^{b+\delta_{\rm Sys}}_{2\sigma}(100~\invfb)$	&4	&4	&2	&1	&1	&0	&0  \tabularnewline\hline
 $n^{s}_{2\sigma}(100~\invfb)$			&8	&8	&6	&5	&5	&4	&4  \tabularnewline\hline
 \hline
\end{tabular}
\label{14TeVAna.TB}
\end{center}
\end{table}

We update the 14 TeV LHC discovery potential to heavy Majorana neutrinos above the $W$ boson threshold decaying to same-sign muons.
Our procedure largely follows the 100 TeV scenario but numerical values are based on Ref.~\cite{Atre:2009rg}.
Signal-wise, we require exactly two same-sign muons (vetoing additional leptons) 
and at least two jets (allowing additional jets) satisfying the cuts listed in Table~\ref{14TeVCuts.TB}.
Differences from the analysis introduced by Ref.~\cite{Atre:2009rg} include:
updated smearing parameterization given in Eqs.~(\ref{jetSmear.EQ}) and (\ref{muSmear.EQ});
an $\not\!\! E_T$ requirement based on the ATLAS detector capabilities given in Ref.~\cite{ATLAS:2012yoa};
cuts on the leading charged lepton and jet; and more stringent requirements on the $W$ and $N$ candidate masses.
These differences sacrifice sensitivity to $m_N\lesssim 100\GeV$ for high-mass reach.
For our NNLO in QCD $K$-factor, we use $K=1.2$, as given in Eq.~(\ref{K.EQ}).
We report the bare heavy neutrino rate after all cuts for representative $m_N$ in the first row of Table~\ref{14TeVAna.TB}.
The total bare rate ranges from {$2-580$} fb for $m_N = 100-700\GeV$.

\begin{figure}[!t]
\begin{center}
\subfigure[]{\includegraphics[scale=1,width=.48\textwidth]{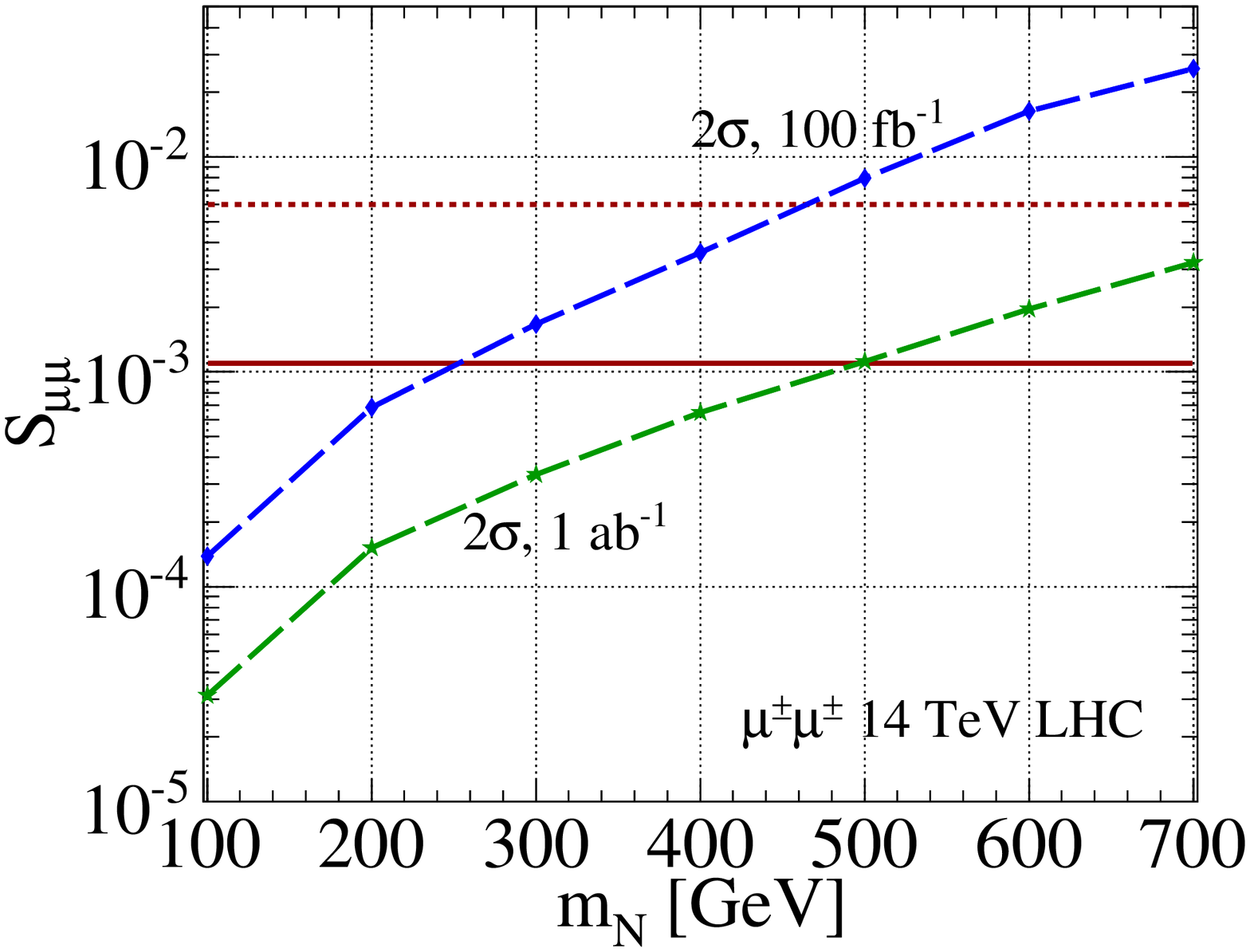}	\label{sMuMuVsMN14TeV.fig}}
\subfigure[]{\includegraphics[scale=1,width=.48\textwidth]{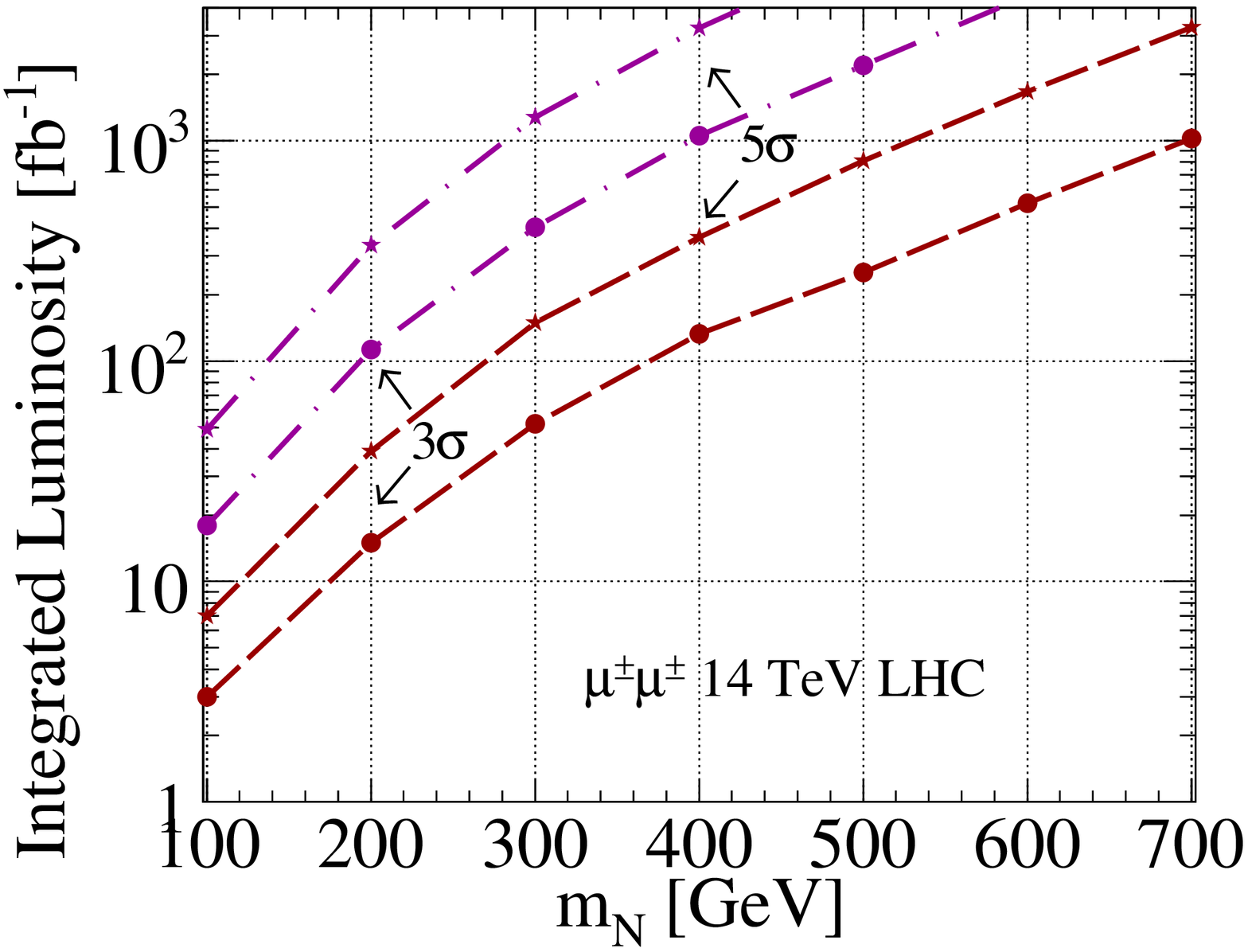}\label{lumiVsMNMuMu14TeV.fig}}
\end{center}
\caption{
At 14 TeV, (a) same as figure~\ref{sMuMuVsMN.fig}; (b) same as figure~\ref{lumiVsMNMuMu.fig}.
}
\label{mumu14TeV.fig}
\end{figure}

As previously discussed or shown, the $t\overline{t}$ background for the dimuon channel is negligible, so we focus on $W^\pm W^\pm$ pairs.
For triboson production, an NLO in QCD $K$ factor of $K=1.8$ is applied~\cite{Binoth:2008kt}.
After all cuts, the expected SM background for representative $m_N$ is given in the second row Table~\ref{14TeVAna.TB}.
After the $m_N$-dependent cut, the expected SM background rate reaches at most {19} ab.
Like the 100 TeV case, a 20\% systematic is introduced into the background.
For the $\mu\mu$ and $e\mu $ channels, respectively, 
The maximum number of background events and requisite number of signal events at a 2$\sigma$ significance after 100$\invfb$ are 
given in the third and fourth rows, respectively, of Table~\ref{14TeVAna.TB}.

%

In figure~\ref{sMuMuVsMN14TeV.fig}, we plot the $2\sigma$ sensitivity to the mixing coefficient $S_{\mu\mu}$ after $100\invfb$ (dash-diamond) 
and 1 ab$^{-1}$ (dash-star).
For the benchmark $m_N = 375\GeV$,
a mixing at the level of {$S_{\mu\mu}=3\times 10^{-3}~(5.5\times10^{-4})$} with 100$^{-1}$ (1 ab$^{-1}$) can be probed;
for $m_N = 500\GeV$, we find sensitivity to be {$S_{\mu\mu}=8\times10^{-3}~(1.1\times10^{-4})$}.
The optimistic (pessimistic) bound is given by the solid (short-dash) horizontal line.
Sensitivity to $S_{\mu\mu}$ at 14 TeV is summarized in Table~\ref{mixingReach.TB}.

In figure~\ref{lumiVsMNMuMu14TeV.fig}, 
we plot as a function of $m_N$ the required luminosity for a $3\sigma$ (circle) and $5\sigma$ (star) discovery in the $\mu\mu$ channel
for the optimistic (red, dash) and pessimistic (purple, dash-dot) mixing scenarios.
With $100\invfb~(1\invab)$ and in the optimistic scenario,
a Majorana neutrino with {$m_N=270~(530)$ GeV} can be discovered at $5\sigma$ significance;
in the pessimistic scenario, the reach is {$m_N=135~(280)$ GeV}. 
In the optimistic (pessimistic) scenario, for the 375 GeV benchmark, a $5\sigma$ discovery can be achieved with {$300~(2600)\invfb$};
for 500 GeV, this is {$810~(6900)\invfb$}.
Sensitivity to $m_N$ at 14 TeV is summarized in Table~\ref{massReach.TB}.


\section{SUMMARY}
\label{sec:summary}
The search for a heavy Majorana neutrino at the LHC is of fundamental importance. 
It is complimentary to the neutrino oscillation programs and, in particular, neutrinoless double-beta decay experiments. 
We have studied the production of a heavy Majorana neutrino at hadron colliders and its lepton-number violating decay as in Eq.~(\ref{ppllnj.EQ}), 
including the NNLO DY contribution, the elastic and inelastic $p\gamma\rightarrow N\ell j$ processes,  and the DIS $pp\rightarrow N\ell jj$ process via $W\gamma^*$ fusion. 
We have determined the discovery potential of the same-sign dilepton signal at a future $100\TeV\,pp$ collider, and updated the results at the 14 TeV LHC.
We summarize our findings as follows:
\begin{itemize}

\item 
Vector boson fusion processes,e.g., $W\gamma \to N \ell$, become increasingly more important at higher collider energies and larger mass scales due to collinear logarithmic enhancements of the cross section. 
At the 14 TeV LHC, the three contributing channels of elastic, inelastic and DIS are comparable in magnitude, 
while at the 100 TeV VLHC, the tendency, in descending importance, is DIS, inelastic, and elastic; see figures~\ref{xsecComb.fig} and ~\ref{xsecComb100TeV.fig}.

 \item We approximately computed the QCD corrections up to NNLO of the DY production of $N\ell$  to obtain the $K$-factor. 
   We found it to span {$1.2-1.5$} for $m_N$ values between  $100\GeV$ and  $1\TeV$ at $14$ and $100~\TeV~pp$ collisions,
   and is summarized in Table~\ref{kFactor.TB}.

   \item The $W\gamma$ fusion processes surpasses the DY mechanism at {$m_{N} \sim 1\TeV \ (770\GeV)$} at the 14 TeV LHC (100 TeV VLHC);
   see figure~\ref{xsecRatio.fig} [\ref{xsecRatio100TeV.fig}].  However, we disagree with the results of Refs.~\cite{Dev:2013wba}, 
 where higher order contributions dominating over the LO DY production at $m_N \geq 200\GeV$ were claimed.
 The discrepancy is attributed to their too low a $p_T^j$ cut that overestimates the contribution of initial-state radiation based on a tree-level calculation.

 \item  We have introduced a systematic treatment for combining initial-state photons from various channels and predict cross sections that are rather stable against the scale choices, typically less than $20\%$.  The exception is the inelastic process, which is rather sensitive to the scale $\lamEl$ where the elastic and inelastic processes are separated. Variation of this scale could lead to about a $30\%$ uncertainty.
  Scale dependence is shown in figure~\ref{scale.fig} and the results summarized in Table~\ref{scale.TB}.
 
\item
We quantified the signal observability by examining the SM backgrounds. 
We conclude that, with the currently allowed mixing {$\vert V_{\mu N}\vert ^2<6\times 10^{-3}$}, 
a $5\sigma$ discovery can be made via the same-sign dimuon channel for {$m_N = 530~(1070)$} GeV at the 14 TeV LHC (100 TeV VLHC) after 1 ab$^{-1}$;
see Table~\ref{massReach.TB}.
Reversely, for $m_N = 500$ GeV and the same integrated luminosity, 
a mixing $\vert V_{\mu N}\vert^2$ of the order {$1.1\times10^{-3}~(2.5\times10^{-4})$} may be probed; see Table~\ref{mixingReach.TB}.
This study represents the first investigation into heavy Majorana neutrino production in 100 TeV $pp$ collisions.

\end{itemize}

\acknowledgments{
We would like to thank
Darin Baumgartel, 
Ayres Freitas,
Jose Kenichi,
Olivier Mattelaer,
Jim Mueller,
Juan Rojo,
Josh Sayre, and Brock Tweedie for valuable discussions.
D.A.~acknowledges the Brazilian agency FAPESP for the support and PITT-PACC for its generous hospitality while doing part of this research. 
The work of T.H.~was supported in part by the U.S.~Department of Energy under Grant No.~DE-FG02-95ER40896 and in part by the PITT-PACC.
R.R.~acknowledges support from the University of Pittsburgh and
the generosity of the Institute for High Energy Physics Center for Future High Energy Physics and the Korea Institute for Advance Science.
}

\hrulefill
\appendix
\label{sec:App}

\section{Elastic Photon PDF}
\label{sec:AppEl}
The elastic photon PDF for a proton is given analytically by~\cite{Budnev:1974de}
\begin{eqnarray}
 \label{elEPA.EQ}
 f_{\gamma/p}^{\rm El }(\xi) &=& \frac{\alpha_{\rm EM}}{\pi}\frac{(1-\xi)}{\xi}
 \left[
 \varphi\left(\frac{{\lamEl}^2}{Q_{0}^2}\right)
 -
  \varphi\left(\frac{Q_{\min}^2}{Q_{0}^2}\right)
 \right],
 \quad \alpha_{\rm EM} \approx 1/137,
 \label{modWWApprox.EQ}
 \\
 Q_{\min}^{2} &=& m_{p}^{2}y, 
 \quad y = \frac{\xi^{2}}{(1-\xi)},
 \quad Q_{0}^2 = 0.71~\text{GeV}^{2},
 \quad m_{p} = 0.938~\text{GeV},
 \\
 \varphi(x) &=& (1+a y) \left[-\log\left(1 + \frac{1}{x}\right) + \sum_{k=1}^{3}\frac{1}{k(1+x)^k}\right]
 + \frac{y(1-b)}{4x(1+x)^3}
 \nonumber\\
 &+& c\left(1+\frac{y}{4}\right)\left[\log\left(\frac{1+x-b}{1+x}\right) + \sum_{k=1}^{3} \frac{b^{k}}{k(1+x)^{k}}\right], 
 \label{varphi.EQ}
 \\
 a&=& \frac{1}{4}(1+\mu_{p}^{2})+\frac{4m_{p}^{2}}{Q_{0}^{2}} \approx 7.16,
 \quad b = 1 -\frac{4m_{p}^{2}}{Q_{0}^{2}} \approx -3.96, 
 \quad c = \frac{\mu_{p}^{2}-1}{b^{4}}\approx 0.028. 
 \end{eqnarray}
 Here, $\lamEl$ is a upper limit on elastic momentum transfers such that $f_{\gamma/p}^{\rm El }=0$ for $Q_{\gamma} >\lamEl$.
In Eq.~(\ref{modWWApprox.EQ}), and later in Eq.~(\ref{WWApprox.EQ}), since $Q_\gamma \ll m_Z$, 
$\alpha(\mu=Q_\gamma)\approx \alpha_{\rm EM} \approx 1/137$ is used.
In the hard scattering matrix elements, $\alpha(\mu = M_Z)$ is used. See Ref.~\cite{Drees:1994zx} for further details.
 
Equation (\ref{elEPA.EQ}) has been found to agree well with data from TeV-scale collisions at $Q_\gamma\sim m_\mu$~\cite{Chatrchyan:2011ci}.
However, applications to cases with larger momentum transfers and finite angles lead to large errors and increase scale sensitivity.
Too large a choice for $\lamEl$ will lead to overestimate of cross sections~\cite{Budnev:1974de}.
However, we observe negligible growth in $f^{\rm El}_\gamma$ at scales well above $\lamEl=1-2\GeV$, in agreement with Ref.~\cite{Sahin:2010zr}.

 Briefly, we draw attention to a typo in the original manuscript that derives Eq.~(\ref{elEPA.EQ}).
This has been only scantly been mentioned in past literature~\cite{deFavereaudeJeneret:2009db,Chapon:2009hh}. 
The sign preceding the ``$y(1-b)$'' term of $\varphi$ in Eq.~(\ref{varphi.EQ}) is erroneously flipped in Eq.~(D7) of Ref.~\cite{Budnev:1974de}.
Both CalcHEP~\cite{Belyaev:2012qa} and MG5\_aMC@NLO~\cite{Alwall:2014hca} have the correct sign in their default PDF libraries.
 
At these scales, the gauge state $\gamma$ is a understood to be a linear combination of 
discrete states: the physical (massless) photon and (massive) vector mesons $(\omega,\phi,...)$, and a continuous mass spectrum,
a phenomenon known as generalized vector meson dominance (GVMD)~\cite{Sakurai:1972wk}.
An analysis of ZEUS measurements of the $F_2$ structure function at $Q^2_\gamma < m_p^2$ and Bjorken-$x\ll1$ concludes that GMVD effects are 
included in the usual dipole parameterizations of the proton's electric and magnetic form factors $G_{\rm E}$ and $G_{\rm M}$~\cite{Alwall:2004wk}.
Thus, the radiation of vector mesons by a proton that are then observed as photons has been folded into Eq.~(\ref{elEPA.EQ}).

\section{Inelastic Photon PDF}
\label{sec:AppIn}
Following the methodology of Ref.~\cite{Drees:1994zx}, the inelastic $N\ell X$ cross section is given explicitly by
\begin{eqnarray}
  \sigma_{\rm Inel}(pp\rightarrow N\ell^{\pm}X &+& \text{anything}) = 
  \sum_{q,q'}
  \int^{1}_{\tau_0}d\xi_{1}\int^{1}_{\tau_0/\xi_1}d\xi_{2}\int^{1}_{\tau_0/\xi_1/\xi_2}dz
  \nonumber\\
  &\times&\left[
  f_{q/p}\left(\xi_1,Q_{f}^2\right) ~  f_{\gamma/q'}\left(z,Q_\gamma^2\right)  ~ f_{q'/p}\left(\xi_2,Q_{f}^2\right)
  \hat{\sigma}\left(q_1\gamma_2\right) + (1\leftrightarrow2)\right],
   \\
 & & \tau_0=m_{N}^{2}/s,	
 \quad	
 \tau = \hat{s} / s = \xi_1\xi_2 z. \nonumber
   \label{inelepa.EQ}
\end{eqnarray}
The Weizs\"acker-Williams photon structure function~\cite{Williams:1934ad,vonWeizsacker:1934sx} is given by
\begin{eqnarray}
 f_{\gamma/q}(z,Q_\gamma^2) = \frac{\alpha_{\rm EM} ~ e_{q}^{2}}{2\pi} 
 \left(\frac{1+(1-z)^2}{z} \right)
 \log\left(\frac{Q_{\gamma}^{2}}{\lamIn}\right),\quad
 \alpha_{\rm EM}\approx 1/137,	
 \label{WWApprox.EQ}
\end{eqnarray}
where $e_{q}^{2} = 4/9 ~ (1/9)$ for up-(down-)type quarks and $\lamIn$ is a low-momentum transfer cutoff.
In DGLAP-evolved photon PDFs~\cite{Martin:2004dh}, $\lamIn$ is taken as the mass of the participating quark.
Ref.~\cite{Drees:1994zx} argues a low-energy cutoff $\mathcal{O}(1-2)$ GeV  so that the 
associated photon is sufficiently off-shell for the parton model to be valid.
As discussed in section~\ref{sec:scale}, taking $\lamIn = \lamEl = \mathcal{O}(1-2)\GeV$ 
allows for the inclusion of non-perturbative phenomena without worry of double counting of phase space.

Fixing $z$ and defining $\xi_\gamma \equiv \xi_2 z$, we have the relationships
\begin{equation}
 \tau_0 = \min\left(\xi_1 \xi_2 z\right) = \min\left(\xi_1 \xi_\gamma \right)
 \implies \min(\xi_\gamma) = \frac{\tau_0}{\xi_1} \text{ for fixed } \xi_1.
\end{equation}
Physically, $\xi_\gamma$ is the fraction of proton energy carried by the initial-state photon.
Eq.~(\ref{inelepa.EQ}) can be expressed into the more familiar two-PDF factorization theorem, i.e., Eq.~(\ref{factTheorem.EQ}),
by grouping together the convolutions about $f_{q'/p}$ and $f_{\gamma/q'}$:
\begin{eqnarray}	
\sum_{q'}
 \int^{1}_{\tau_0/\xi_1}d\xi_{2}\int^{1}_{\tau_0/\xi_1/\xi_2}dz 
 ~ f_{\gamma/q'}(z ) 
 ~ f_{q'/p}(\xi_2 )
 &=& 
 \sum_{q'}
 \int^{1}_{\tau_0/\xi_1}\frac{d\xi_{\gamma}}{z}\int^{1}_{z_{\min}}dz 
~ f_{\gamma/q'}(z )
 ~ f_{q'/p}\left(\frac{\xi_\gamma}{z} \right)
 \\
 &=& 
 \int^{1}_{\tau_0/\xi_1} d\xi_\gamma ~f_{\gamma/p}^{\rm Inel}(\xi_\gamma)
 \\
 f_{\gamma/p}^{\rm Inel}\left(\xi_\gamma,Q_\gamma^2,Q_{f}^2\right)
 &\equiv& 
 \sum_{q'} \int^{1}_{z_{\min} = \xi_\gamma}\frac{dz}{z}  
 ~ f_{\gamma/q'}\left(z, Q_\gamma^2 \right)  
 ~ f_{q'/p}\left(\frac{\xi_\gamma}{z},Q_f^2 \right).~~~~~
\end{eqnarray}
The minimal fraction $z$ of energy that can be carried away by the photon from the quark corresponds to when the quark has 
the maximum fraction $\xi_2$ of energy from its parent proton. Thus, for a fixed $\xi_\gamma$, we have 
\begin{equation}
 1 = \max(\xi_2) = \max\left(\frac{\xi_\gamma}{z}\right) = \frac{\xi_\gamma}{\min(z)} \implies \min(z) = \xi_\gamma.
\end{equation}
The resulting expression is
\begin{eqnarray}
  \sigma_{\rm Inel}(pp\rightarrow N\ell^{\pm}X) = 
  \sum_{q}
  \int^{1}_{\tau_0}d\xi_{1} \int^{1}_{\tau_0/\xi_1} d\xi_2 
  \left[
  ~f_{q/p}\left(\xi_1,Q_{f}^2\right)
  ~f_{\gamma/p}^{\rm Inel}\left(\xi_2,Q_\gamma^2,Q_{f}^2\right) 
  \hat{\sigma}\left(q_1\gamma_2\right) + (1\leftrightarrow2)\right]
\end{eqnarray}

Real, initial-state photons from inelastic quark emissions can be studied in MG5 by linking the appropriate Les Houches accord PDFs (LHAPDF) 
libraries~\cite{Whalley:2005nh} and using the MRST2004QED~\cite{Martin:2004dh} or NNPDF QED~\cite{Ball:2013hta} PDF sets.
With this prescription, sub-leading (but important) photon substructure effects~\cite{Drees:1984cx}, 
e.g., $P_{g\gamma}$ splitting functions, are included in evolution equations.

\section{Poisson Statistics}
\label{sec:stats}
To determine the discovery potential at a particular significance, 
we first translate significance into a corresponding confidence level (CL),\footnote{We use $\sigma$-sensitivity and CL interchangeably in the text.} e.g.,
\begin{equation}
 2\sigma \leftrightarrow 95.45\%~\text{CL},\quad
  3\sigma \leftrightarrow 99.73\%~\text{CL},\quad
   5\sigma \leftrightarrow 99.99994\%~\text{CL}.
\end{equation}
Given an given integrated luminosity $\mathcal{L}$, SM background rate $\sigma_{\rm SM}$, and CL, say 95.45\% CL,
we solve for the maximum number of background-only events, denoted by $n^{b}$, using  the Poisson distribution:
\begin{equation}
 0.9545 = \sum_{k=0}^{n^b} P\left(k \vert \mu^b = \sigma_{\rm SM}\mathcal{L} \right)
	= \sum_{k=0}^{n^b} \frac{(\sigma_{\rm SM}\mathcal{L})^k}{k!} e^{-\sigma_{\rm SM}\mathcal{L}}.
\end{equation}
The requisite number of signal events at a 95.45\% CL (or $2\sigma$ significance) 
is obtained by solving for the mean number of signal events $\mu^s$ such that a mean number of total expected events $(\mu^s + \mu^b)$
will generate $n^{b}$ events only $4.55\%(=100\%-95.45\%)$ of the time, i.e., find $\mu^s$ such that
\begin{equation}
 P\left(k=n^b \vert \mu = \mu^s + \mu^b \right) = \frac{(\mu^s + \mu^b)^{n^b}}{(n^b)!} e^{-(\mu^s + \mu^b)} = 0.0455.
 \label{poiDis.EQ}
\end{equation}
The 2$\sigma$ sensitivity to nonzero $S_{\ell\ell}$ is then
\begin{equation}
 S_{\ell \ell'}^{2\sigma} = \frac{\mu^s}{\mathcal{L} \times \sigma_{\rm Tot~0}}.
\end{equation}
For fixed signal $\sigma_{s}$ and background $\sigma_{\rm SM}$ rates, $\mu^s + \mu^b = (\sigma_{s}+\sigma_{\rm SM})\times\mathcal{L}$.
The required luminosity for a $2\sigma$ discovery can then be obtained by solving Eq.~(\ref{poiDis.EQ}) for $\mathcal{L}$.

\hrulefill




\end{document}